\documentclass{desyprocA4}
\usepackage{units}
\usepackage{color}
\usepackage{rotating,multirow,booktabs}
\usepackage{array}
\newcolumntype{R}[1]{>{\raggedleft\hspace{0pt}}m{#1}}
\definecolor{dblack}   {rgb}{0.00, 0.00, 0.00}  
\definecolor{mgrey}    {rgb}{0.45, 0.45, 0.50}  
\definecolor{rred}     {rgb}{1.00, 0.00, 0.00}  
\definecolor{dgreen}   {rgb}{0.05, 0.65, 0.00}  
\definecolor{bblue}    {rgb}{0.10, 0.10, 0.96}  
\def\to{\rightarrow}

\def\bi{\begin{itemize}}
 \def\ei{\end{itemize}}

\def\c1p{C1^\prime}
\def\msq3{\overline{m}_{\tilde{q}}(3)}

\def\tH{\tilde H}

\def\tH{\widetilde H}

\def\tB{\widetilde B}
\def\tW{\widetilde W}
\def\be{\begin{equation}}  
\def\ee{\end{equation}}  
\def\bea{\begin{eqnarray}}  
\def\eea{\end{eqnarray}}

\def\twpm{\widetilde\chi^\pm}
\def\tz{\widetilde\chi^0}
\newcommand{\drbar}{\overline{\rm DR}}

\newcommand{\mneu}[1]{M_{\widetilde{\chi}^0_{#1}}}
\newcommand{\mcha}[1]{M_{\widetilde{\chi}^\pm_{#1}}}

\newcommand{\gev}{\ \mathrm{GeV}}

\newcommand{\wt}{\widetilde}
\newcommand{\eqn}{equation}
\newcommand{\MeV}{{\ensuremath\rm MeV}}
\newcommand{\GeV}{{\ensuremath\rm GeV}}
\newcommand{\TeV}{{\ensuremath\rm TeV}}
\newcommand{\mO}{\mathcal{O}}

\newcommand{\neu}[1]{\widetilde{\chi}^0_{#1}}

\newcommand{\eeto}    {\ensuremath{ {\, e}^+ {e}^- \to}}
\DeclareMathOperator{\sign}{sign}

\def\beq{\begin{equation}}
\def\eeq#1{\label{#1}\end{equation}}
\def\eeqn{\end{equation}}

\newenvironment{Eqnarray}%
   {\arraycolsep 0.14em\begin{eqnarray}}{\end{eqnarray}}
\def\beqa{\begin{Eqnarray}}
\def\eeqa#1{\label{#1}\end{Eqnarray}}
\def\eeqan{\end{Eqnarray}}

\newcommand{\ra}            {\ensuremath{ \rightarrow     }}

\begin{document}
%------------------------------------
\title{Tackling light higgsinos at the ILC}

%for single authors the superscripts are optional
\author{{\slshape Mikael Berggren$^1$, Felix Br\"ummer$^1$, Jenny List$^1$, Gudrid Moortgat-Pick$^2$, Tania Robens$^3$, Krzysztof Rolbiecki$^{4,5}$, Hale Sert$^{1,2}$ }\\[1ex]
$^1$DESY, Notkestra{\ss}e 85, 22607 Hamburg, Germany\\
$^2$University of Hamburg, Physics Department, Luruper Chaussee 149, 22761 Hamburg, Germany\\
$^3$IKTP, TU Dresden, Zellescher Weg 19, 01069 Dresden, Germany\\
$^4$IFT-UAM/CSIC, C/ Nicol\'{a}s Cabrera 13--15, 28049 Madrid, Spain\\
$^5$IFT, University of Warsaw, ul.\ Ho\.{z}a 69, 00681 Warsaw, Poland 
}
% please enter the contribution ID for the DOI
%\contribID{xy}

% TO THE CONFERENCE EDITORS: 
% please update the following information      
% before sending the template to the authors
%\confID{4980}  % if the conference is on Indico uncomment this line
%\desyproc{LC-xxx, DESY 13-098}
\desyproc{DESY 13-098}
%\acronym{PLHC2010} % if you want the Acronym in the page footer uncomment this line
\doi  % if there is an online version we will register DOIs

\maketitle

%\pubdate

 \begin{abstract}
In supersymmetric extensions of the Standard Model, higgsino-like charginos and
neutralinos are preferred to have masses of the order of the electroweak scale by
naturalness arguments. 
Such light $\widetilde{\chi}^0_1$, $\widetilde{\chi}^0_2$ and 
$\widetilde{\chi}^{\pm}_1$ states can be almost mass degenerate, and their
decays are then difficult to observe at colliders. 
In addition to the generic naturalness argument,
light higgsinos are well motivated from a top-down perspective.
For instance, they arise naturally in certain models
of hybrid gauge-gravity mediation. 
In the present analysis, we study
two benchmark points which have been derived in the framework of such a model, which exhibit mass differences of $\mO(\GeV)$ in the higgsino sector.
For chargino pair and neutralino associated production with initial-state photon radiation, 
we simulate the detector response and determine how accurately the small mass differences, 
the absolute masses and the cross sections can be measured at the International Linear Collider. 
Assuming that 500 fb$^{-1}$ has been collected at each of two
beam-polarisations 
$P(e^+,e^-)=(\pm 30\%,\mp 80\%)$,
we find that the mass differences
can be measured to 40--300 MeV,
the cross sections to 2--5\%, and the absolute masses to 1.5--3.3 GeV,
where the range of values correspond to the different scenarios and
channels.
Based on these observables we perform 
a parameter fit in the MSSM, from which we infer that the higgsino mass parameter
$\mu$ can be measured to a precision of about $\Delta\mu=2$--$7$~GeV. 
For the electroweak
gaugino mass parameters $M_1$, $M_2$, which are chosen in the multi-TeV range,
a narrow region is compatible with the measurements.
For both parameters independently, we can determine a lower bound.  
 \end{abstract}

\section{Introduction}
\label{sec:intro}

The LHC experiments have discovered a Standard Model (SM)-like Higgs
 boson with $m_h\approx 126$~GeV~\cite{Aad:2012tfa,Chatrchyan:2012ufa}.
 However, it is not clear whether this newly discovered particle is a 
pure SM Higgs boson, or whether
 it is part of an extension of the SM, which might also include 
multiple or composite Higgs-like states. Precise measurements of the Higgs
 couplings and branching ratios will be required to resolve this
 question. Another, more direct way to gain insights about physics beyond
the SM (BSM) would obviously be to discover more new particles directly.

A well-motivated extension of the SM is the Minimal
Supersymmetric SM (MSSM), a fully renormalisable theory of
 high predictive power. In this model,  the
U(1), SU(2) and SU(3) gauge couplings can unify, and a dark matter candidate
can be naturally accommodated.
Furthermore, supersymmetry stabilises the Higgs potential with respect to
quadratically divergent radiative corrections. In the MSSM the Higgs mass
is no longer a free parameter but becomes strongly constrained. 

If nature is supersymmetric at the electroweak
scale, a large number of new superpartner mass parameters is introduced
whose values depend on the unknown mechanism of supersymmetry breaking
at high energies. This poses a challenge for any phenomenological analysis. 
Due to the large number of unknown parameters it is very difficult to make reliable 
predictions for SUSY background processes, and in many cases the worst 
background contributions for SUSY signals originate in fact from SUSY processes.

The experiments at the LHC have not yet found any hints for
supersymmetry in squark and gluino searches~\cite{bib:atlas,bib:cms},
which points towards a rather heavy coloured superpartner spectrum with
soft masses above a \unit{TeV}.  However, there are large regions in the
MSSM parameter space which lead to a SM-like Higgs with $m_h=126$~GeV, a
light electroweak SUSY sector, and a rather heavy coloured SUSY sector
beyond the current LHC discovery range. In general, if the electroweak
sector is light enough, it should be possible to discover
neutralinos and charginos via Drell--Yan production at the
LHC~\cite{ATLAS-future,CMS-future}, and the LHC experiments have already put limits on the electroweak-ino parameter space under several simplified model assumptions \cite{LHC-Drell-Yan,CMS-EW}. However, in cases where these light states
are very close in mass and their decay products are therefore very soft,
it will be almost impossible to observe such processes at the LHC. 
An example is given by scenarios in which the higgsino mass parameter 
$\mu$ is much smaller than the electroweak gaugino masses $M_1$ and $M_2$
(current bounds require only $|\mu|\gtrsim\unit[100]{GeV}$ in order to avoid 
the limits from LEP chargino searches~\cite{Beringer:1900zz}). 
Then, the new light states are a chargino and two neutralinos which are 
higgsino-like and almost mass degenerate. The chargino and the heavier 
neutralino therefore decay only into extremely soft SM particles and the 
invisible lightest neutralino. 

The most ``natural'' versions of the phenomenological MSSM which are still 
not ruled out are indeed characterised by higgsino masses of a few 100 
\unit{GeV}, together with sub-\unit{TeV} third-generation squarks and a
moderately heavy gluino, while the rest of the spectrum can be in the multi-TeV
range \cite{Papucci:2011wy}. Models with light higgsinos are also attractive for 
cosmology, either in combination with gravitino dark matter or with a 
higgsino-like neutralino as a non-thermally produced lightest supersymmetric
particle (LSP); see e.g.~\cite{Allahverdi:2012wb}.

While the light higgsino scenario is extremely challenging for the 
LHC, it may well be observable in the clean
environment at an $e^+e^-$ Linear Collider, for instance, at the
International Linear Collider (ILC)~\cite{Chen:1995yu, Baer:2011ec, Baer:2012uy}. 
At the ILC, the striking feature of a  
precisely known initial state allows to successfully apply the initial-state radiation (ISR) method,  
i.e.\ to select only those signal events that are accompanied by a 
sufficiently hard photon.
In this context, polarised beams help to significantly enhance the signal
cross section, at least in the chargino case.

The subject of the present work is the phenomenology of light
quasi-degenerate higgsinos at the ILC. We study their production and decay, and 
investigate in how far one can derive the fundamental MSSM 
parameters without assuming any specific SUSY breaking scheme
if only the light higgsinos are kinematically accessible.
We study the processes $e^+ e^-\to
\widetilde{\chi}^+_1\widetilde{\chi}^-_1$ and $e^+e^-\to
\widetilde{\chi}^0_1\widetilde{\chi}^0_2$ at $\sqrt{s}=500$~GeV with polarised
beams and initial-state radiation at the ILC. We simulate the 
detector response for two MSSM benchmark points 
with different ${\cal O}(1$ GeV) mass splittings between the chargino
and neutralinos, leading to different final state SM particles.
 The uncertainties on masses, mass differences and 
cross sections as determined by our simulation study are then used as inputs for a fit
in order to reconstruct the MSSM parameters 
$M_1$, $M_2$, $\mu$ and $\tan\beta$. Our overall goal is to determine both how 
well the chargino and neutralino masses can be measured at the ILC, and what 
could be inferred from such a measurement about the underlying SUSY model.

In the following chapters we first give an overview of the theoretical
arguments in favour of light higgsinos and the phenomenological 
aspects of such scenarios in the MSSM.
The expected SM background will also be discussed, including 
beam conditions and detector capabilities at the ILC.
In Section 3 we define two benchmark scenarios on which our study will be
based and provide the details for the detector simulation. In
Section 4 we discuss the expected experimental results at the ILC. In Section
5 we present our results on reconstructing the fundamental parameters
 and the precision we expect to achieve with our analysis strategy.

\section{Theoretical context and analysis strategy}\label{theo-context}
There are two main reasons why higgsinos are distinguished among the
MSSM superpartners of the SM particles. First, the size of the higgsino
mass parameter $\mu$ is not necessarily related to the scale of SUSY 
breaking, as it is allowed by unbroken supersymmetry
(it is indeed the only dimensionful MSSM parameter with this property). 
It can therefore be generated
independently of the soft supersymmetry breaking terms, and at leading order does not
receive any soft-term-dependent radiative corrections in the evolution from the high scale. Second,
$\mu$ enters the tree-level Higgs potential and is therefore directly
connected to the electroweak scale. For instance, the tree-level $Z$ 
boson mass is given in terms of the Higgs soft masses $m^2_{H_u,H_d}$, 
the ratio $\tan\beta$ of Higgs vacuum expectation values, and $\mu$ as
\begin{equation}
m_Z^2\;=\;2\frac{m_{H_u}^2
 \tan^2\beta-m_{H_d}^2}{1-\tan^2\beta}-2\,|\mu|^2\;\approx\; -2\,(m_{H_u}^2 +|\mu|^2)\,.
\end{equation}
To avoid finely tuned large cancellations among these parameters, it
follows that $\mu$ should not be too far above the electroweak scale,
even if some or all of the squark, slepton and gaugino masses are much
larger. 
These two arguments are the main motivation for investigating models where
 the higgsinos are light, while the other MSSM superpartners may be
much heavier. 

\subsection{Light higgsinos from UV-scale models }\label{sec:model}

From a top-down perspective, light higgsinos are equally well
motivated. For instance, they are generic in certain models of
hybrid gauge-gravity mediation~\cite{Brummer:2011yd,Brummer:2012zc},
whose main features we will now briefly recapitulate. In gauge-mediated
SUSY breaking, some of the hidden sector states carry SM gauge
charges. These messenger fields acquire large supersymmetric masses $M$,
and messenger loops generate soft terms of the order
 \begin{equation}
  m_{\rm soft}\sim\frac{n}{16\pi^2}\frac{F}{M}\,,
 \end{equation}
 where $\sqrt{F}$ is the scale of SUSY breaking, and $n$ roughly counts
 the number of contributing messenger states. In addition to these,
 generically there are also gravity-mediated contributions to soft terms
 as well as a gravity-mediated effective $\mu$ parameter, generated by
 gravitationally suppressed interactions between the hidden sector and
 the MSSM. They are of the order of the gravitino mass,
 \begin{equation}
 m_{3/2}\sim\mu\sim\frac{F}{M_{\rm Planck}}\,.
 \end{equation}
 In models where the MSSM is embedded in a more fundamental theory, for
 instance in superstring compactifications, the details of supersymmetry
 breaking are often unknown, while the spectrum of sub-Planckian states
 can still be calculated. In phenomenologically promising models, these
 typically consist of the MSSM along with a multitude of additional
 exotic states, some of which are charged under the SM gauge groups
 (see e.g.~\cite{Nilles:2008gq, Maharana:2012tu} and references therein). These exotic
 fields can be given large masses and act as gauge mediation
 messengers. A natural scale for them to decouple is the scale of the Grand
 Unification,
 \begin{equation}
  M\sim M_{\rm GUT}=\unit[2\cdot 10^{16}]{GeV}\approx\frac{1}{16\pi^2}M_{\rm Planck}\,.
 \end{equation}
 With some simple assumptions about the couplings of the messengers to
 the SUSY-breaking fields, and choosing parameters such that $m_{3/2}$ is
 of the order of the electroweak scale, one can obtain a realistic soft
 term spectrum. For large messenger multiplicities $n\gtrsim 10$, the
 soft terms are predominantly generated by gauge mediation and can be in
 the multi-TeV range, while $\mu$ is induced by gravity mediation and of
 the order of $\unit[$100$-$200$]{GeV}$.

The benchmark points we are using for the present analysis are derived
from such hybrid gauge-gravity mediation models with large messenger
numbers. However, we stress that these are by no means the only
possibility to obtain light higgsinos from high-scale models, and that
our results on the light higgsino phenomenology are independent
of a specific SUSY breaking model.

%\vspace{3mm}

\subsection{Phenomenology}\label{sec:pheno}
In a more general context, scenarios with a rather heavy coloured sector 
and a mass-degenerate light chargino/neutralino sector can easily be 
constructed in the unconstrained MSSM, where the soft SUSY-breaking parameters 
are taken as independent. One simply chooses sufficiently large values for the 
squark and gluino masses (and possibly also for the slepton masses).
 
At the tree level the chargino/higgsino sector~\cite{Haber:1984rc} 
is determined by the U(1) bino mass parameter $M_1$, the SU(2) wino mass
parameter $M_2$, the higgsino mass parameter $\mu$ and $\tan\beta$.  The
neutralino mass matrix in the
$(\widetilde{B},\widetilde{W}^0,\widetilde{H}_d^0,\widetilde{H}_u^0)$ basis is given by
\begin{equation}\label{eq:Y}
Y =\left( \begin{array}{cccc}
M_1 & 0 & -m_Z \cos\beta\sin\theta_W & m_Z\sin\beta\sin\theta_W \\
0   & M_2 & m_Z\cos\beta\cos\theta_W & -m_Z\sin\beta\cos\theta_W \\
-m_Z\cos\beta\sin\theta_W & m_Z\cos\beta\cos\theta_W & 0 & -\mu \\
m_Z\sin\beta\sin\theta_W & -m_Z\sin\beta\cos\theta_W & -\mu & 0 \end{array} \right)\,.
\end{equation}
Since $Y$ is complex symmetric, it can be diagonalised by one unitary matrix $N$ 
via $\mathbf{M}_{\widetilde{\chi}^0}=N^*Y N^\dag$ (where the matrix 
${\bf M}_{\widetilde{\chi}^0}$ contains the neutralino masses $M_{\widetilde\chi^0_i}$
on the diagonal). The mixing matrix $N$ relates the  neutralino mass eigenstates
to the neutral gauge eigenstates. For full analytical solutions of the system see~\cite{neut-analyt,Choi:2001ww}.

The chargino mass matrix in the $(\widetilde{W}^+,\widetilde{H}^+)$ basis, 
\begin{equation}\label{eq:X}
X=
\left( \begin{array}{cc}
M_2 & \sqrt{2} m_Z\cos\theta_W\sin\beta  \\
\sqrt{2} m_Z\cos\theta_W\cos\beta  & \mu
\end{array} \right)\;,
\end{equation}
is not symmetric and therefore has to be diagonalised
via the bi-unitary transformation 
$\displaystyle\mathbf{M}_{\widetilde{\chi}^+}=U^* X V^\dag$. $U$ and $V$
relate the chargino mass eigenstates to the charged gauge eigenstates. In general, $M_1$, $M_2$ and $\mu$ are complex-valued parameters. By a suitable rotation of the gaugino and higgsino fields, $M_2$ can be made real and positive without loss of generality. In the following we assume that also $M_1$ and $\mu$ are real (positive or negative) so no new sources of CP violation are introduced. 

One can easily classify three limiting cases:
in the case $|M_1| < M_2\ll |\mu|$, one obtains $|M_1|\approx M_{\widetilde{\chi}^0_1}$
with a bino-like $\tz_1$,
$M_2\approx M_{\widetilde{\chi}^0_2}, M_{\widetilde{\chi}^{\pm}_1}$ with wino-like
$\tz_2$ and $\twpm_1$,
and higgsino-like heaviest states with masses
$|\mu|\approx  M_{\widetilde{\chi}^0_{3,4}}$,  $M_{\widetilde{\chi}^{\pm}_2}$,
cf.\ Eqs.~(\ref{eq:Y}) and (\ref{eq:X}). By contrast,
the hierarchy $M_2<|M_1|\ll |\mu|$ leads to the lightest two states being mass degenerate 
and wino-like, $M_2\approx M_{\widetilde{\chi}^0_1}, M_{\widetilde{\chi}^{\pm}_1}$
but a bino-like $\tz_2$ with mass $|M_1|\approx  M_{\widetilde{\chi}^0_2}$,
while $\tz_{3,4}$ and $\twpm_2$ remain higgsino-like and heavy as before. 

Choosing $|\mu|\ll |M_1| < M_2$ (or $|\mu|\ll M_2 < |M_1|$) leads to mass-degenerate 
lightest states with
$|\mu|\approx M_{\widetilde{\chi}^0_{1,2}}, M_{\widetilde{\chi}^{\pm}_1}$
but $|M_1|\approx M_{\widetilde{\chi}^0_{3}}$ (or $|M_1|\approx M_{\widetilde{\chi}^0_{4}}$)
and $M_2\approx M_{\widetilde{\chi}^{\pm}_2}, M_{\widetilde{\chi}^0_{4}}$ 
(or $M_2\approx M_{\widetilde{\chi}^{\pm}_2}, M_{\widetilde{\chi}^0_{3}}$). The lighter neutralinos 
and the lightest chargino are then mostly higgsino-like and have unsuppressed couplings to 
the SM gauge bosons. This is the scenario which we are investigating.

It is instructive to explicitly write down the tree-level masses of the three
light states in
the case that $M_1$ and $M_2$ are large. They are found by diagonalising the
mass matrices of Eqs.~\eqref{eq:Y} and \eqref{eq:X}, and given by
\begin{equation}\label{eq:masses}
 \begin{split}
M_{\tz_{1,2}}&=\eta_{1,2}\left(|\mu|\mp\frac{m_Z^2}{2}\left(1\pm\sin2\beta\,\sign(\mu)\right)
\left(\frac{\sin^2\theta_W}{M_1}+\frac{\cos^2\theta_W}{M_2}\right)\right)\,,\\
  M_{\twpm_1}&=|\mu|-\sin 2\beta\;\sign(\mu)\;\cos^2\theta_W\;\frac{m_Z^2}{M_2}\,,\\
 \end{split}
\end{equation}
up to terms suppressed by higher powers of $M_i$, see e.g.~\cite{Gunion:1987yh}. %The neutralino mass eigenvalues are
%given by $\eta_{\tilde{\chi}^0_{1}}M_{\tz_{1}}$, $\eta_{\tilde{\chi}^0_{2}}M_{\tz_{2}}$, where 
The factors
$\eta_{1,2}$ denote the sign of the mass eigenvalue, that depends on the CP-quantum number of the corresponding neutralino~\cite{Haber:1984rc,neut-analyt}.

The wino and bino admixtures are small, of the order $m_Z/M_{1,2}$. For instance, for $\mu>0$,  the mass eigenstates are decomposed as
\begin{equation}\label{higgsinomixing}
\begin{split}
 \chi^0_1=&\;\frac{1}{\sqrt{2}}\left(\tH_d^0-\tH_u^0\right)+\frac{\sin\beta+\cos\beta}{\sqrt{2}}\frac{m_Z}{M_1}\sin\theta_W\;\tB-\frac{\sin\beta+\cos\beta}{\sqrt{2}}\frac{m_Z}{M_2}\cos\theta_W\;\tW^0\,,\\
 \chi^0_2=&\;\frac{1}{\sqrt{2}}\left(\tH_d^0+\tH_u^0\right)-\frac{|\sin\beta-\cos\beta|}{\sqrt{2}}\frac{m_Z}{M_1}\sin\theta_W\;\tB+\frac{|\sin\beta-\cos\beta|}{\sqrt{2}}\frac{m_Z}{M_2}\cos\theta_W\;\tW^0\,,\\
 \chi^+_1=&\;\tH_u^+-\sqrt{2}\,\sin\beta\frac{m_W}{M_2}\;\tW^+\,,\\
 \chi^-_1=&\;\tH_d^--\sqrt{2}\,\cos\beta\frac{m_W}{M_2}\;\tW^-\,,
\end{split}
\end{equation}
where we have only written the respective leading-order terms of the Weyl gaugino and higgsino components.
For multi-TeV gaugino masses, the gaugino admixtures are of the order of a few percent, and the mass splittings are of the order of $1$--$10$~GeV. Explicitly, the tree-level mass difference between the lightest neutralino and the lightest chargino reads, again for $\mu>0$,
\begin{equation}\label{eq:chadiff}
M_{\widetilde{\chi}_1^\pm} - M_{\widetilde{\chi}_1^0} =  \frac{m_Z^2}{2}
\left[\sin2\beta\left(\frac{\sin^2\theta_W}{M_1}-\frac{\cos^2\theta_W}{M_2}\right)
 +\left(\frac{\sin^2\theta_W}{M_1}+\frac{\cos^2\theta_W}{M_2}\right)
 + \mathcal{O}\left(\frac{\mu}{M_i^2}\right)\right] \,,
\end{equation}
which for $\tan\beta\gg 1$ further simplifies to
\begin{equation}
 M_{\widetilde{\chi}_1^\pm} - M_{\widetilde{\chi}_1^0} =  \frac{m_Z^2}{2}\left( \frac{\sin^2\theta_W}{M_1} + \frac{\cos^2\theta_W}{M_2} \right) + \mathcal{O}\left(\frac{\mu}{M_i^2},\frac{1}{\tan\beta}\right)\,.
\end{equation}
For the neutralinos, the mass difference is
\begin{equation}\label{eq:neudiff}
M_{\widetilde{\chi}_2^0} - M_{\widetilde{\chi}_1^0} = m_Z^2 \left( \frac{\sin^2\theta_W}{M_1} + \frac{\cos^2\theta_W}{M_2} \right) + \mathcal{O}\left(\frac{\mu}{M_i^2}\right)\,.
\end{equation}
Remarkably, it does not depend on $\tan\beta$ in the leading
approximation. The corresponding expressions for $\mu<0$ are similar and 
the phenomenology of that parameter range has been discussed in~\cite{Ambrosanio:1996gz}. In our case the sign of $\mu$ has negligible impact on observables. Equations~\eqref{eq:masses}--\eqref{eq:neudiff} hold for positive and negative values of $M_1$. It is interesting to note that negative values of $M_1$ allow for cancellations of wino and bino contributions to neutralino masses. Generally, they will result in different mass patterns depending on the sign of $M_1$. In particular, for low values of $\tan\beta$ the lightest neutralino can be heavier than the light chargino, i.e.\ $M_{\widetilde{\chi}_1^\pm} < M_{\widetilde{\chi}_1^0}$. Both possibilities, $M_1 > 0$ and $M_1 < 0$, will be analysed in Section~\ref{sec:paradet}.

Comparing with the analytical formulae of the neutralino (chargino)
masses and couplings \cite{Choi:2001ww,Choi:2000hb}, it is obvious that one can express the
neutralino and chargino sector as functions of $\sin 2 \beta$ and $\cos
2\beta$. In the large $\tan\beta$ limit, $|\cos2\beta| \to 1$ and
$\sin2\beta \to 2/\tan\beta$, therefore all neutralino and chargino
masses as well as the cross sections are only weakly dependent on
$\tan\beta$. This weak dependence is also reflected in our final results, cf.\ Section~\ref{sec:paradet}.

Radiative corrections to the chargino and neutralino observables, see e.g.~\cite{Bharucha:2012nx}
(and references therein for a more complete list of studies), can be sizeable and increase the mass splittings by several hundreds of MeV. These corrections will lift the degeneracy between higgsino states. If the tree-level mass difference is substantially less than a GeV, a correction of 200~MeV~\cite{Aoife-priv} will have a profound impact on the lifetime and decay branching ratios. This may also be exploited to better determine the underlying MSSM parameters, see~\cite{Bharucha:2012ya}. A detailed analysis of NLO corrections in this regard, is, however, 
beyond the scope of this study and in the following we will assume tree-level relations.

\begin{figure}[t]
%  \begin{center}
\centering
\includegraphics[width=0.6\textwidth]{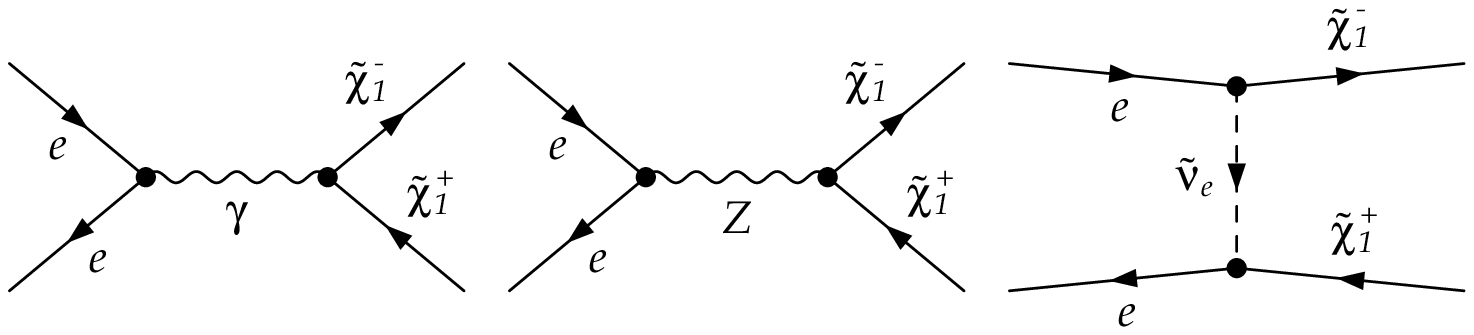}
\includegraphics[width=0.6\textwidth]{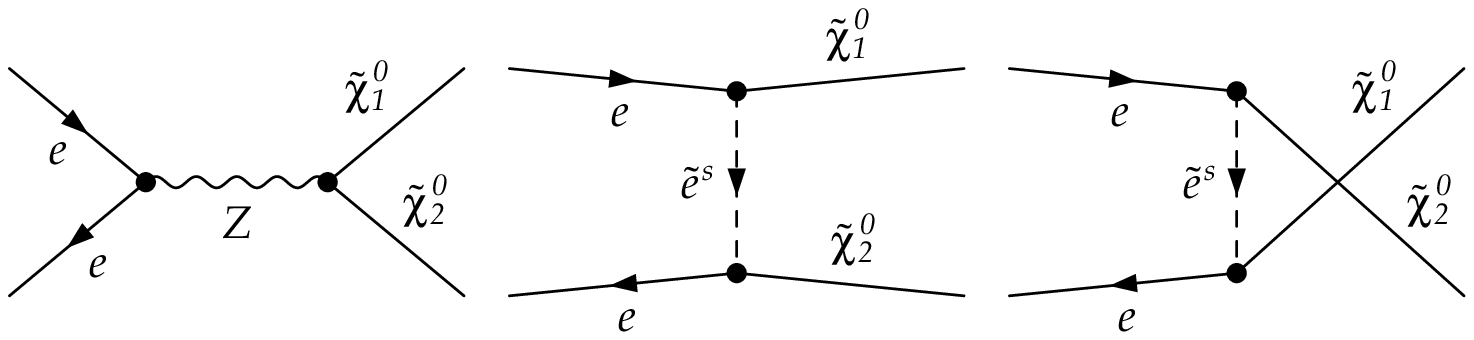}
% \end{center}
  \caption{Chargino (upper row) and neutralino (lower row) production graphs at tree level. Note that $t$- and $u$-channel selectron/sneutrino exchange graphs are negligible in our scenario due to a small Yukawa coupling between sleptons and higgsinos. \label{fig:chaprod}}
\end{figure}

At the ILC higgsino production proceeds via $Z$ or $\gamma$ exchange in
the $s$-channel, see Fig.~\ref{fig:chaprod}. In general, $t$- and $u$-channel exchange 
of selectrons or sneutrinos is also possible, but in the higgsino
case they are highly suppressed due to the small Yukawa couplings. Hence, even
if the selectrons and sneutrinos were relatively light, they would not
affect any observables.
Higgsinos will be produced in $\widetilde\chi_1^0\widetilde\chi_2^0$ or
$\widetilde\chi_1^+\widetilde\chi_1^-$ pairs (the
$Z\widetilde\chi^0_1\widetilde\chi^0_1$ and $Z\widetilde\chi^0_2\widetilde\chi^0_2$
couplings vanish in the pure higgsino limit).\footnote{This is a
  consequence of the fact that $\tH_u^0$ and $\tH_d^0$ have opposite
  weak isospin, and that both $\tz_{1}$ and $\tz_2$ contain equally
  large $\tH_u^0$ and $\tH_d^0$ admixtures as seen from
  Eq.~\eqref{higgsinomixing}.} Both $\widetilde\chi^0_2$ and
$\widetilde\chi^\pm_1$ will undergo three-body decays into
$\widetilde\chi^0_1$ via off-shell $Z$ and $W$ bosons~\cite{three-body}. The decay
$\widetilde\chi^0_2\,\rightarrow\, W^{\pm *}\widetilde\chi^\mp_1$ is also
possible but kinematically highly suppressed. The detector signature comprises
exclusively a few low-energetic leptons or hadrons, whose
softness poses a major challenge to experiment. For small mass
splittings, the radiative decay
$\widetilde\chi^0_2\,\rightarrow\,\widetilde\chi^0_1\gamma$ via a $W$-chargino
loop can also become very important~\cite{Haber:1988px}. More details
on higgsino production and decay will be given below when we discuss
the details of our simulation.

\subsection{Standard Model background}
For the discussion of the SM processes at the ILC
it is convenient 
to classify them
by the initial state and the number of final state fermions.
Due to the very strong
fields between the colliding bunches,
there is a sizeable photon component from back-scattered
synchrotron radiation in the ILC beams~\cite{Yokoya:1991qz,Chen:1991wd}.
Therefore,
the initial states can be $e^+e^-$, but also $\gamma\gamma$, or
$\gamma e^\pm$. For  $e^+e^-$ and $\gamma\gamma$, the final state has
an even number of fermions: 2f, 4f, 6f, ... , while
the $\gamma e^\pm$ initial state gives rise to final states with
an odd number of fermions: 1f, 3f, 5f, ...
In this notation, $\gamma$
can represent both real beam photons or virtual ones. For initial
virtual photons, the final state is understood to also contain the
electron and/or positron that emitted the photon (the ``beam-remnant"), 
in addition to
the fermion(s).

The signal comprises events with a few low-energy charged
particles and no other activity.
Possible SM processes giving this topology are
of two types. Either ({\it i}) events with large amounts of energy
in invisible neutrinos, or ({\it ii})  events where a large fraction
of the energy escapes with particles going in un-instrumented
regions of the detector; in an ILC detector, the only such
regions are the
holes in the low-angle calorimeter through which the beam pipes enter
the detector. %\TeV

The first background type mainly contains events with two $\tau$s,
possibly together with additional neutrinos, 
i.e.\ the subset of the 
classes \eeto~2f and \eeto~4f where 
one of the fermion pairs
are $\tau$ leptons, and the other one (if present) is a neutrino pair.
These channels do not have
cross sections much larger than the signal.
In addition, only a small fraction of the
$\tau$s will decay to the very soft visible system which is
typical for the signal. Hence, while it is not negligible,
this background type is expected to be manageable
without too strong cuts on the event-topology.
 
The potentially most severe background processes are of type ({\it ii}),
and are expected
to be the high cross section multi-peripheral process 
$\eeto e^+e^-\gamma^*\gamma^* \rightarrow  e^+e^- f\bar{f}$,
i.e.\ the $\gamma\gamma \rightarrow$ 2f class.
In such events, the final beam-remnant electrons and positrons
carry nearly the full beam energy in most cases, have
low transverse momentum,
and therefore leave the detector unseen through the outgoing
beam pipe. 
The $f\bar{f}$-pair, on the other hand,
can be emitted at any angle, and tends to have low momentum.
In other words, these events have a large resemblance to
the signal.
However, if one of the incoming electrons emits an ISR
photon at an angle and energy high enough to be detected,
this electron will recoil against the photon, changing its
direction sufficiently to deflect it into the acceptance of the
detector.
In order to reduce
this background to acceptable levels,
we therefore require that the accepted signal candidate
events should contain a detected ISR photon.
Also the other component of the $\gamma\gamma \rightarrow$ 2f class,
where the
initial photons are real photons from the photon component
of the beams, is suppressed by this requirement. This is simply due 
to the fact that, in this case, there are no initial
charged particles that could emit an ISR photon.

Instead, the most severe remaining background is expected to arise from the 
$e\gamma \rightarrow$ 3f class,
in particular from the $t$-channel diagram with incoming
real photons. 
This process is similar to Compton scattering, with the difference
that the scattered photon is virtual, and gives rise to
an $f\bar{f}$-pair. 
In this case, the hard ISR photon does not recoil against the beam electron, 
but against the $f\bar{f}$-system, like in the signal events.
However, in the event as a whole, a large amount of energy is carried away by the 
{\it exchanged} electron once it has been
brought back on-shell by the beam-photon. This electron will receive most of
the photon's energy, and closely follow its direction in the lab frame,
i.e.~along the beam pipe.
As the second incoming particle is neutral in this case, no ISR will be
emitted from it, and thus it cannot be deflected into the detector acceptance either.

\section{Benchmark points and simulation}
For our simulation we chose two benchmark points with the lightest Higgs mass which is compatible with the LHC discovery (within the theory uncertainty), namely $124$ and $127$ GeV. We will refer to them as dM1600 and dM770, according to their chargino--LSP mass difference. In the following sections we will discuss their low-scale spectrum and the higgsino production and decay modes at the ILC as well as their implementation into our simulation. We especially comment on the decay
modes and branching ratios in the nearly mass-degenerate $\wt{\chi}_1^{\pm},\,\wt{\chi}_{1,2}^{0}$ system.

Beyond the two specific mass differences considered here, it has been shown recently that the
ISR technique and the selection strategy described in the following are also sensitive in case of larger mass differences of a few GeV~\cite{Berggren:2013bua}. As soon as the chargino--LSP mass difference becomes even larger, analyses without ISR requirement start to gain sensitivity and eventually take over. For mass differences smaller than the ones considered here, striking signatures like displaced vertices, kinks, anomalous ionisation etc.\ will make searches experimentally more easy. In this sense, the benchmarks chosen here fall into the experimentally most challenging regime.

\begin{table}
\begin{\eqn*}
\begin{array}{lccccccccc}\toprule
&n_1&n_2&n_3&m_{\text{GM}}&m_A&\tan\beta&\mu^{\drbar}&M_{1}^{\drbar}&M_{2}^{\drbar}\\ \midrule
{\mathrm{dM1600}} &20&28&11&190\,\GeV&2160\,\GeV&43.81&160\,\GeV&1.72\,\TeV&4.33\,\TeV\\
{\mathrm{dM770}} &46&46&20&250\,\GeV&4050\,\GeV&47.66&160\,\GeV&5.37\,\TeV&9.47\,\TeV\\ \bottomrule
\end{array}
\end{\eqn*}
\caption{\label{tab:highspec} Input parameters and running gaugino masses for the benchmark points. 
Here
 $n_{1,2,3}$ refers to the messenger indices for $\mathrm{U}(1)_Y$, $\mathrm{SU}(2)_L$, and $\mathrm{SU}(3)_C$, respectively, $m_{\text{GM}}$ is the characteristic gauge-mediated soft mass per messenger pair. $\mu_{\drbar}$ is the running higgsino mass parameter in the $\drbar$ scheme evaluated at the soft mass scale (as defined e.g.~in \cite{Allanach:2001kg});
 it generally differs from the on-shell $\mu$-parameter~\cite{Fritzsche:2002bi}
by a few percent. Moreover, $m_A$ is the pole mass of the pseudoscalar Higgs, and $\tan\beta$ is the ratio of Higgs vacuum expectation values. For details about the precise definitions of $n_i$ and $m_{\rm GM}$ see~\cite{Brummer:2012zc}. We also give the $\drbar$ values of $M_1$ and $M_2$ at the soft mass scale, which
are determined at the high scale from the input parameters by the usual minimal gauge mediation relations.}
\end{table}

\subsection{Spectrum generation, higgsino production and decay  }
\label{sec:proddecay}

\subsubsection*{Spectrum generation}

The general tree-level spectrum of our model was calculated using the
spectrum generator {\tt SOFTSUSY}~\cite{Allanach:2001kg}. The model we are considering is characterised by the high-scale parameters specified in Table~\ref{tab:highspec}. 
The parameters in the gaugino, squark, and slepton sector at the high scale are obtained from Eqs.~(3) and (4) in~\cite{Brummer:2012zc}, while the Higgs sector parameters are computed 
such that both the model predictions at the high scale are matched and electroweak symmetry breaking at the proper scale is obtained.  With the input parameters 
in Table~\ref{tab:highspec}, we obtain a spectrum which is mostly at the multi-\TeV\ scale, with only the two lightest neutralinos as well as the light chargino masses within current collider ranges, and the SM-like Higgs boson with a mass within a $\pm 2\,\GeV$ theory uncertainty interval from the LHC central value of $125.5\,\GeV$. 

The physical chargino and neutralino masses are calculated by {\tt SOFTSUSY} using the running mass parameters 
$\mu^{\drbar}$, $M_{1}^{\drbar}$ and $M_{2}^{\drbar}$ as given in Table~\ref{tab:highspec}, which 
includes one-loop corrections~\cite{Pierce:1996zz}. The light chargino and neutralino spectrum, 
and the underlying on-shell $\mu$ and $M_{1,2}$ parameters\footnote{The calculation of 
on-shell parameters follows Refs.~\cite{Bharucha:2012nx} and \cite{Fritzsche:2002bi}. Three 
particles from the chargino/neutralino sector are chosen to be on-shell, e.g.\ $\wt{\chi}^{\pm}_1$,
 $\wt{\chi}^{\pm}_2$ and $\wt{\chi}^{0}_3$. The most convenient choice should include 
 ``representative'' states for each of the components, bino, wino and higgsino. This choice 
 guarantees that the NLO corrections will be small~\cite{Bharucha:2012nx}. Next, for fixed 
 $\tan{\beta}$ and using tree-level relations, see Eqs.~\eqref{eq:Y} and \eqref{eq:X}, one calculates on-shell MSSM parameters. The masses of the remaining charginos and/or neutralinos will be shifted by appropriately calculated radiative corrections.} are given by 
\begin{eqnarray}
{\mathrm{dM1600}} &&M_{1}\,=\,1.70\,\TeV,\;M_{2}\,=\,4.36\,\TeV,\;\mu\,=\,165.89\,\GeV,\;\tan\beta|_{m_Z}\,=\,44, \label{124}\\
&&M_{\wt{\chi}^{\pm}_1}\,=\,165.77\,\GeV,\;M_{\wt{\chi}^{0}_1}\,=\,164.17\,\GeV,\;M_{\wt{\chi}^{0}_2}\,=\,166.87\,\GeV,\;m_h\,=\,124\,\GeV;\nonumber\\
{\mathrm{dM770}} &&M_{1}\,=\,5.30\,\TeV,\;M_{2}\,=\,9.51\,\TeV,\;\mu\,=\,167.40\,\GeV,\;\tan\beta|_{m_Z}\,=\,48, \label{127}\\
&&M_{\wt{\chi}^{\pm}_1}\,=\,167.36\,\GeV,\;M_{\wt{\chi}^{0}_1}\,=\,166.59\,\GeV,\;M_{\wt{\chi}^{0}_2}\,=\,167.63\,\GeV,\;m_h\,=\,127\,\GeV.\nonumber
\end{eqnarray}
Note that the above masses include higher-order corrections, and therefore differ from the tree-level predictions in Eq.~\eqref{eq:masses} at the sub-permil level.\footnote{These small differences in absolute masses, however, translate into sizeable corrections of the mass differences between the lightest chargino/second-lightest neutralino and the LSP. We accordingly correct for these higher-order contributions in the fit used for the determination of the fundamental parameters $M_1,\,M_2,\,\mu,$ and $\tan\beta$; see Section \ref{sec:paradet} for further details.}
As stated before, the two lightest neutralinos are nearly completely higgsino-like, with negligible gaugino admixtures. The mass splittings are very small with $M_{\wt{\chi}^{\pm}_1}-M_{\wt{\chi}^{0}_1} = 770\,\MeV\; (1.6\,\GeV)$ and $M_{\wt{\chi}^{0}_2}-M_{\wt{\chi}^{0}_1} = 1.04\,\GeV\;(2.7\,\GeV)$ in the dM770 (dM1600) scenario. This has interesting consequences especially for the decay, which we will discuss in the following.\\

\subsubsection*{Production}

The production cross sections have been calculated with \texttt{Whizard 1.95}~\cite{Kilian:2007gr} using the ILC set-up provided by the ILC Generator Group~\cite{bib:ilc-whizard}. It includes soft initial-state photon emission using an electron structure function~\cite{Skrzypek:1990qs} as well as the beam energy-spectra corresponding to the ILC Technical Design Report (ILC TDR)~\cite{bib:ILCTDR}. 

Due to the small mass difference, only very few and soft visible particles will result from the higgsino decay. For reasons mentioned above, a realistic analysis will therefore need to require an additional photon from initial-state radiation to be seen in the detector. 
The ISR photon, in addition to its usefulness to suppress background, provides an elegant way to determine the $\wt{\chi}^{\pm}_{1}$ and $\wt{\chi}^{0}_{2}$ masses from their recoil against the photon. This has already been applied successfully for near-degenerate winos in AMSB scenarios~\cite{Hensel:2002bu}.

We therefore determine the observable cross sections with one final state photon included in the hard matrix element. This photon is required to have a minimal invariant mass of $4$\,\GeV\ with the corresponding beam electron and an energy of more than $10\;\GeV$, corresponding to the cut used later at the analysis stage. We have verified that the effect of double-counting with the soft initial-state photon emission is negligible.

%%%%%%%%%%%%%%%%%%%%%%%%%%%%%%%%%%%%%%%%%%%%%%%%%%%%%%%%%%%%
\begin{figure}[htb]
%  \begin{center}
\centering
\includegraphics[width=0.49\textwidth]{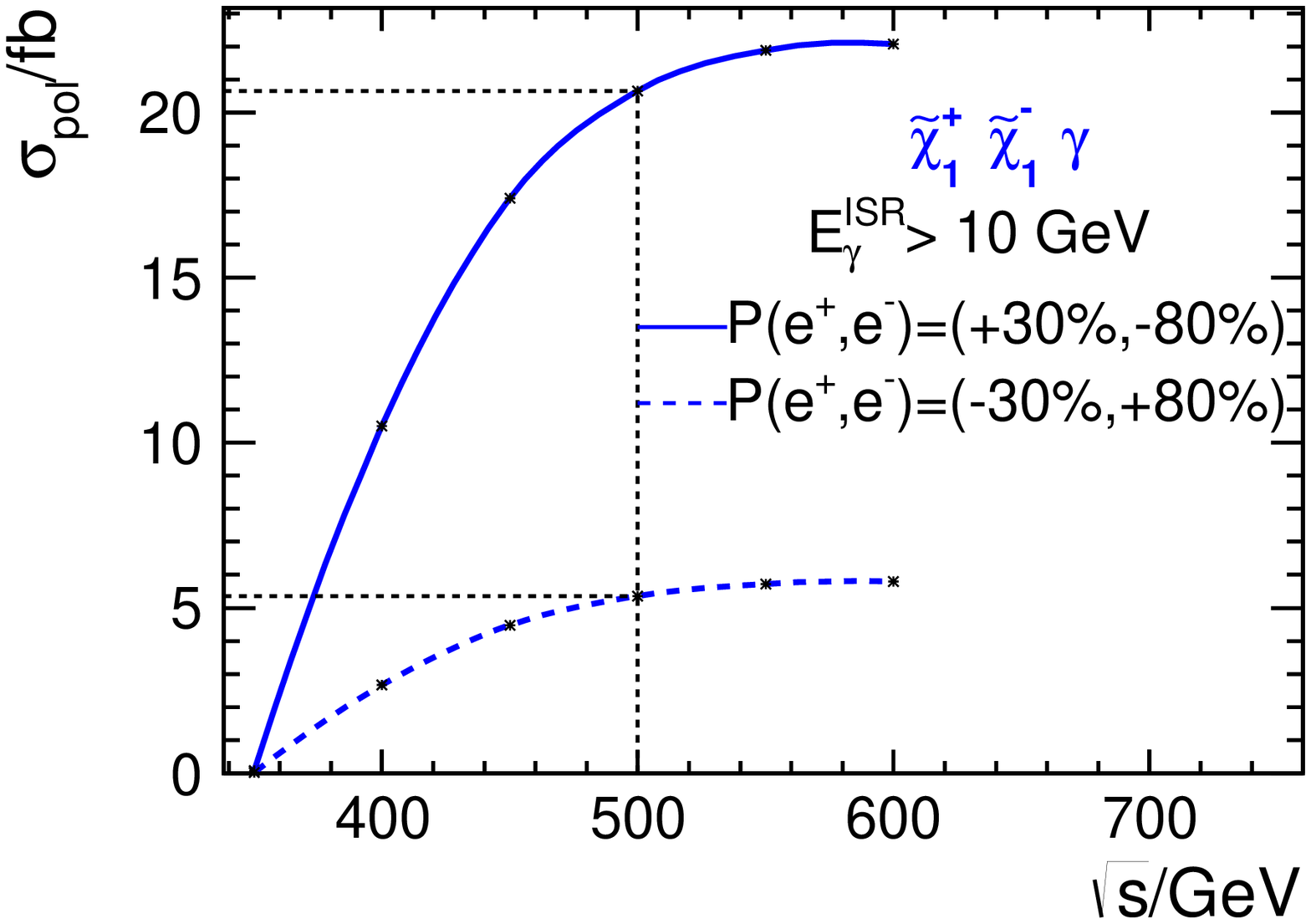}
\hspace{0.1cm}
\includegraphics[width=0.49\textwidth]{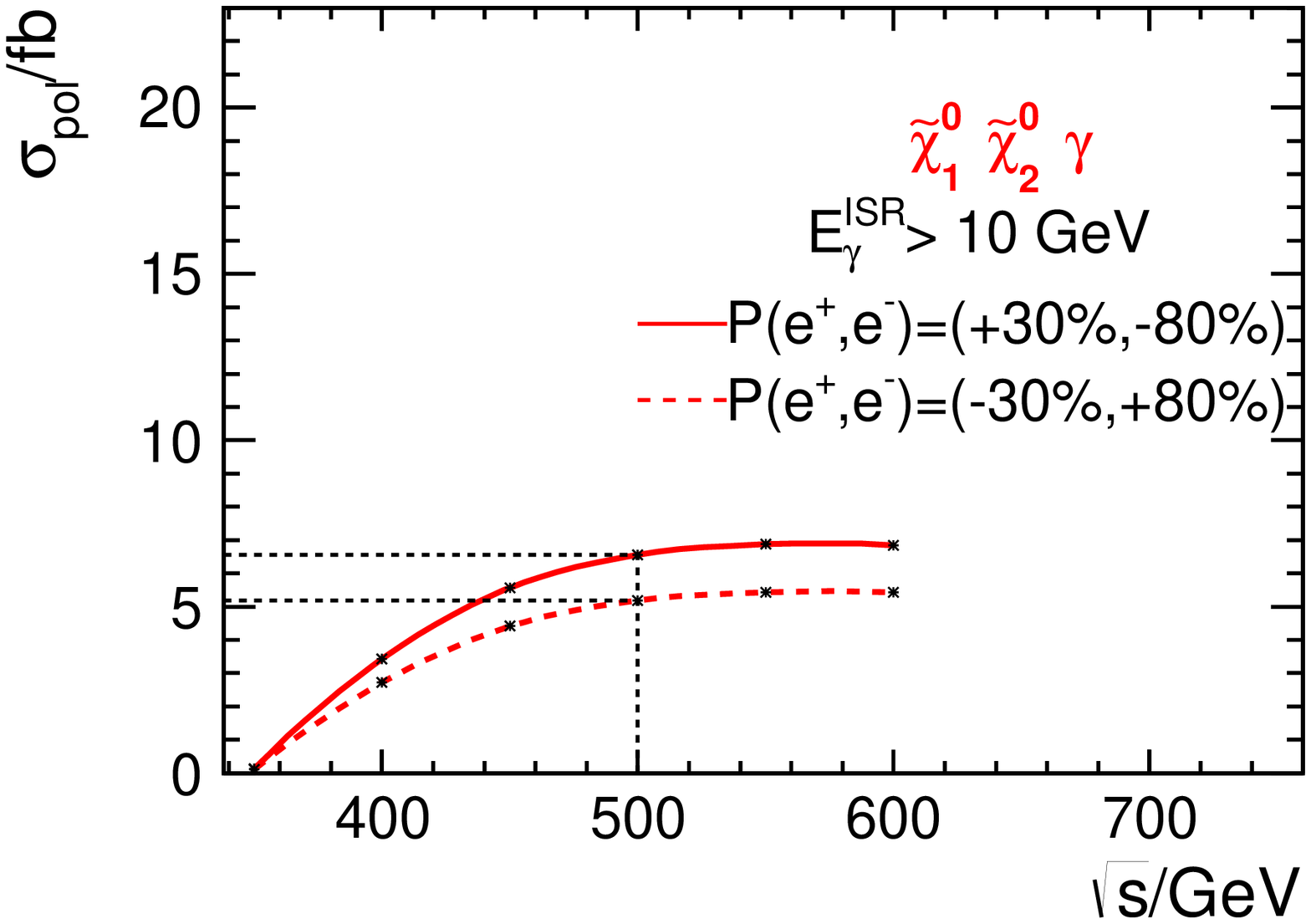}
%  \end{center}
  \caption{Chargino (left) and neutralino (right) production cross section in scenario dM770 for $P(e^+,e^-) = (-30\%,+80\%)$ and  $P(e^+,e^-) = (+30\%,-80\%)$ as a function of the centre-of-mass energy.}
\label{fig:xsec_roots}
\end{figure}
\begin{figure}[htb]
%  \begin{center}
\centering
\includegraphics[width=0.49\textwidth]{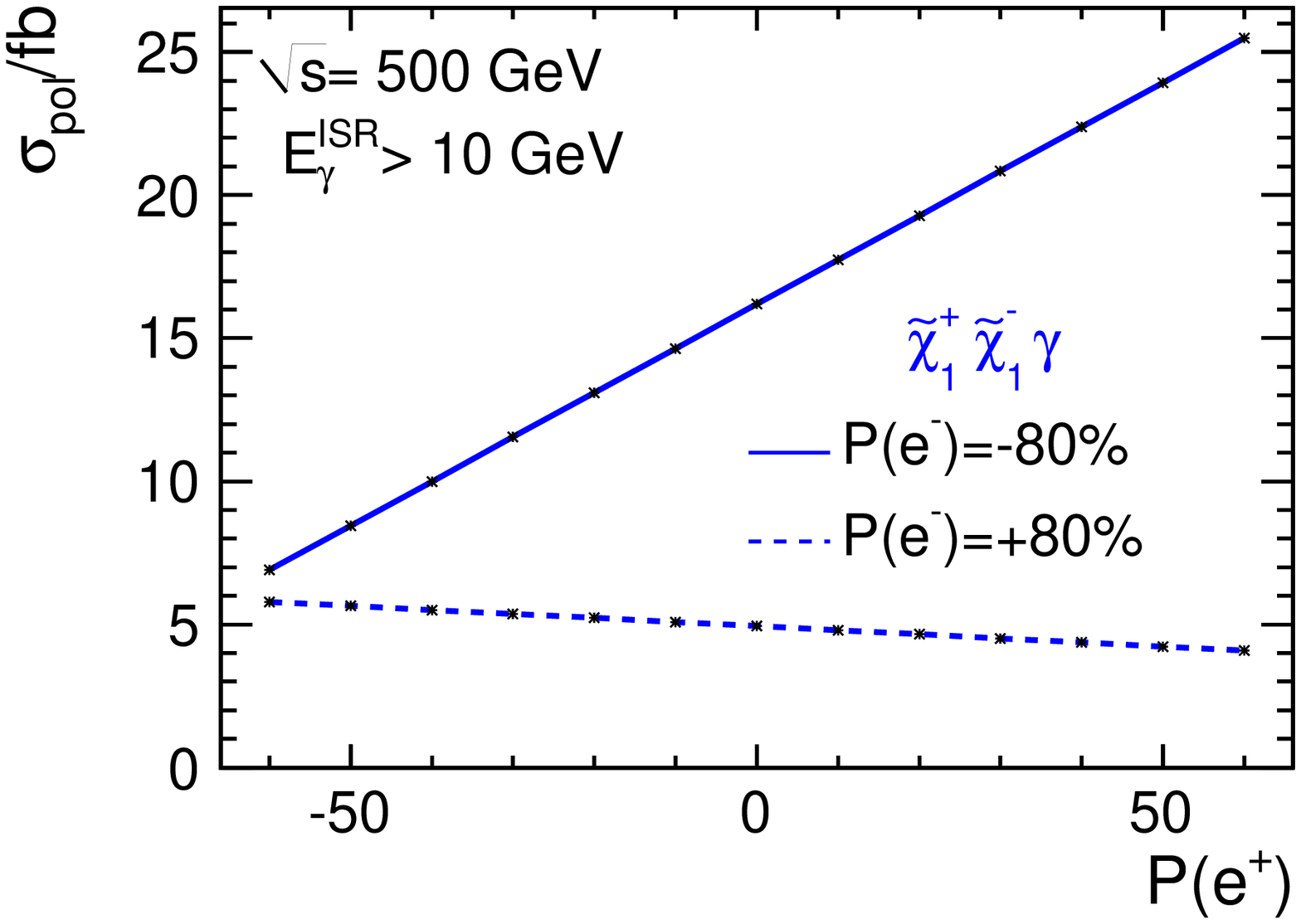}
\hspace{0.1cm}
\includegraphics[width=0.49\textwidth]{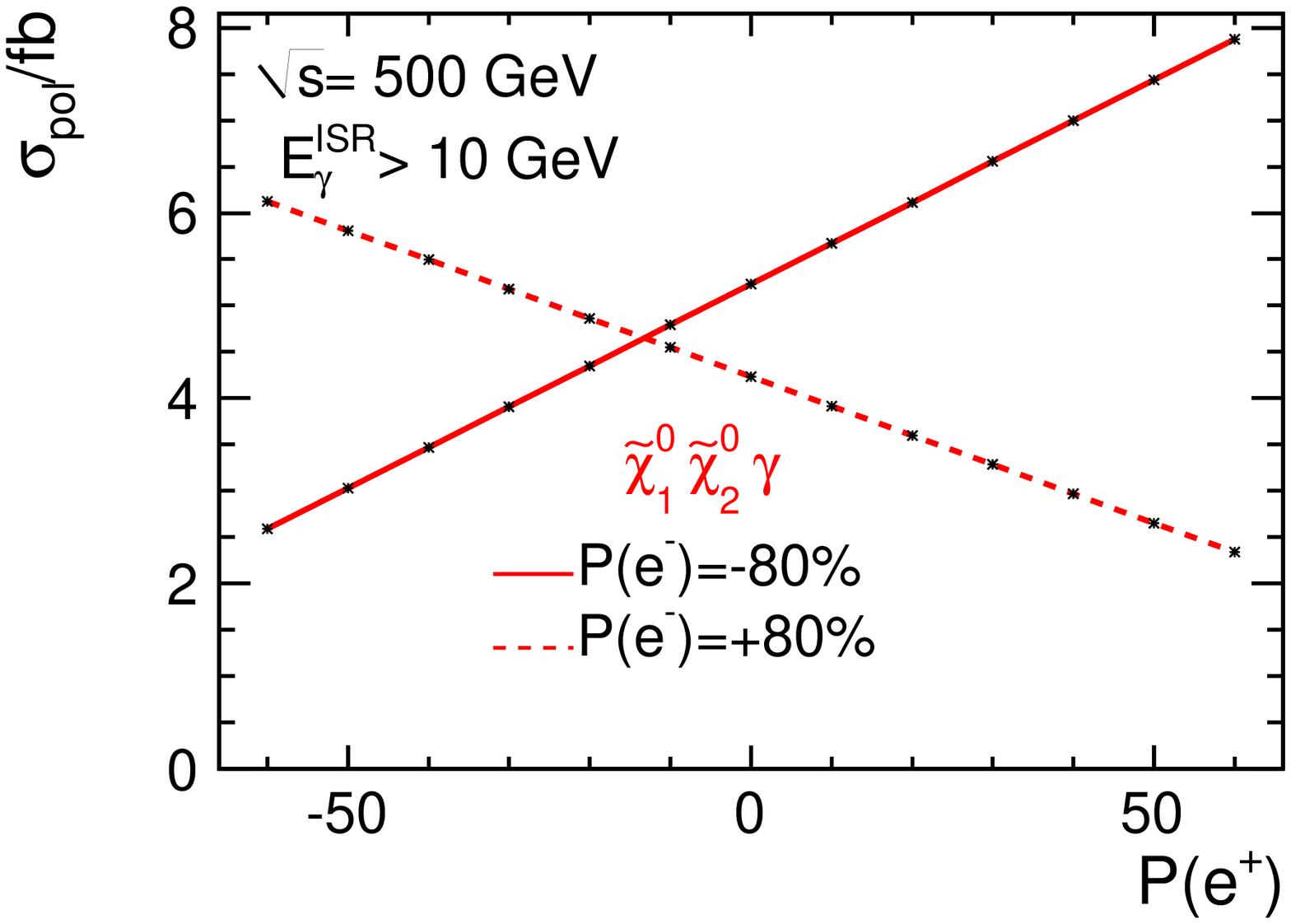}
%\end{center}
  \caption{Chargino (left) and neutralino (right) production cross section in scenario dM770 at $\sqrt{s}=500$~GeV as a function of the positron polarisation for $P(e^-)=+80\%$ and 
$P(e^-)=-80\%$.}
\label{fig:xsec_pol}
\end{figure}

%%%%%%%%%%%%%%%%%%%%%%%%%%%%%%%%%%%%%%%%%%%%%%%%%%%%%%%%%%%%
Figure~\ref{fig:xsec_roots} shows the observable chargino and
neutralino cross sections in the dM770 scenario as a function of the
centre-of-mass energy, and for two different configurations of the
polarisation of the beams, with the positron beam polarised to
$-(+)30\%$, the electron beam to $+(-)80\%$.\footnote{In the
  following, the notation $P(e^+,e^-) = (-30\%,+80\%)$ etc.\ will be
  used to denote the assumed beam polarisation.}
Figure~\ref{fig:xsec_pol} shows the same cross sections for a fixed
centre-of-mass energy of $500\,\GeV$ as a function of the positron
polarisation. The chargino production can be strongly enhanced by
choosing the appropriate beam helicities, whereas the differences are
significantly smaller in the neutralino case. Beyond the signal
enhancement, beam polarisation can be used to prove that the observed
process is mediated purely by $s$-channel exchange of a $Z$-boson, i.e.\ by 
analysing the corresponding scaling factors between the polarised and 
unpolarised cross sections together  with the order of magnitude of the
 cross section, see below and cf.\ discussion in~\cite{MoortgatPick:2005cw}.
 Effects from beam polarisation at
an $e^+\,e^-$ collider on chargino/neutralino pair-production have
already been studied
in~\cite{Choi:2001ww,Choi:2000hb,three-body}\footnote{For (fully
  differential) chargino production at NLO,
  cf.~\cite{Bharucha:2012ya,Fritzsche:2004nf,Oller:2005xg,Kilian:2006cj,Robens:2006np}.}
and have successfully been exploited for full parameter determination
in~\cite{Desch:2003vw}.

%%%%%%%%%%%%%%%%%%%%%%%%%%%%%%%%%%%%%%%%%%%%%%%%%%%%%%%%%%%%
%%%%%%%%%%%%%%%%%%%%%%%%%%%%%%%%%%%%%%%%%%%%%%%%%%%%%%%%%%%%

As mentioned, higgsinos are produced by $Z$ $(\gamma,Z)$ exchange in
the $s$-channel for neutralinos (charginos) respectively, cf.\ the
diagrams shown in Fig.~\ref{fig:chaprod}.  It is then straightforward
to explain the differences in production cross sections in
Figs.~\ref{fig:xsec_roots} and~\ref{fig:xsec_pol} by considering only
these diagrams.  For neutralino production, the ratio of
$(e^+_R\,e^-_L\rightarrow\,\wt{\chi}^0_1\,\wt{\chi}^0_2)/(e^+_L\,e^-_R\rightarrow\,\wt{\chi}^0_1\,\wt{\chi}^0_2)\,\sim\,1.3$
for fully polarised beams can directly be obtained from the ratio of
the respective fermion couplings to the $Z$ boson, given by
$L^2/R^2=[1-1/(2\sin^2\theta_W)]^2\sim 1.3$, with
$L=-\frac{1}{2}+\sin^2\theta_W$, $R=\sin^2\theta_W$.  For arbitrarily
polarised beams the scaling factor between polarised and unpolarised
cross section is given by
\begin{equation}
\sigma_{\rm pol}/\sigma_{\rm unpol}
=\left(1+\frac{R^2-L^2}{R^2+L^2} P_{\rm eff}\right) \cdot \mathcal{L}_{\rm eff}/\mathcal{L}=
(1-0.151 P_{\rm eff}) \cdot \mathcal{L}_{\rm eff}/\mathcal{L}\,,
\end{equation}
 where $\mathcal{L}_{\rm eff}=(1-P(e^-)P(e^+))\,\mathcal{L}$ denotes the normalised effective
 number of interactions and $P_{\rm eff}=
 \frac{P(e^-)-P(e^+)}{1-P(e^-)P(e^+)}$
the effective polarisation (for details see~\cite{MoortgatPick:2005cw}). 
It is expected to achieve high $|P_{\rm eff}|\sim 95\%$ and 
$\mathcal{L}_{\rm eff}/\mathcal{L}\sim 1.5$ for centre-of-mass energies of $\sqrt{s}\ge 350$~GeV
at the ILC~\cite{Desy123} which leads to scaling factor of about 1.2
 (1.7) with $P(e^-,e^+)=(+80\%,-60\%)$  ($P(e^-,e^+)=(-80\%,+60\%)$) 
configuration of polarised $(e^-,e^+)$-beams.

In case of chargino pair production, the large difference in the cross
sections stems from positive (in LR configuration) or negative 
(in RL configuration) interference 
of the $\gamma$ and
$Z$ $s$-channel contributions, which is again due to the structure of
the fermion-$Z$ couplings. These characteristics (which are independent
of additional photon radiation) are well reflected in the total cross
sections displayed in Figures~\ref{fig:xsec_roots} and~\ref{fig:xsec_pol}.

\subsubsection*{Decay widths and branching ratios}
Due to the quite pronounced mass degeneracy in the $\wt{\chi}^\pm_1,\,\wt{\chi}^0_1$ system, the $\wt{\chi}^\pm_1$ decays require a specific treatment, which we will now briefly describe.
In case of small mass splittings, a perturbative treatment on parton level, followed by standard hadronisation, leads to large uncertainties for slight variations of the masses of the partonic decay products. In the scenarios considered here, this especially plays a role in the decays of the chargino, where the mass difference can be in the sub-\GeV~range. In fact, the same situation is encountered in the SM for the decay of $\tau$s into hadrons. In this case, a more appropriate description then follows an effective theory approach using hadronic currents. For the case of supersymmetry, such a description has been implemented in the event generator \texttt{Herwig++} for $\wt{\chi}^\pm_1$ decays~\cite{Grellscheid:2007tt,Bahr:2008pv}. We therefore used \texttt{Herwig++} to generate the according branching ratios and decays widths, they are given in Table~\ref{tab:chi1}. On the other hand, the branching ratios of the $\wt{\chi}^0_2$ are calculated at parton-level followed by hadronisation,
with the results listed in Table~\ref{tab:neu2}.\footnote{In principle, the mass difference for the $\wt{\chi}^\pm_1$-LSP in dM1600 scenario and the $\wt{\chi}^0_2$-LSP in dM770 scenario are of similar order. However, since for neutral current decays there is no analogue to the charged current 
$\tau$-decay in the the SM,
no dedicated code is yet available which treats the decays of $\wt{\chi}^0_2$ using hadronic currents. The effects of including such an improved description are in the line of future work.} The decay widths of loop-induced processes have been cross checked with available analytic results~\cite{Haber:1988px}. They agree within $5\,\%$ with the \texttt{Herwig++} values.

\begin{table}[t]
\renewcommand{\arraystretch}{1.25}
%\begin{center}
\centering
\begin{tabular}{lcc} \toprule
$\wt{\chi}^+_1$ decay mode& BR(dM1600) & BR(dM770)\\ \midrule
$e \nu \neu{1}$                    & $17.3\%$ & $ 15.0\%$ \\
$\mu \nu \neu{1}$                  & $16.6\%$ & $ 13.7\%$ \\
$\pi^+\neu{1}$                     & $16.5\%$ & $ 60.4\%$ \\
$\pi^+ \pi^0 \neu{1}$              & $28.5\%$ & $  7.3\%$ \\
$\pi^+ \pi^0 \pi^0 \neu{1}$        & $ 7.5\%$ & $ 0.03\%$ \\
$\pi^+ \pi^+ \pi^- \neu{1}$        & $ 7.1\%$ & $ 0.03\%$ \\
$\pi^+ \pi^+ \pi^- \pi^0 \neu{1}$  & $ 2.4\%$ & $   -   $ \\
$\pi^+ \pi^0 \pi^0 \pi^0 \neu{1}$  & $ 0.5\%$ & $   -   $ \\
$K^+ \neu{1}$                      & $ 1.2\%$ & $  3.5\%$ \\ 
$K^0 \pi^+ \neu{1}$                & $ 1.0\%$ & $ 0.03\%$ \\ 
$K^+ \pi^0 \neu{1}$                & $ 0.5\%$ & $ 0.02\%$ \\ \bottomrule
\end{tabular}
\caption{\label{tab:chi1} Chargino $\wt{\chi}^+_{1}$ decay modes according to \texttt{Herwig++ 2.6.0}. 
}
%\end{center}
\end{table}

\begin{table}[t]
\renewcommand{\arraystretch}{1.25}
%\vspace{1cm}
%\begin{center}
\centering
\begin{tabular}{lcc}
\toprule
$\neu{2}$ decay mode& BR(dM1600) & BR(dM770)\\ \midrule
$\gamma\neu{1}$ & $23.6\%$  & $74.0\%$ \\
$\nu\bar{\nu}\neu{1} $ & $21.9\%$  & $9.7\%$ \\
$e^+e^-\neu{1}$ & $3.7\%$  & $1.6\%$ \\
$\mu^+\mu^-\neu{1}$ & $3.7\%$  & $1.5\%$ \\
hadrons $+\neu{1}$ & $44.9\%$ & $12.7\% $ \\ \midrule
$\wt\chi_1^{\pm} + X$ & $1.9\%$  & $0.4\%$ \\\bottomrule
\end{tabular}
%\end{center}
\caption{\label{tab:neu2} Neutralino $\neu{2}$ decay modes according to \texttt{Herwig++ 2.6.0}. 
}
\end{table}

\subsection{Event generation and detector simulation}
\label{sec:evtgeneration}

\subsubsection*{Event generation}
The simulation part of this study has been performed for a centre-of-mass energy of $\sqrt{s}=500\,\GeV$.
Signal events for the two benchmark scenarios have been generated with \texttt{Whizard}, using the same set-up as for the cross section calculation, described in Section~\ref{sec:proddecay}. The charginos and neutralinos were
generated by \texttt{Whizard}, with the subsequent decays simulated by \texttt{Pythia}~\cite{bib:pythia}.
The decay branching fractions used by \texttt{Pythia} were those 
given in Table~\ref{tab:chi1} and~\ref{tab:neu2}. As for the cross section calculation, one photon was included in the hard matrix element, and it was required to have a minimal invariant mass of $4$\,\GeV\ with the corresponding beam electron. The generator-level cut on the photon energy was, however, not applied in the event generation step, in order to be able to estimate the experimental acceptance after reconstruction.

%%%%%%%%%%%%%%%%%%%%%%%%%%%%%%%%%%%%%%%%%%%%%%%%%%%%%%%%%%%%
\begin{figure}[htb]
%  \begin{center}
\centering
\includegraphics[width=0.49\textwidth]{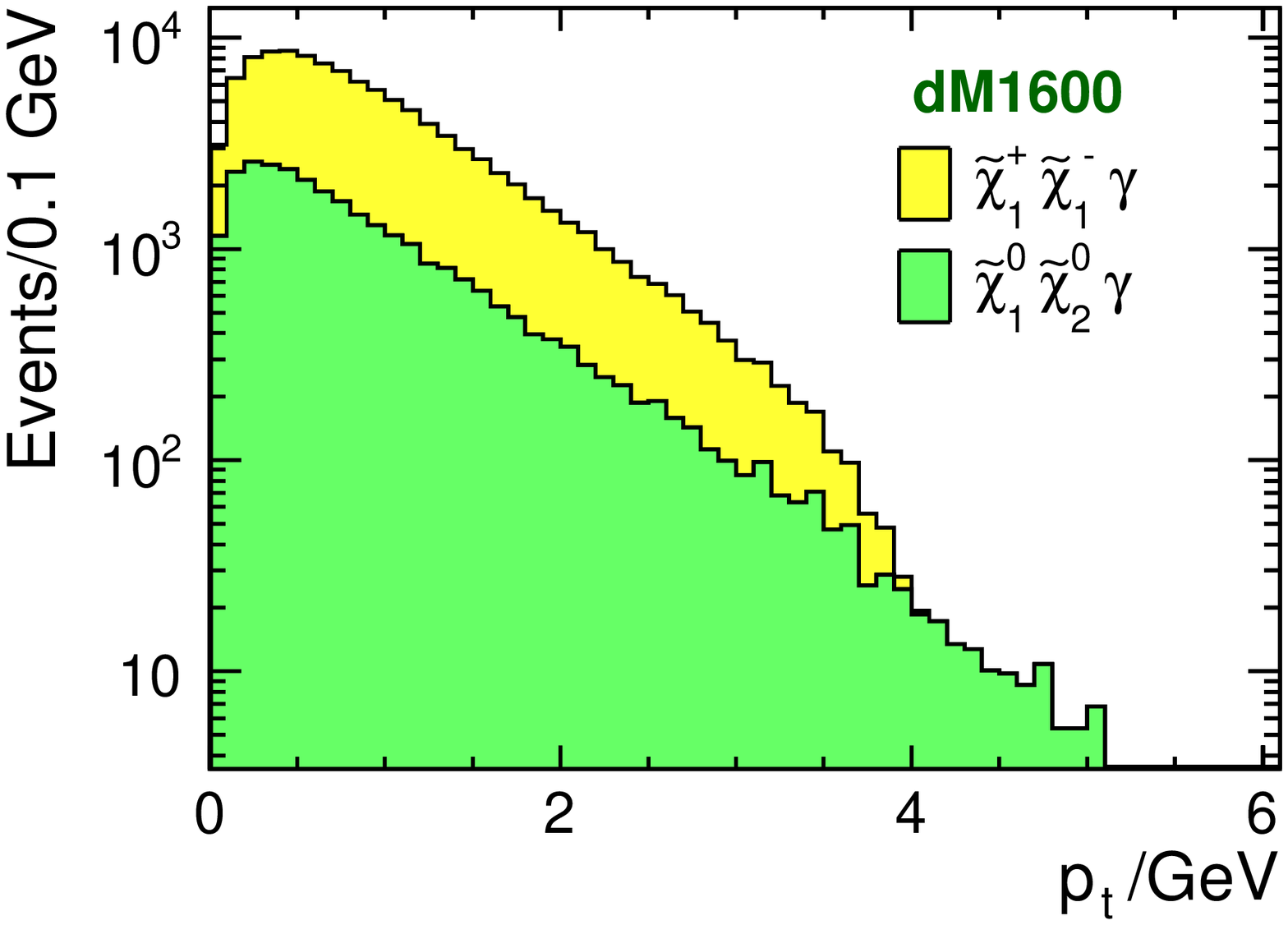}
\hspace{0.1cm}
\includegraphics[width=0.49\textwidth]{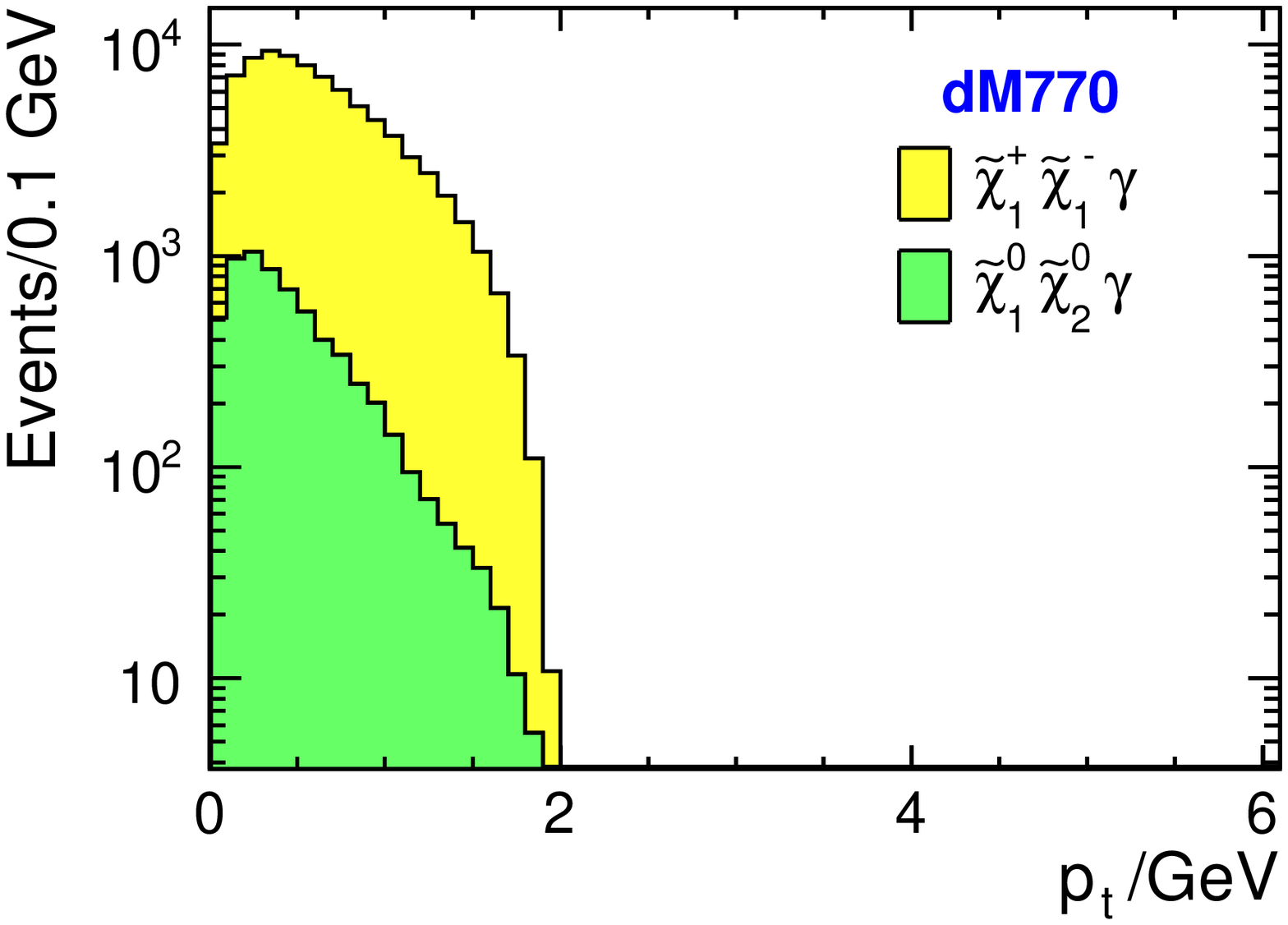}
%\end{center}
  \caption{$p_t$ spectrum of visible chargino and neutralino decay products at generator level for 
  the two scenarios. Left: dM1600, right: dM770. }
\label{fig:pt_decayprod}
\end{figure}
%%%%%%%%%%%%%%%%%%%%%%%%%%%%%%%%%%%%%%%%%%%%%%%%%%%%%%%%%%%%

Figure~\ref{fig:pt_decayprod} shows the transverse momentum spectrum of the visible decay products of chargino and neutralino events at generator level for both benchmark scenarios. Especially in the dM770 scenario, the visible decay particles of the higgsinos are extremely soft, with $p_t<2\,\GeV$. While for larger momenta the tracking at the ILC is almost $100\%$ efficient~\cite{bib:ILDDBD}, the situation is more challenging for sub-GeV particles, especially due to the presence of random hits and additional tracks caused by $e^+e^-$-pair background from beamstrahlung. Therefore we will discuss the tracking efficiency in more detail in the following subsection.

The event samples centrally
produced for the detector benchmarking in context of the 
ILC TDR~\cite{bib:ilc-whizard} were used as SM background. 
Hence, the same \texttt{Whizard} version was used both for
signal and background generation.
Apart from some processes with very high cross sections, the available statistics corresponds at least to an integrated luminosity of $500$~fb$^{-1}$. 
A notable exception was the $e\gamma \ra$ 3f class, for which only an integrated luminosity of about $100$~fb$^{-1}$ was available. This was taken care of by assigning appropriate weights to these events. 
Furthermore, radiative Bhabha events and
pure photon final states deserve special comments in the context of this analysis. 
Radiative Bhabha events, i.e.\ $e^+e^- \ra e^+e^- \gamma$, have been generated in a restricted phase space requiring an invariant mass of at least $4\,\GeV$ between any two of the final state particles, a momentum transfer of $Q>4\,\GeV$ between the incoming and outgoing electron (positron) and 
between the incoming lepton and the photon, resulting in a minimum acolinearity of the final state leptons. 
Pure photon final states have not been included in the production. However it has been shown in a previous full simulation study of a pure photon plus missing energy signature that the contribution after a very similar selection is negligible~\cite{Bartels:2012ex}. Also not included are final states with only neutrinos and photons as well as QED Compton events. They could in principle contribute if overlaid with one of the low-$p_t$ $\gamma\gamma \ra $ hadrons events of which on average $1.2$ occur per bunch-crossing for TDR machine parameters.
However, it is rather improbable that such an overlay will mimic the exclusive decay modes of the 
higgsinos studied here. A quantitative estimate of the impact of overlay events is only meaningful in full, {\tt Geant4}-based~\cite{Agostinelli:2002hh} simulation and thus beyond the scope of
this study.

\subsubsection*{Detector simulation}

A detailed description of the ILD detector can be found in~\cite{bib:ILDDBD}. The most important detector capabilities for this analysis comprise the tracking efficiency at very low momenta, the photon reconstruction and the hermeticity of the detector, including the ability to veto high energetic electrons in the forward calorimeters (LumiCal, LHCal and BeamCal). The tracking system is based on a Time Projection Chamber (TPC), augmented with silicon trackers on the inside, the outside and in the forward direction, as well as for vertexing. The transverse momentum resolution
($\Delta(1/p_t)=\Delta(p_t)/p^2_t$) is expected
to be $2.0 \times 10^{-5}\,\GeV^{-1}$ asymptotically, worsening to 
$9.0 \times 10^{-5}\,\GeV^{-1}$ at $10\,\GeV$, and 
to $9.0 \times 10^{-4}\,\GeV^{-1}$ at $1\,\GeV$.
In the low-angle region, charged particles will be efficiently
detected down to $\theta=7^{\circ}$ ($|\cos{\theta}|=0.993$), while the only
region not in the acceptance of the calorimetric system are
the holes in the BeamCal for the beam pipes. Around the outgoing beam pipe, the radius of the hole
is $20\,$mm at $z=3550\,$mm, corresponding to $5.6\,$mrad.
Since the crossing angle of the beams is $14\,$mrad and the hole for the
incoming beam pipe has $R=16\,$mm, the lower edge of the acceptance increases to $18.5\,$mrad at $\phi \approx 180^{\circ}$. The electromagnetic calorimeter (ECal) is a highly granular SiW sampling calorimeter with a transverse cell size of 5~mm $\times$ 5~mm and 20 layers. In test-beam measurements with a prototype detector a resolution of $(16.6\pm 0.1)/\sqrt{E(\mathrm{GeV})}\oplus (1.1 \pm 0.1)\%$  has been achieved~\cite{bib:ILDLOI}.

We simulate the response of the ILD detector using the fast detector simulation {\tt SGV}~\cite{bib:SGV}. {\tt SGV} derives the full covariance matrix of the track parameter reconstruction from the ILD detector geometry, taking into account point resolutions, positions of measurement planes and material effects like multiple scattering, bremsstrahlung and pair conversion. The calorimeter response has been parametrised to model the performance of the ILD particle flow reconstruction. 

The response of the BeamCal is modelled taking into account the large energy depositions from beam backgrounds. The efficiency to tag an electron or photon on top of the beam background has been studied in full simulation and parametrised as function of the energy of the particle and the local background energy density at its impact point on the BeamCal surface.

%%%%%%%%%%%%%%%%%%%%%%%%%%%%%%%%%%%%%%%%%%%%%%%%%%%%%%%%%%%%
\begin{figure}[htb]
%  \begin{center}
\centering
\includegraphics[width=0.49\textwidth]{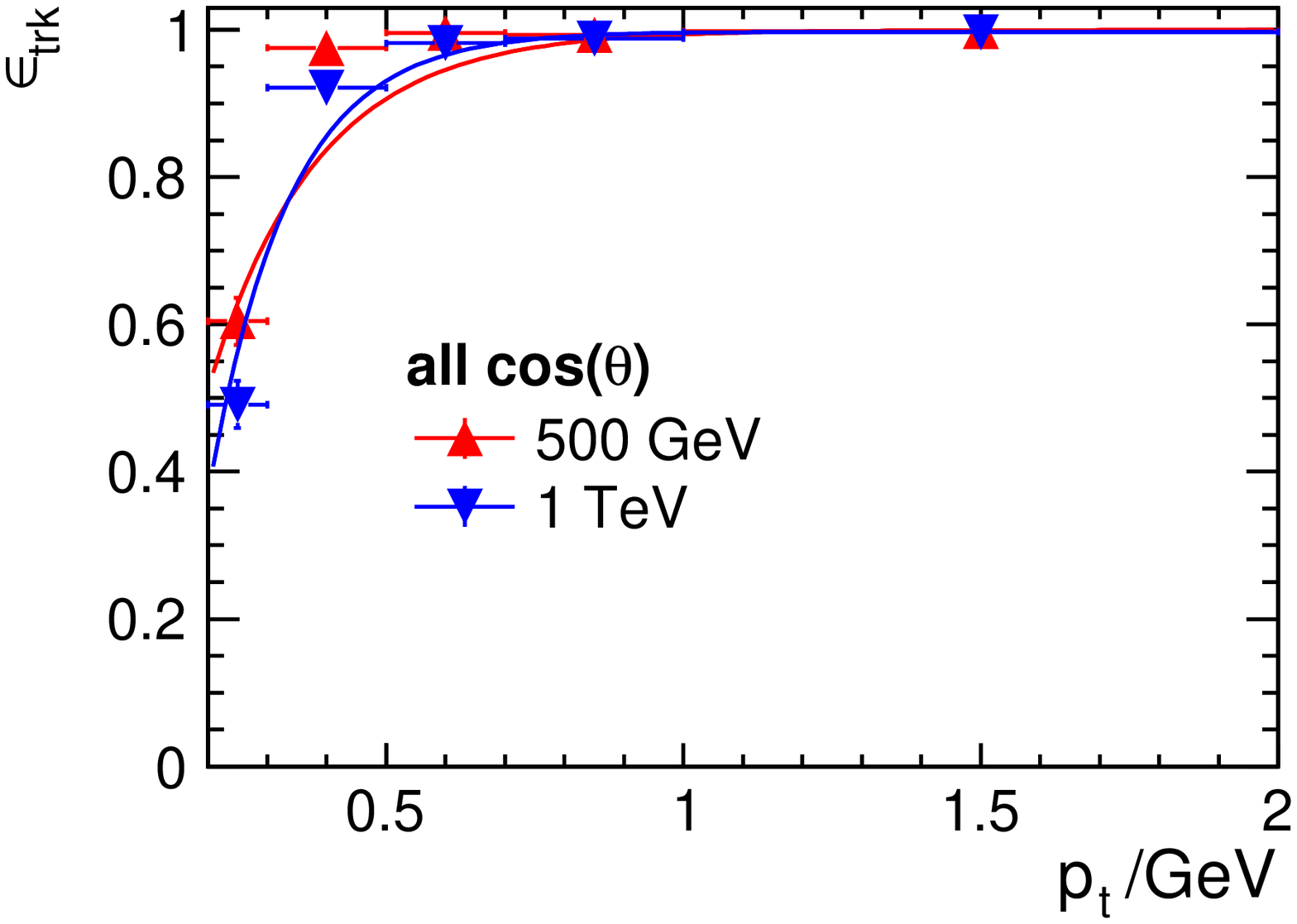}
\hspace{0.1cm}
\includegraphics[width=0.49\textwidth]{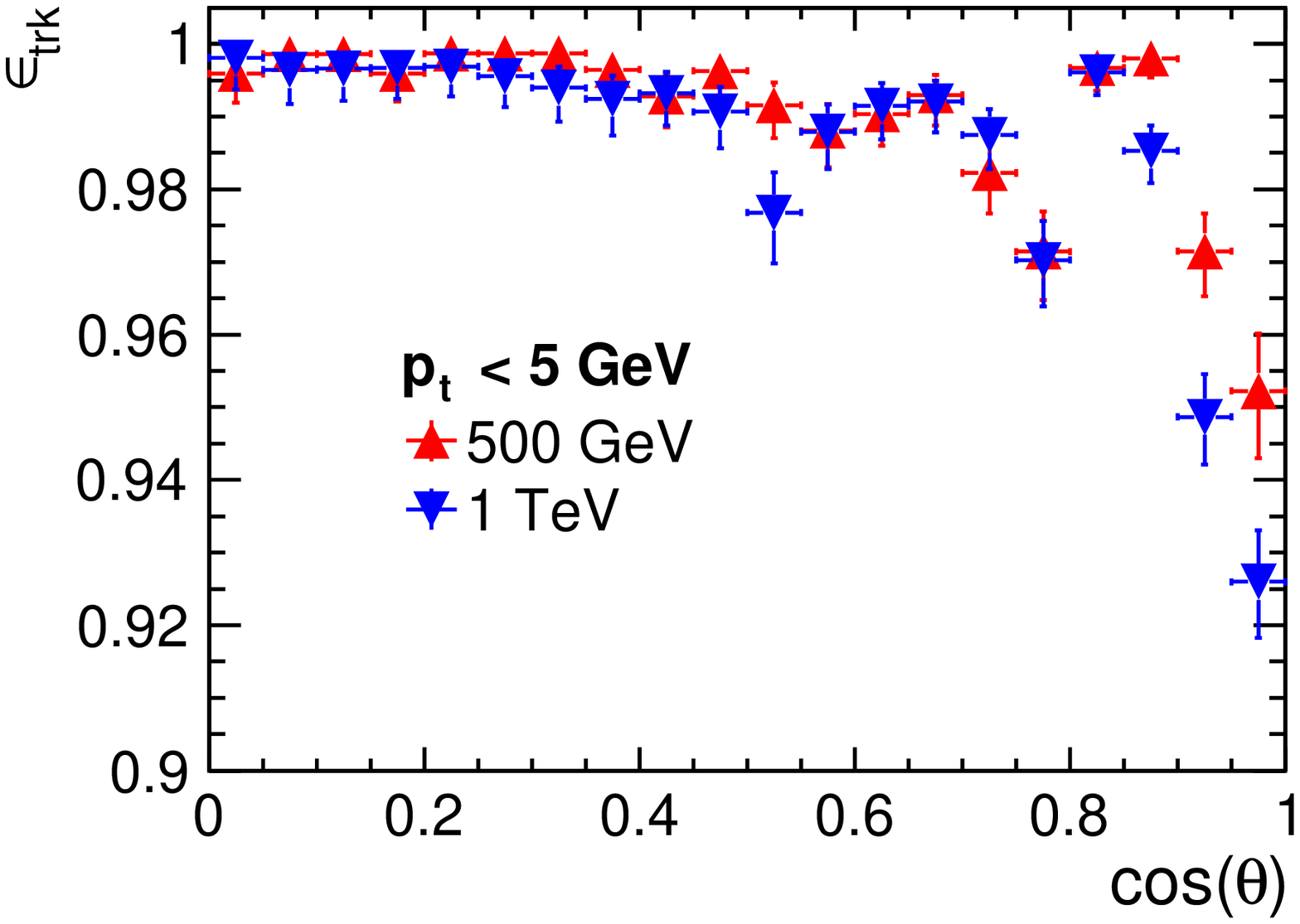}
%  \end{center}
  \caption{
  Tracking efficiency in $t\bar{t}$ events in the presence of pair background from beamstrahlung for $\sqrt{s}=500\,\GeV$ and $\sqrt{s}=1\,\TeV$, from full simulation of the ILD detector, assuming TDR beam parameters. Left: as a function of the transverse momentum $p_t$, Right: as a function of the absolute value of the cosine of the polar angle $|\cos{\theta}|$ for tracks with  $p_t < 5\,\GeV$.}
\label{fig:ILD_trackeffi}
\end{figure}
%%%%%%%%%%%%%%%%%%%%%%%%%%%%%%%%%%%%%%%%%%%%%%%%%%%%%%%%%%%%

A special challenge for this analysis is the tiny mass gap, which leads to extremely low momenta of the visible higgsino decay products. Beyond what was done for the TDR studies, the tracking efficiency in the presence of $e^+e^-$ pair background from beamstrahlung was determined in full simulation and reconstruction of the ILD detector. 
In order to have a conservative estimate,
$e^+e^-$ pairs generated with \texttt{GuineaPig}~\cite{bib:GuineaPig} 
were  overlaid on events with much higher particle-multiplicity 
($t\bar{t}$ events) than what is expected from the signal studied here.
These events were treated by \texttt{Mokka}~\cite{bib:Mokka}, the full {\tt Geant4}-based simulation of the ILD detector. 
For momenta larger than $1\,\GeV$, track finding efficiency was found
to be $99.5\%$. For low-$p_t$ tracks, the track finding efficiency is shown as a function of $p_t$ and as function of $\cos{\theta}$ in Figure~\ref{fig:ILD_trackeffi}. It can be seen that for beam-background conditions as expected at $\sqrt{s}=500\,\GeV$, the tracking system is nearly fully efficient down to $p_t$ values around $400\,\MeV$, dropping to about $60\%$ for $p_t$ between $200$ and $350\,\MeV$.\footnote{For 
$p_t < 200$~MeV, the tracking efficiency has not been studied and therefore we set it to zero here.} For all angles, 
the average tracking efficiency for all tracks with $p_t<5\,\GeV$ was found to be
larger than $99\%$ in the barrel region, dropping slightly in the transition region between the barrel and the end-cap, but staying larger than $95\%$ everywhere in the geometric acceptance of the tracking system.

In {\tt SGV}, we implemented four parametrisations of the efficiency as function of $p_t$ for different polar angle ranges and applied it to the signal and the dominating background classes, namely the $e\gamma \rightarrow 3$f and the $\gamma \gamma \rightarrow 2$f processes, which are basically the only type of events remaining at the selection stage where the tracking efficiency for low-energetic particles enters.

\section{Experimental precision at the ILC   \label{sec:expprec}}

In this section, we describe the event selection and the extraction of masses and cross sections in the two benchmark scenarios. All numbers refer to an integrated luminosity of $\int{{\cal L} dt} = 500\,$fb$^{-1}$ each for two configurations of the beam polarisation, namely $P(e^+,e^-)=(+30\%, -80\%)$ and $P(e^+,e^-)=(-30\%, +80\%)$. With the current machine design, collecting $1\,$ab$^{-1}$ 
of data at $\sqrt{s}=500\,\GeV$ takes $4$ years of design ILC operation, based on typical machine availability times and efficiencies. 

As a first step, a common preselection discriminating higgsino events against the majority of 
SM background processes is performed. Based on this, two different sets of cuts are applied for selecting
either chargino or neutralino events and measuring the respective masses and cross sections.

\subsection{Preselection}
\label{subsec:presel}
The aim of the preselection is to suppress the majority of SM background, 
while not yet discriminating charginos from neutralinos. The key features of the 
signal are that the events only contain a few low-energetic but centrally produced 
decay products of the higgsinos, and the hard ISR photon.
Thus, the following selection criteria are applied:
\begin{itemize}
\item There should be no significant activity in the BeamCal beyond the beam background in order to suppress events with high energetic electrons scattered under a small angle, e.g.~from $\gamma\gamma$ and $e\gamma$ type of processes.
\item The total number of reconstructed particles $N_{\mathrm{RP}}$ should be less than $15$.
\item There should be exactly one reconstructed photon with $E_{\mathrm{ISR}}>10\,\GeV$ and a polar angle $\theta_{\mathrm{ISR}}$ fulfilling $|\cos{\theta_{\mathrm{ISR}}}|<0.993$. The latter condition ensures that the photon is within the acceptance of the tracking system in order to be able to distinguish photons and electrons.
\item Any other reconstructed particle in the event is required to be at least $20^{\circ}$ 
away from the beam axis, and thus within the acceptance of the TPC, and to have an energy $E_{\mathrm{soft}}$ of less than $5\,\GeV$.\footnote{For charged particles, the energy is reconstructed from the momentum measured in the tracking system assuming the $\pi^{\pm}$ mass.}
\item Finally, the missing energy $E_{\mathrm{miss}}$ should be larger than $300\,\GeV$. This rejects the class of events where either the beam electron or positron is scattered under very small angles such that it escapes through the beam pipe with an energy of roughly $250\,\GeV$.   
\item In order to ensure that the missing energy is not due to particles escaping along the beam pipe, the missing momentum vector is required to point well into the acceptance region of the detector by fulfilling $|\cos{\theta_{\mathrm{miss}}}|<0.992$.   
\end{itemize}

The resulting cut flow is displayed for $P(e^{+},e^{-})=(+30\%,-80\%)$ in Table~\ref{tab:presel_cutflow}. 
At this stage of the selection, the SM background is reduced by four orders of magnitude and is dominated by $e\gamma \rightarrow 3$f and $\gamma\gamma \rightarrow 2$f processes. The biggest loss 
in the signals is due to the ISR photon requirement, since at event generation only a minimal 
invariant mass of the photon with respect to the beam electron/positron of $4\,\GeV$ was required,
 while the preselection demands a photon energy of at least $10\,\GeV$, the latter being more
  stringent.

\begin{table}[ht!]
\renewcommand{\arraystretch}{1.25}
\centering
    \begin{tabular}{l cc cc ccc } \toprule
               & \multicolumn{2}{c}{dM1600} &  \multicolumn{2}{c}{dM770} &  \multicolumn{3}{c}{Standard Model}  \\ \midrule
               &  $\widetilde{\chi}_1^{+} \widetilde{\chi}_1^{-} \gamma $   & $\widetilde{\chi}_1^{0} \widetilde{\chi}_2^{0} \gamma $ &  $\widetilde{\chi}_1^{+} \widetilde{\chi}_1^{-} \gamma $   & $\widetilde{\chi}_1^{0} \widetilde{\chi}_2^{0} \gamma $ & $ee \rightarrow 2,4,6$f  & $e\gamma \rightarrow 3,5$f &  $ \gamma\gamma \rightarrow 2,4$f  \\ \midrule
%                                            cha124      neu124     cha127     neu127           ee                          eg                 gg
no cut 	        		           & $38672$ &  $24250$  & $38130$   & $23940$  & $2.6434\times10^{7}$  & $8.8820\times10^{7} $ & $9.7554\times10^{8} $  \\ 
BeamCal veto                               & $38591$ &  $24187$  & $38054$   & $23874$  & $2.6284\times10^{7}$  & $8.8178\times10^{7} $ & $9.6757\times10^{8} $  \\ 
$N_{\mathrm{RP}} <$ 15                     & $38591$ &  $24185$  & $38054$   & $23874$  & $6.4968\times10^{6}$  & $6.5811\times10^{7} $ & $6.6308\times10^{8} $  \\ 
$N_{\mathrm{ISR}} = 1$                     & $30058$ &  $9551 $  & $29675$   & $9317 $  & $3.1640\times10^{6}$  & $1.5074\times10^{7} $ & $1.7752\times10^{7} $  \\ 
$|\cos{\theta_{\mathrm{soft}}}| < 0.9397$    & $21501$ &  $7318 $  & $23117$   & $7458 $  & $7.1453\times10^{5}$  & $4.5646\times10^{6} $ & $4.7083\times10^{6} $  \\ 
$E_{\mathrm{soft}} < 5\,\GeV$              & $20611$ &  $6615 $  & $22156$   & $7110 $  & $9092              $  & $5.9732\times10^{5} $ & $1.2390\times10^{6} $  \\  
$E_{\mathrm{miss}} > 300\,\GeV$            & $20611$ &  $6615 $  & $22156$   & $7110 $  & $6462              $  & $1.5822\times10^{5} $ & $4.6306\times10^{5} $  \\ 
$|\cos{\theta_{\mathrm{miss}}}|<0.992$     & $19872$ &  $6365 $  & $21558$   & $6872 $  & $5731              $  & $1.1837\times10^{5} $ & $3.3051\times10^{5} $  \\ \bottomrule
                                         
    \end{tabular}                             
\caption{\label{tab:presel_cutflow} Cut-flow table of the preselection for an integrated luminosity of 
$500\,$fb$^{-1}$ and $ P(e^{+},e^{-})=(+30\%,-80\%)$. }                         
  \end{table}

\subsection{Chargino selection}
In order to discriminate chargino-pair events from the associated neutralino production, we select semi-leptonic
events, i.e.\ events with one chargino decaying leptonically and the other hadronically. This signature does not
appear in the associated neutralino production, since the $\wt{\chi}^0_2$ decays via a virtual $Z$ boson and thus either fully leptonically or fully hadronically. The main background contributions to the semi-leptonic signature are $\tau$-pairs, either from $e^+e^-\rightarrow \tau^+\tau^-$, from $e^{\pm} \gamma \rightarrow e^{\pm} \tau^+\tau^- $ or from $\gamma \
\gamma \rightarrow \tau^+\tau^- $.

The cleanest semi-leptonic signature arises when the hadronic chargino decays to a single charged pion. In the 
dM770 scenario, this signature occurs in as many as $35\%$ of the chargino-pair events, cf.\ Table~\ref{tab:chi1}.
In the dM1600 scenario, the branching ratio to single pion is significantly lower due to the larger mass difference,
so that only $11\%$ of the chargino pairs give the semi-leptonic single pion signature. Therefore the $\pi^{\pm} \pi^0$
 channel is included for the dM1600 scenario by allowing the charged pion to be accompanied by two photons, raising
  the fraction of events with the target signature to $30\%$. For the dM770 scenario, the gain in signal is too
  small to balance the penalty of larger background, and thus only the single pion mode is allowed in that scenario.

A very discriminative variable against $\tau$-events can be constructed from hadronic chargino decay:
\begin{eqnarray}
  E^*_{\pi} = \frac{(\sqrt{s} - E_{\gamma}) E_{\pi} + \vec{p}_{\pi} \cdot \vec{p}_{\gamma}}{\sqrt{s'}}.
  \label{eqn:boostedEnergy}
\end{eqnarray}
The reduced centre-of-mass energy $\sqrt{s'}$ of the system recoiling against the ISR photon is defined as
 $s' = s - 2 \sqrt{s} E_{\gamma}$, where $\sqrt{s} = 500\,\GeV$ is the nominal centre-of-mass energy.
$E_{\pi}$ and $\vec{p}_{\pi}$ are the energy and 3-momentum of the pion in the
laboratory frame, while $\vec{p}_{\gamma}$ is the 3-momentum of the ISR photon in that frame.
$E^*_{\pi}$ is related to the energy of the pion(s) in the rest frame of the chargino pair, which is only a 
few GeV for pions from chargino decays, but large for $\tau$ decays.\footnote{In case of the additional $\pi^0$, 
the reconstructed four-momenta of the two photons are added to the charged pion four-momentum.} Therefore, we 
require $E^*_{\pi} < 3\,\GeV$. As can be seen from Table~\ref{tab:cha1sel_cutflow}, this cut is fully efficient 
in the dM770 scenario, but not in the dM1600 scenario. The reason is that the values of $E^*_{\pi}$ are 
closely related to the mass difference between the chargino and the LSP, which is nearly twice as large in 
the dM1600 scenario. However the losses occur in the less relevant regime far from the endpoint. 
At this stage, the background is dominated by contributions from $\gamma\gamma$ events, in particular 
$\gamma\gamma \ra \tau^+\tau^-$, which are typically back-to-back in the transverse plane and have large 
$\sqrt{s'}$. Therefore the combined requirement of $\Phi_{\text{acop}} < 2$ \mbox{or} $\sqrt{s'} < 480\,\GeV$, where
$\Phi_{\text{acop}}$ is the acoplanarity angle between the leptonic and the hadronic system, reduced this background
by a factor of four.

\begin{table}[ht!]
\renewcommand{\arraystretch}{1.25}
\centering
    \begin{tabular}{ l cc ccc } \toprule
               & \multicolumn{2}{c}{dM1600} &   \multicolumn{3}{c}{Standard Model}  \\ \midrule
               &  $\widetilde{\chi}_1^{+} \widetilde{\chi}_1^{-} \gamma $   & $\widetilde{\chi}_1^{0} \widetilde{\chi}_2^{0} \gamma $ & $ee \rightarrow 2,4,6$f  & $e\gamma \rightarrow 3,5$f &  $ \gamma\gamma \rightarrow 2,4$f  \\ \midrule
%                              		     	         cha       neu      ee         eg          gg
after preselection     				     & $19872$ & $6365$ & $5731$ & $1.1837\times10^{5} $ & $3.3051\times10^{5} $ \\ 
$l^{\pm} \pi^{\pm} (\pi^0)$      		             &  $5509$ & $134$  & $38$   &  $6197$  &  $13991$  \\ 
$E^*_{\pi} < 3\,\GeV$            	             &  $4435$ & $103$  & $0 $   &  $2635$  &  $6162$  \\ 
$\Phi_{\text{acop}} < 2$ \mbox{or} $\sqrt{s'} < 480\,\GeV$  &  $3813$ &  $97$  & $0$    &  $2564$  &  $1452$  \\ \midrule
$E_{\mathrm{miss}} > 350\,\GeV$ 		     &  $3812$ & $97$  & $ 0$	 &  $1016$  &  $511$  \\ \bottomrule
%\hline
\multicolumn{6}{c}{} \\  \toprule                                     
               & \multicolumn{2}{c}{dM770} &  \multicolumn{3}{c}{Standard Model}  \\ \midrule
               &  $\widetilde{\chi}_1^{+} \widetilde{\chi}_1^{-} \gamma $   & $\widetilde{\chi}_1^{0} \widetilde{\chi}_2^{0} \gamma $ & $ee \rightarrow 2,4,6$f  & $e\gamma \rightarrow 3,5$f &  $ \gamma\gamma \rightarrow 2,4$f  \\ \midrule
%                                			   cha        neu       ee       eg         gg
after preselection     				        & $21558$ & $6872 $ & $5731$ & $1.1837\times10^{5} $ & $3.3051\times10^{5} $ \\ 
$l^{\pm} \pi^{\pm}$            				& $5489$  &  $38$   &  $19$  &  $2478$ &  $6754$  \\ 
$E^*_{\pi} < 3\,\GeV$          				& $5489$  &  $38$   &  $ 0$  &  $1465$ &  $4755$  \\ 
$\Phi_{\text{acop}} < 2$ \mbox{or} $\sqrt{s'} < 480\,\GeV$ & $4600$ & $36$  &  $ 0$  &  $1417$ &  $782$  \\ \midrule
$E_{\mathrm{miss}} > 350\,\GeV$ 			& $4599$  &  $36$   &  $ 0$  &  $ 536$ &  $218$  \\ \bottomrule
                                       
    \end{tabular}                             
\caption{\label{tab:cha1sel_cutflow} Cut-flow tables of the chargino selection in the two 
scenarios for an integrated luminosity of $500\,$fb$^{-1}$ and $ P(e^{+},e^{-})=(+30\%,-80\%)$ 
following the preselection (cf.\ Tab.~\ref{tab:presel_cutflow}). For the dM1600 scenario, also the two-pion decay mode is included, therefore also the SM expectation changes. The chargino mass measurement is performed after the cut on the acoplanarity angle and the reduced centre-of-mass energy in both scenarios. The respective last line is applied for the cross section measurement and the determination of the mass difference.}                         
  \end{table} 
  
Table~\ref{tab:cha1sel_cutflow} shows that at this level only a very small contribution from 
$\widetilde{\chi}_1^{0} \widetilde{\chi}_2^{0}$ events remains in the sample. This is important 
since we can assume the ability to model electroweak SM processes to any necessary precision by the time an ILC is 
actually running, whereas background contributions from other new physics processes, in our case the neutralino production, are 
a priori unknown.

\subsubsection*{Chargino mass reconstruction}
\begin{figure}[htb]
%  \begin{center}
\centering
\includegraphics[width=0.49\textwidth]{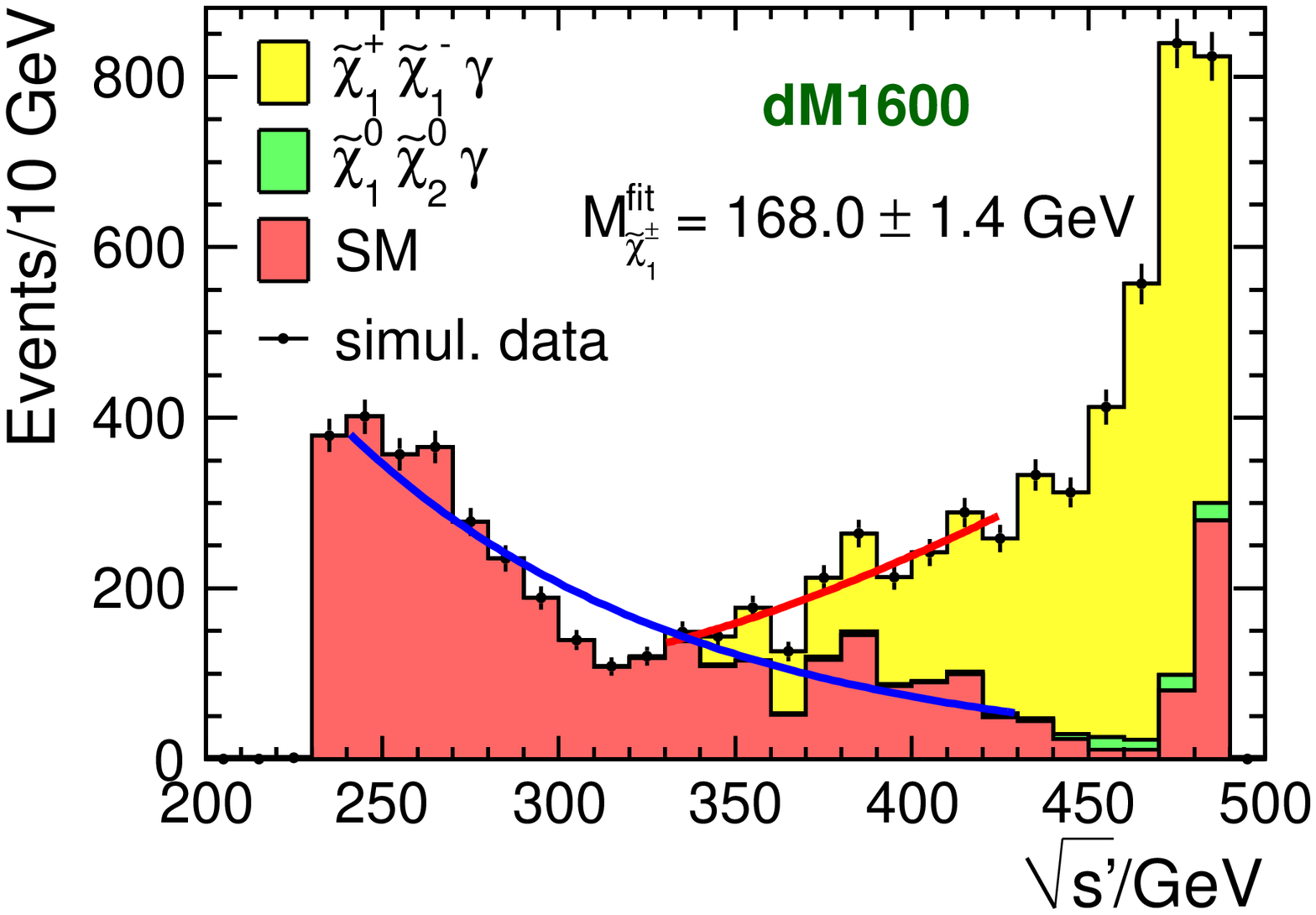}
\hspace{0.1cm}
\includegraphics[width=0.49\textwidth]{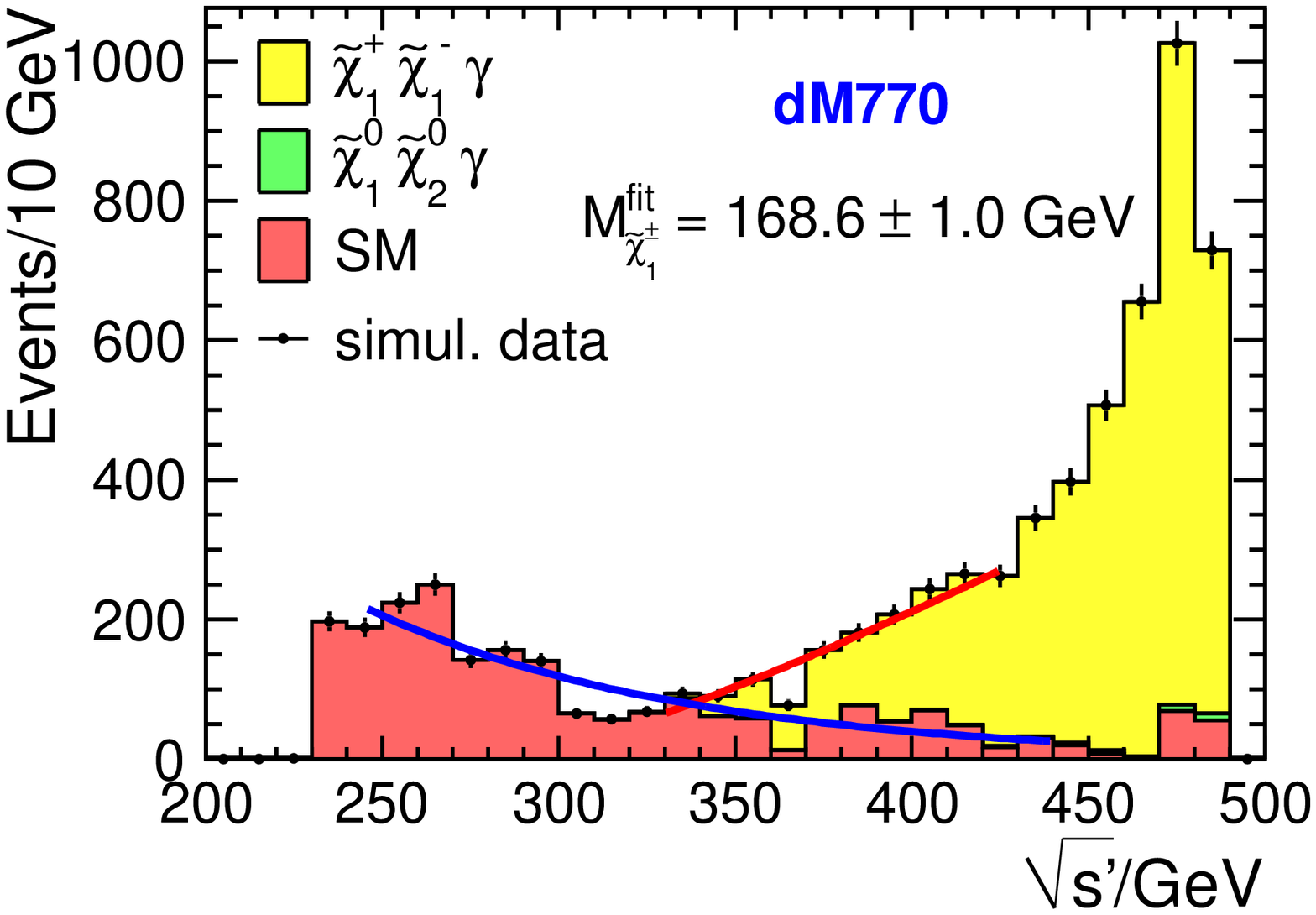}
%  \end{center}
  \caption{Distribution of the reduced centre-of-mass energy ($\sqrt{s'}$)
of the system recoiling against the hard ISR photon for all events passing the chargino selection for an integrated luminosity of $500\,$fb$^{-1}$ with $P(e^+,e^-)=(+30\%,-80\%)$. $M_{\wt \chi^{\pm}_1}$ is determined from a linear fit to the distribution near the endpoint. Left: dM1600 scenario; Right: dM770 scenario.}
\label{fig:recoilMcha1}
\end{figure}

The chargino mass is reconstructed using the reduced  centre-of-mass energy,
$\sqrt{s'}$, introduced above.  At the threshold, where the value of $s'$ is denoted $s'|_\text{thresh}$, the chargino pair is produced nearly at rest, 
and $\sqrt{s'|_\text{thresh}}$ is twice the chargino mass, thus:
\begin{eqnarray}
 M_{\wt \chi^{\pm}_1} = \frac{1}{2} \sqrt{s'|_\text{thresh}} =  \frac{1}{2} \sqrt{s - 2 \sqrt{s} E_{\gamma}}\,.
\end{eqnarray}
Figure~\ref{fig:recoilMcha1} shows the resulting $\sqrt{s'}$ distributions for both scenarios for an 
integrated luminosity of $500\,$fb$^{-1}$ with $P(e^+,e^-)=(+30\%,-80\%)$. The onset of the signal is 
clearly visible on top of the SM background, which stretches out to lower $\sqrt{s'}$ values. 
The cut-off in the SM background near $\sqrt{s'} = 230\,\GeV$ is due to the cut on 
$E_{\mathrm{miss}} > 300\,\GeV$. This is chosen on purpose so that a signal-free region is 
available to fix the background level, here by fitting an exponential function with two free 
parameters $f(x)=p_1 \cdot e^{-p_2\cdot x}$ to the SM prediction only (blue line). 
In a second step, a straight line is added on top of the background to model the signal 
contribution and fitted to the simulated data in the endpoint region (red line). The two parameters 
of the SM background function are fixed to the values obtained from 
the SM-only fit in the wider $\sqrt{s'}$ window. It has been verified that the 
results are stable against reasonable variations of the fit 
ranges or the bounds on the background parameters. The chargino mass is fitted to 
$168.0 \pm 1.4\,\GeV$ in the dM1600 scenario and to $168.6 \pm 1.0\,\GeV$ in the dM770 case. 

%%%%%%%%%%%%%%%%%%%%%%%%%%%%%%%%%%%%%%%%%%%%%%%%%%%%%%%%%%%%

%%%%%%%%%%%%%%%%%%%%%%%%%%%%%%%%%%%%%%%%%%%%%%%%%%%%%%%%%%%%
 
The fitted central values agree within $1.6$ ($1.2$) standard deviations with the respective input 
masses of $M_{\wt{\chi}^{\pm}_1}\,=\,165.77\,\GeV$
($M_{\wt{\chi}^{\pm}_1}\,=\,167.36\,\GeV$) in the dM1600 (dM770) scenario. 
Since the relation between $\sqrt{s'}$ and the chargino mass is only approximate, e.g.~due to the 
approximation of equal chargino energies, but even more so due to the beam energy-spectrum, an exact 
agreement is not necessarily expected. Therefore we investigated the dependence of the fitted mass 
on the input mass by simulating signal samples with different chargino masses. 
In order to minimise a possible bias due to changes in the acceptance, all higgsino masses were varied 
simultaneously, so that e.g.\ the momentum distribution of the decay products does not change 
significantly. 

\begin{figure}[htb]
%  \begin{center}
\centering
\includegraphics[width=0.49\textwidth]{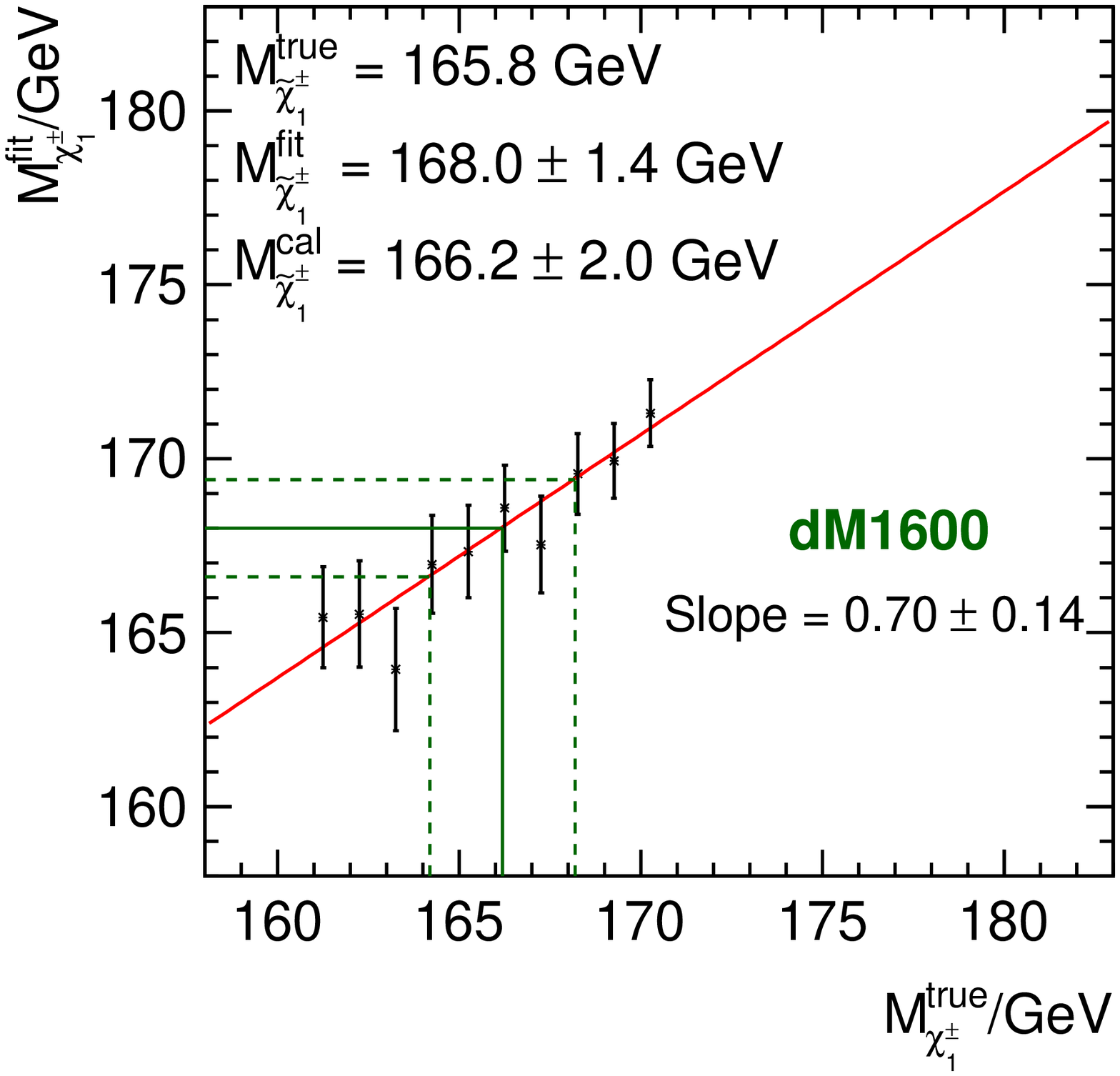}
\hspace{0.1cm}
\includegraphics[width=0.49\textwidth]{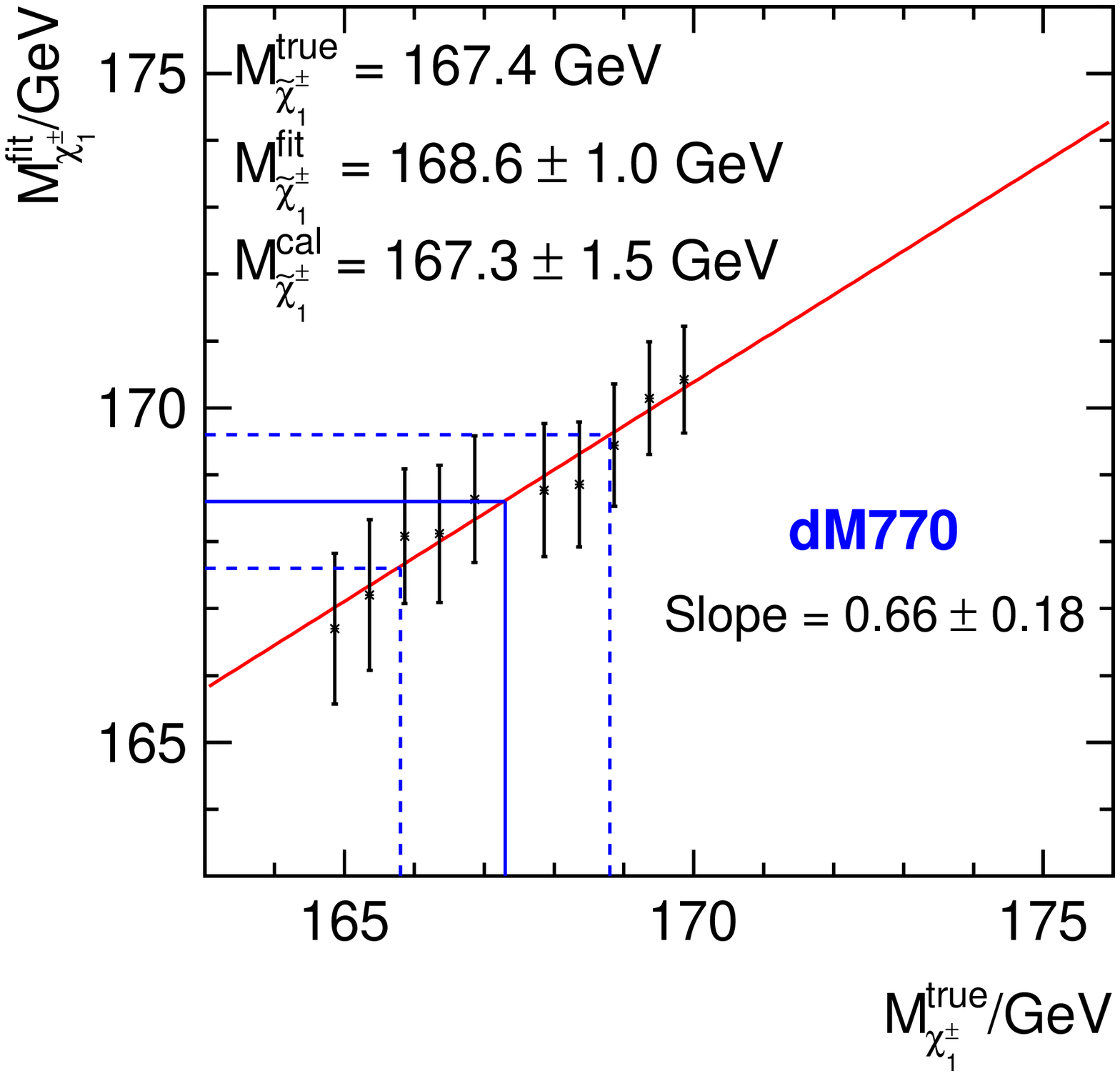}
%\end{center}
  \caption{Calibration of the $\sqrt{s'}$ method for the chargino mass determination. 
  Left: dM1600 scenario; Right: dM770 scenario.}
\label{fig:recoilMcalcha1}
\end{figure}

Figure~\ref{fig:recoilMcalcha1} shows the fitted mass as function of the true mass for both scenarios.
They clearly display a linear behaviour, which can easily be used to calibrate the reconstruction 
method. The calibrated mass (and its uncertainty) can be found on the $x$-axis as a projection of 
the fitted
 values on the $y$-axis as indicated by the lines.\footnote{The uncertainty on the
calibration curve itself is not propagated to the final result, since its
origin, namely the limited amount of available MC statistics especially
for the SM background, will not be an issue in a real ILC measurement.}

%%%%%%%%%%%%%%%%%%%%%%%%%%%%%%%%%%%%%%%%%%%%%%%%%%%%%%%%%%%%
%%%%%%%%%%%%%%%%%%%%%%%%%%%%%%%%%%%%%%%%%%%%%%%%%%%%%%%%%%%%

At a first sight it might appear worrisome that both scenarios have different correlations between 
fitted and true masses. 
However one needs to remember that in the dM1600 case, an additional decay channel is included and 
thus different selections are applied in the two scenarios, which in this case changes the behaviour 
of the reconstruction. A substantial component of the uncertainty is correlated between all masses. This is due to the remaining background, and is most visible in the dM770 scenario, where the signal rate is higher
and thus statistical fluctuations between the different signal samples at the different masses are 
smaller than in the dM1600 case. In the absence of any background, the statistical error on an individual mass determination would shrink to about $0.5\,\GeV$. 

After applying the calibration, we find an excellent agreement with the input mass values, at
the price of the uncertainties increasing by $\sim 50\%$:
\begin{eqnarray}
{\mathrm{dM1600}} &:& \, M_{\wt{\chi}^{\pm}_1}^{\mathrm{cal}}\,=\,166.2\pm 2.0\,\GeV \quad (M_{\wt{\chi}^{\pm}_1}^{\mathrm{true}}\,=\,165.8\,\GeV)\,, \label{mass-char-124}\\
{\mathrm{dM770}} &:& \, M_{\wt{\chi}^{\pm}_1}^{\mathrm{cal}}\,=\,167.3\pm 1.5\,\GeV \quad (M_{\wt{\chi}^{\pm}_1}^{\mathrm{true}}\,=\,167.4\,\GeV)\,.\label{mass-char-127} 
\end{eqnarray}

\subsubsection*{Reconstruction of the chargino--LSP mass difference}

After a clear observation of the signal in the $\sqrt{s'}$ spectrum and the determination of the 
chargino mass from the endpoint, the background can be further reduced by tightening the missing 
energy cut to $E_{\mathrm{miss}} > 350\,\GeV$. The resulting event count is displayed for both 
scenarios  in the respective last line of Table~\ref{tab:cha1sel_cutflow}.

%%%%%%%%%%%%%%%%%%%%%%%%%%%%%%%%%%%%%%%%%%%%%%%%%%%%%%%%%%%%
\begin{figure}[htb]
%  \begin{center}
\centering
\includegraphics[width=0.49\textwidth]{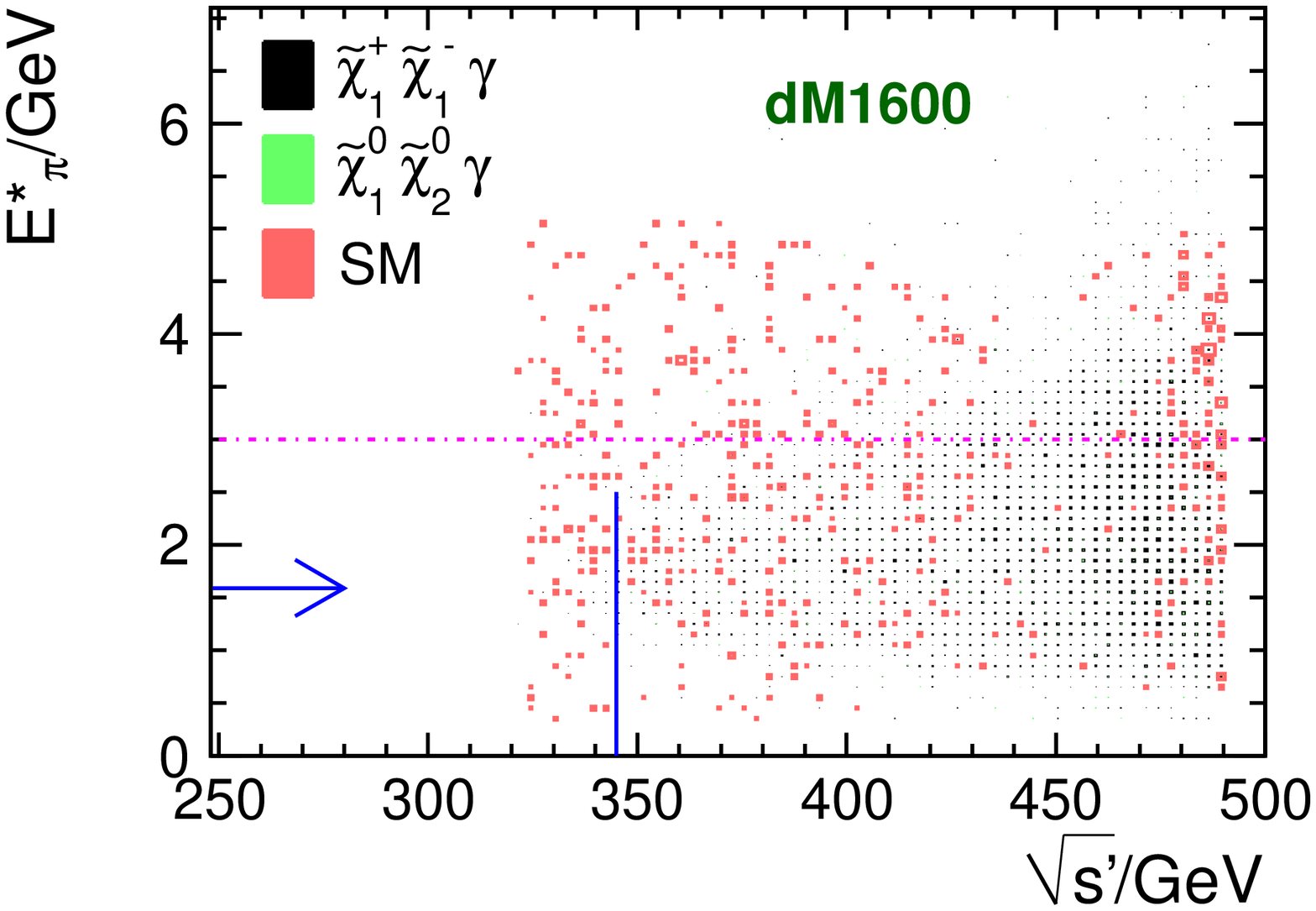}
\hspace{0.1cm}
\includegraphics[width=0.49\textwidth]{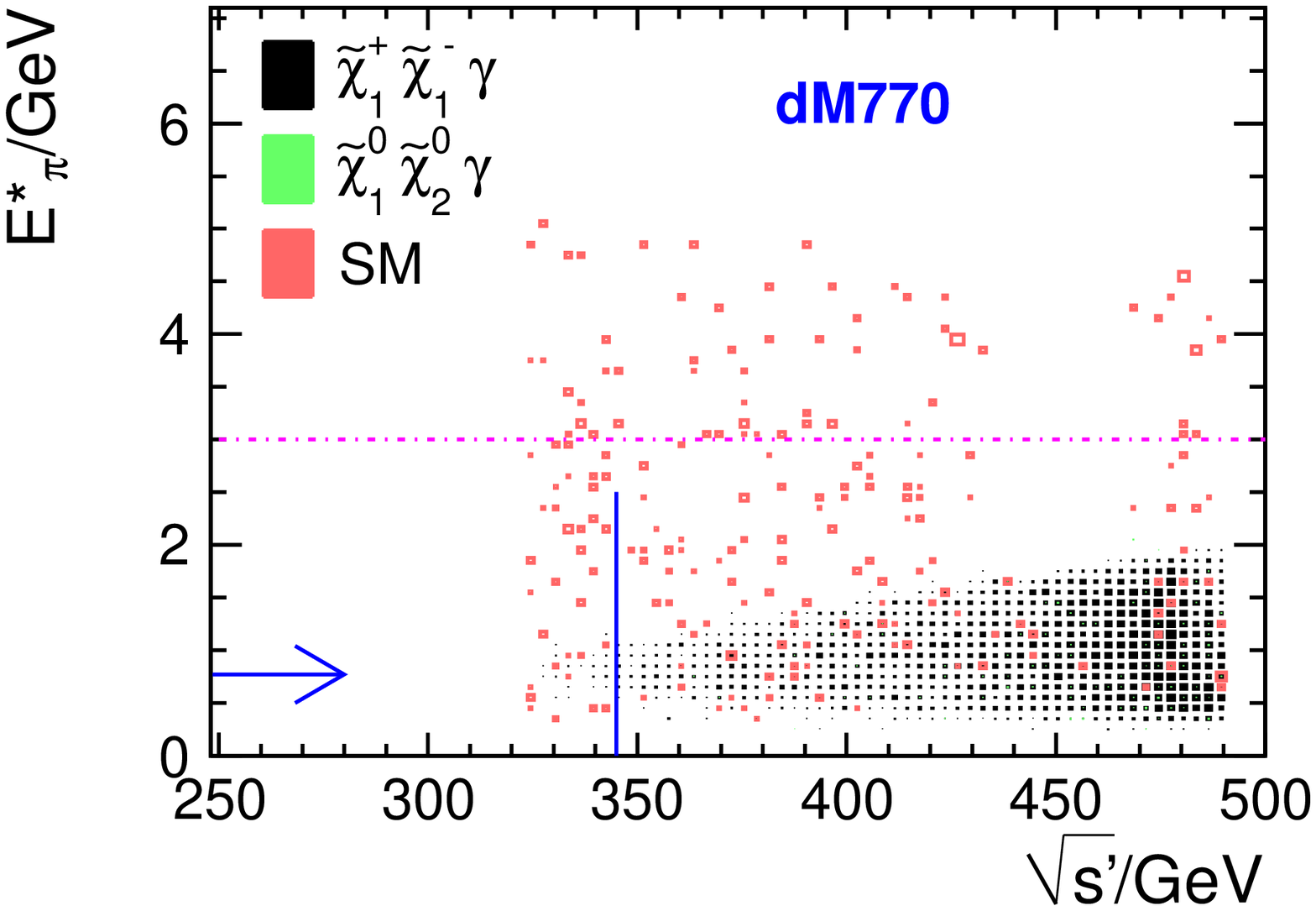}
\vspace{0.1cm}
\includegraphics[width=0.49\textwidth]{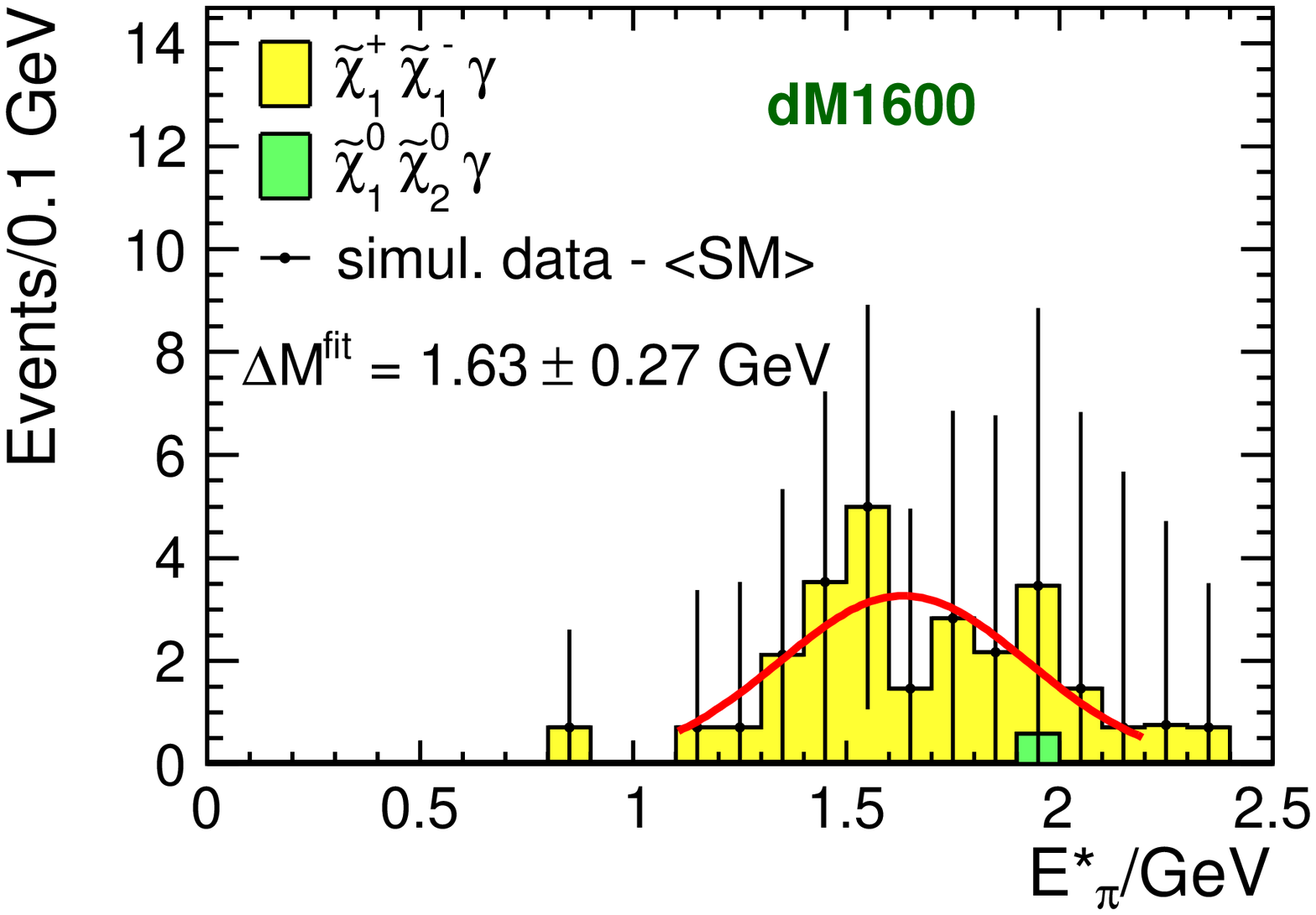}
\hspace{0.1cm}
\includegraphics[width=0.49\textwidth]{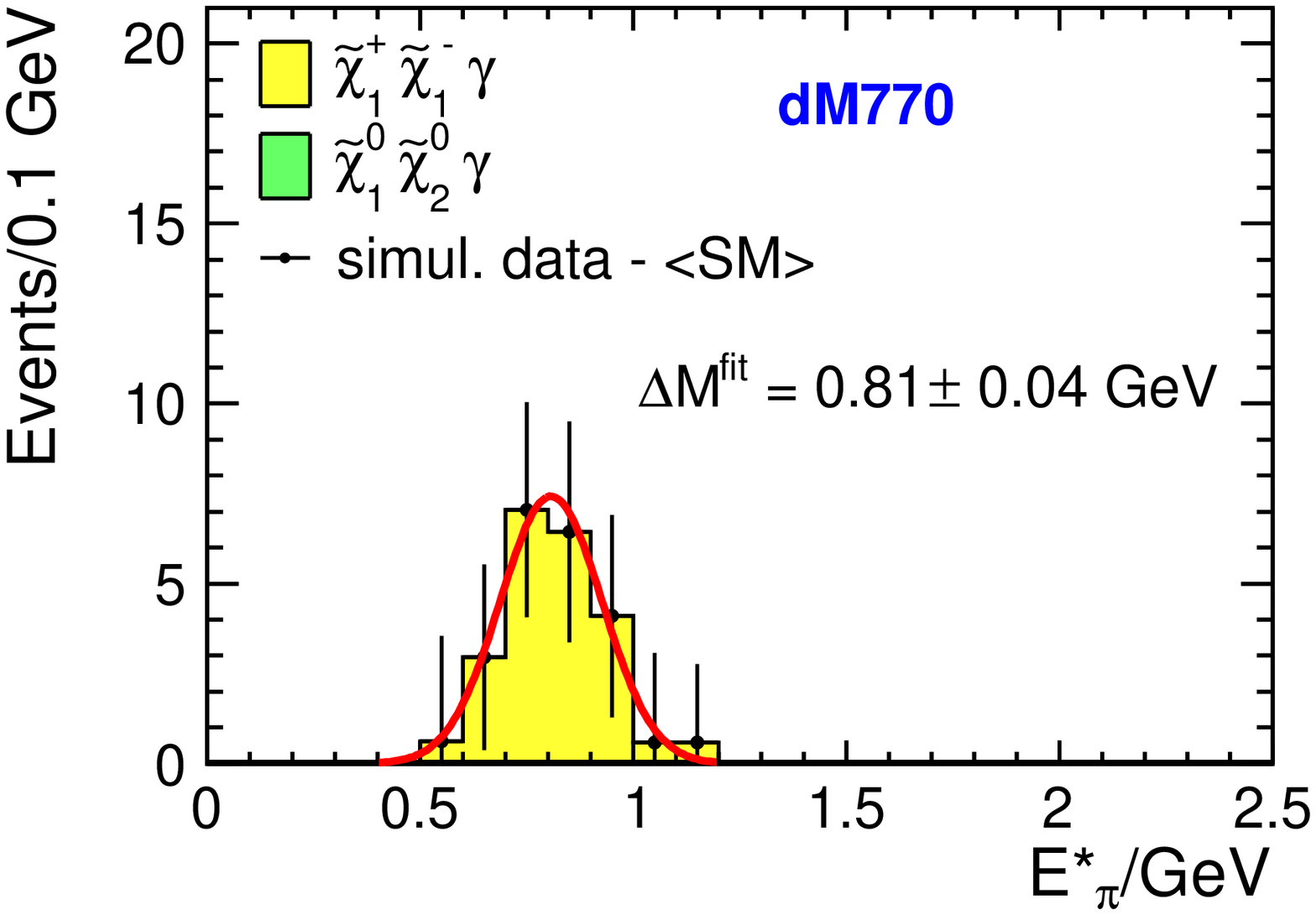}
%\vspace{0.1cm}
%\includegraphics[width=0.49\textwidth]{Eboost_wbkg_mh127.eps}
%  \end{center}
  \caption{Measurement of the chargino--LSP mass difference for an integrated luminosity of 
  $500\,$fb$^{-1}$ with $P(e^+,e^-)=(+30\%,-80\%)$. The upper row shows $E^*_{\pi}$ vs. $\sqrt{s'}$. 
  The horizontal dashed line indicates the cut on $E^*_{\pi}$, the vertical line the cut on $\sqrt{s'}$.
  The arrow indicates the input value of the chargino--LSP mass difference.
  The lower row the $E^*_{\pi}$ after cutting on $\sqrt{s'}$ . Left: dM1600 scenario; Right: dM770 
  scenario.}
\label{fig:deltaMcha1}
\end{figure}
%%%%%%%%%%%%%%%%%%%%%%%%%%%%%%%%%%%%%%%%%%%%%%%%%%%%%%%%%%%%

The mass difference between the chargino and the LSP is very small in both scenarios, 
namely $1.6\,\GeV$ ($770\,\MeV$) in the dM1600 (dM770) scenario. It is nevertheless accessible 
via the energy of the visible chargino decay products in the chargino-pair rest frame, calculated
 according to the definition in Eq.~(\ref{eqn:boostedEnergy}). If the charginos are actually 
 produced at rest, $E^*_{\pi}$ can only take one value, {\it viz}:
\begin{eqnarray}
E^*_{\pi}=\frac{(\mcha{1}-\mneu{1})(\mcha{1}+\mneu{1})+m^2_\pi}{2 \mcha{1}} = \frac{\Delta(M)\Sigma(M) + m^2_\pi}{2\mcha{1}} =
\frac{1}{1/\Delta(M) + 1/\Sigma(M)} +\frac{m^2_\pi}{2\mcha{1}}\,,
\end{eqnarray}
wherein $\Delta(M)$ and $\Sigma(M)$ are the difference and sum of the chargino and the LSP masses, respectively. 
As both the mass difference and $m_\pi$ are several orders of magnitude smaller than the higgsino masses,
$E^*_{\pi}$ is equal to the mass difference itself to a very
good approximation.
This is illustrated by the upper row of plots in Figure~\ref{fig:deltaMcha1}, which show the selected 
events in the $E^*_{\pi}$--$\sqrt{s'}$ plane for both scenarios. The signal events form a distinctive 
triangular region, with a tip just at the mass difference indicated by the blue arrow. We select 
the tip of the triangle by requiring $\sqrt{s'}< 345$~GeV, as indicated by the blue line, and then determine
the tip position from the projection of the remaining events onto the $E^*_{\pi}$ axis.
In a more complex analysis the whole two-dimensional $\sqrt{s'}$ distribution could be exploited to determine 
the mass difference, potentially even together with the cross section.

The resulting $E^*_{\pi}$ distributions are displayed in the lower row of Figure~\ref{fig:deltaMcha1}, 
again for both scenarios. The mass difference is then reconstructed as the mean of a Gaussian fitted 
to the distribution. Due to the limited amount of available SM events in the simulation, we subtract 
the SM background, assuming that precise theoretical predictions and sufficient MC statistics would 
be available for a real ILC measurement. However, the error bars include the full expected statistical 
fluctuations from the background. Especially in the case of the dM1600 scenario, the remaining 
signal is rather small, but still has a significance of more than $5\sigma$.
The mass difference is reconstructed as
\begin{eqnarray}
{\mathrm{dM1600}} &:& \Delta M_{\wt{\chi}^{\pm}_1-\widetilde{\chi}^0_1}^{\mathrm{rec}}\,=\,1630\pm 270\,\MeV \quad (\Delta M_{\wt{\chi}^{\pm}_1-\widetilde{\chi}^0_1}^{\mathrm{true}}\,=\,1600\,\MeV)\,, \label{diff-124}
\\
{\mathrm{dM770}} &:& \Delta M_{\wt{\chi}^{\pm}_1-\widetilde{\chi}^0_1}^{\mathrm{rec}}\,=\,810\pm 40\,\MeV \quad \quad (\Delta M_{\wt{\chi}^{\pm}_1-\widetilde{\chi}^0_1}^{\mathrm{true}}\,=\,770\,\MeV)\,. \label{diff-127}
\end{eqnarray}
These values agree with the input mass differences within the uncertainty.

\subsubsection*{Measurement of the polarised chargino cross sections}

The achievable precision on the polarised cross sections is estimated based on the event counts in 
the final row of Table~\ref{tab:cha1sel_cutflow}, i.e.\ after requiring $E_{\mathrm{miss}} > 350\,\GeV$,
according to
\begin{equation}
\frac{\delta \sigma}{\sigma} = \frac{1}{\sqrt{\epsilon \cdot \pi \cdot \sigma \cdot \int {\mathcal{L}} dt}}\,.
\end{equation} 
Table~\ref{tab:polXsecCha1} displays the corresponding efficiencies ($\epsilon$) and purities ($\pi$) 
as well as the 
resulting precision on the cross sections assuming an integrated luminosity of $500\,$fb$^{-1}$ 
for each polarisation configuration. Since the efficiencies in this case refer to the total number 
of produced chargino events regardless of their decay channel, we display for information the 
branching ratios 
of the decay modes to which the selections are tailored, as well as the assumed production 
cross sections.

\begin{table}[ht!]
\renewcommand{\arraystretch}{1.25}
\centering
%\footnotesize
\begin{tabular}{l cc cc } \toprule
& \multicolumn{2}{c}{$P(e^{+},e^{-})=(+30\%,-80\%)$} &  \multicolumn{2}{c}{$P(e^+,e^-)=(-30\%, +80\%)$} \\ \midrule
& \quad dM1600 \quad \quad & dM770 & \quad dM1600 \quad \quad & dM770  \\ \midrule
 $\int {\mathcal{L}} dt$ & $500\,$fb$^{-1}$ 	& $500\,$fb$^{-1}$ 	&  $500\,$fb$^{-1}$ 	&  $500\,$fb$^{-1}$\\ 
 $\sigma$ & $ 78.7\,$fb  	& $77.0\,$fb   		&  $20.4\,$fb   	&  $19.9\,$fb   \\ 
 BR of selected mode(s)  & $30.5\%$ 		& $34.7\%$ 		&  $30.5\%$ 		&  $34.7\%$ \\ 
\midrule
 efficiency $\epsilon$   & $9.9\%$ 		& $12.1\%$ 		&  $9.5\%$  		&  $12.2\%$\\ 
 purity $\pi$            & $70.1\%$ 		& $85.3\%$ 		&  $36.4\%$ 	 	&  $56.1\%$\\ 
\midrule
 $\delta \sigma/\sigma$ & $1.9\%$ & $1.6\%$ & $5.3\%$  		& $3.8\%$   \\ \bottomrule
 \end{tabular} 
\caption{\label{tab:polXsecCha1} Efficiency, purity, and relative statistical precision on 
the visible cross section for charginos. The cross section
values given here are higher than those displayed in Fig.~\ref{fig:xsec_roots} and~\ref{fig:xsec_pol} since
in the event generation step no generator-level cut on the energy of the ISR photon was applied, 
cf. Sec.~\ref{sec:evtgeneration}.}
\end{table}

For the signal-enriching choice of $P(e^{+},e^{-})=(+30\%,-80\%)$, the cross section can be determined 
to $1.9\%$ ($1.6\%$) in the dM1600 (dM770) scenario. For $P(e^+,e^-)=(-30\%, +80\%)$, the signal 
cross sections are lower, while the dominating SM background processes do not depend on the 
polarisation. Therefore the resolution worsens to $5.3\%$ and $3.8\%$ in the two scenarios, 
respectively. These statistical uncertainties refer to the cross section for the selected semi-leptonic final state.
The branching ratio is determined completely by the mass difference. Therefore the total production cross section
can be extracted, where the additional uncertainty on the branching ratio from the limited knowledge of the
mass difference needs to be included in addition. The corresponding numbers will be given in Section~\ref{sec:paradet}.

\subsection{Neutralino selection}

The neutralino events are selected by the decay to photon and the LSP. For the dM770 scenario, this is 
the dominant decay by far, occurring in $74\%$ of the events. In the dM1600 case, hadronic decay modes 
dominate, but the branching ratio for the very clean radiative decay is still sizeable with $24\%$. 
Therefore, after the common preselection described in Section~\ref{subsec:presel}, events
with only photons in the final state are selected by applying a veto against reconstructed leptons and
hadrons. All ``soft'' photons, i.e.\ all photons apart from the hard ISR photon candidate, are required
 to fulfil $ |\cos\theta_{\gamma_\text{soft}}| < 0.85$. If more than one ``soft'' photon is reconstructed in addition to the hard ISR photon, then the candidate with the highest transverse momentum is considered as the neutralino decay photon.  In analogy to $E_{\pi}^*$ in the chargino case, the variable $E_{\gamma_\text{soft}}^*$ is defined as
\begin{eqnarray}
  E^*_{\gamma_\text{soft}} = \frac{(\sqrt{s} - E_{\gamma}) E_{\gamma_\text{soft}} + \vec{p}_{\gamma_\text{soft}} \cdot \vec{p}_{\gamma}}{\sqrt{s'}},
  \label{eqn:boostedEnergyPhoton}
\end{eqnarray}
and $ E_{\gamma_\text{soft}}^* > 0.5\,\GeV$ is required at the final selection step. 
The resulting event counts are summarised in Table~\ref{tab:neu2sel_cutflow}.

%%%%%%%%%%%%%%%%%%%%%%%%%%%%%%%%%%%%%%%%%%%%%%%%%%%%%%%%%%%%
\begin{table}[ht!]
\renewcommand{\arraystretch}{1.25}
\centering
    \begin{tabular}{l cc cc ccc } \toprule
               & \multicolumn{2}{c}{dM1600} &  \multicolumn{2}{c}{dM770} &  \multicolumn{3}{c}{Standard Model}  \\ \midrule
               &  $\widetilde{\chi}_1^{+} \widetilde{\chi}_1^{-} \gamma $   & $\widetilde{\chi}_1^{0} \widetilde{\chi}_2^{0} \gamma $ &  $\widetilde{\chi}_1^{+} \widetilde{\chi}_1^{-} \gamma $   & $\widetilde{\chi}_1^{0} \widetilde{\chi}_2^{0} \gamma $ & $ee \rightarrow 2,4,6$f  & $e\gamma \rightarrow 3,5$f &  $ \gamma\gamma \rightarrow 2,4$f  \\ \midrule
%                    			 cha124    neu124    cha127    neu127    ee       eg     gg
after preselection     		        & $19872$ & $6365$ & $21558$ & $6872$ & $5731$ & $1.1837\times10^{5} $ & $3.3051\times10^{5} $ \\ 
Photon final state 			&  $53$   & $1733$ & $155$   & $5224$ & $399$  & $1217$ & $2254$  \\ 
$|\cos\theta_{\gamma_\text{soft}}| < 0.85$ 	&  $38$   & $1467$ & $120$   & $4538$ & $233$  & $800$ & $1145$  \\                                        
$ E_{\gamma_\text{soft}}^* > 0.5\,\GeV$ 		&  $19$   & $1395$ & $22$    & $4095$ & $109$  & $242$ & $413$  \\  \midrule
$ E_{\mathrm{miss}} > 350\,\GeV$ 	&  $19$   & $1395$ & $22$    & $4095$ & $90$  & $180$ & $384$  \\ \bottomrule
    \end{tabular}                             
\caption{\label{tab:neu2sel_cutflow} Number of events passing the final neutralino selection, 
following the preselection in Tab.~\ref{tab:presel_cutflow}, for an 
integrated luminosity of $500\,$fb$^{-1}$ and $ P(e^{+},e^{-})=(+30\%,-80\%)$.}                         
  \end{table} 
%%%%%%%%%%%%%%%%%%%%%%%%%%%%%%%%%%%%%%%%%%%%%%%%%%%%%%%%%%%%

%%%%%%%%%%%%%%%%%%%%%%%%%%%%%%%%%%%%%%%%%%%%%%%%%%%%%%%%%%%%
%
\subsubsection*{Neutralino mass reconstruction}
%
%%%%%%%%%%%%%%%%%%%%%%%%%%%%%%%%%%%%%%%%%%%%%%%%%%%%%%%%%%%%
As in the chargino case, the mass of the $\widetilde{\chi}_2^0$ is reconstructed from $\sqrt{s'}$ 
determined using the ISR photon. As opposed to the chargino case, the masses of the two produced 
neutralinos are not equal, but their difference is smaller than the resolution of the $\sqrt{s'}$ 
method. Therefore
we use the approximation of two equal mass particles being produced and calibrate the method in 
the end. 

%%%%%%%%%%%%%%%%%%%%%%%%%%%%%%%%%%%%%%%%%%%%%%%%%%%%%%%%%%%%
\begin{figure}[htb]
%  \begin{center}
\centering
\includegraphics[width=0.49\textwidth]{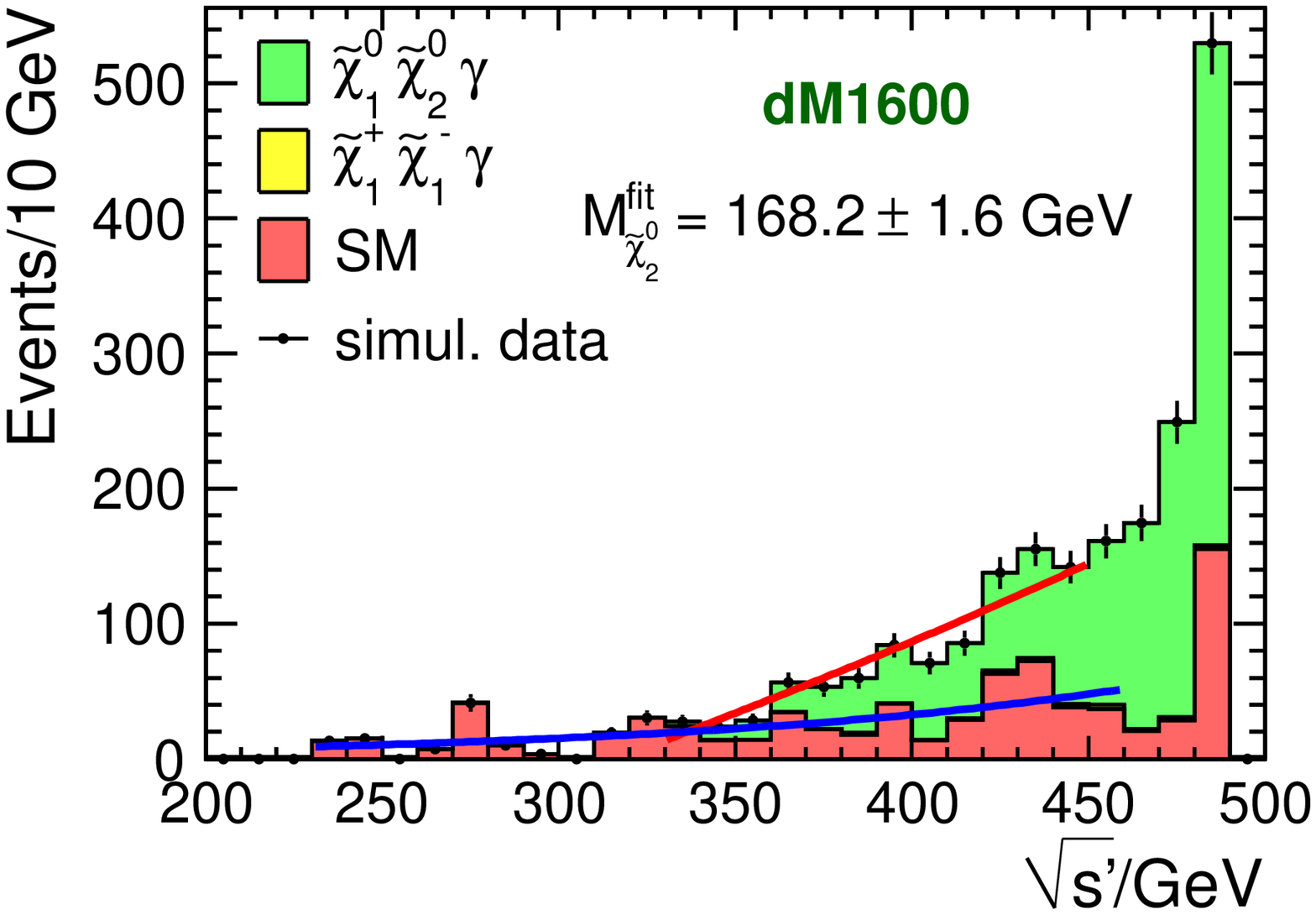}
\hspace{0.1cm}
\includegraphics[width=0.49\textwidth]{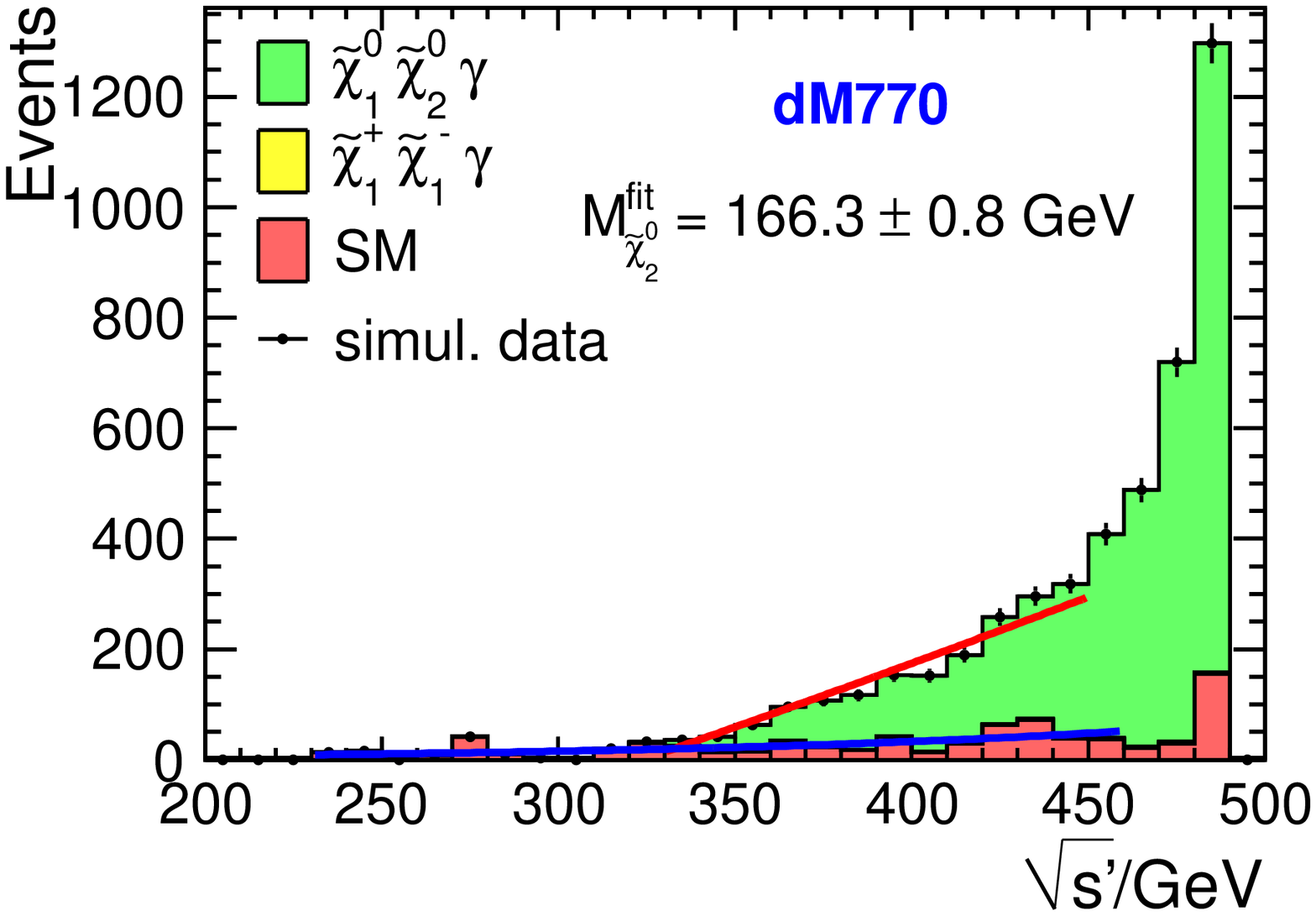}
%  \end{center}
  \caption{Distribution of the reduced centre-of-mass energy ($\sqrt{s'}$)
of the system recoiling against the hard ISR photon for all events passing the neutralino selection 
for an integrated luminosity of $500\,$fb$^{-1}$ with $P(e^+,e^-)=(+30\%,-80\%)$. $M_{\wt \chi^0_2}$ is 
determined from fitting the sum (red curve) of a straight line for the signal and the background 
parametrisation (blue curve) to the distribution near the endpoint. Left: dM1600 scenario; 
Right: dM770 scenario.}
\label{fig:recoilMneu2}
\end{figure}

%%%%%%%%%%%%%%%%%%%%%%%%%%%%%%%%%%%%%%%%%%%%%%%%%%%%%%%%%%%%

Figure~\ref{fig:recoilMneu2} shows the $\sqrt{s'}$  distribution obtained in both scenarios. 
The signal is clearly visible above the background, which has only a negligible contribution from 
chargino production. Like in the chargino case, the SM background is fitted first with an exponential 
(blue line). In a second step, a straight line is added on top of the background to model 
the signal contribution and fitted to the simulated data in the endpoint region (red line). 
Again the parameters of the SM-background function are fixed to 
the values obtained from the SM-only fit in the wider $\sqrt{s'}$ window. The neutralino mass 
is fitted to $168.2 \pm 1.6\,\GeV$ in the dM1600 scenario and to $166.3 \pm 0.8\,\GeV$ in the dM770 case.
Both numbers agree with the input values within $1$--$1.5\,\sigma$.

%%%%%%%%%%%%%%%%%%%%%%%%%%%%%%%%%%%%%%%%%%%%%%%%%%%%%%%%%%%%
\begin{figure}[htb]
%  \begin{center}
\centering
\includegraphics[width=0.49\textwidth]{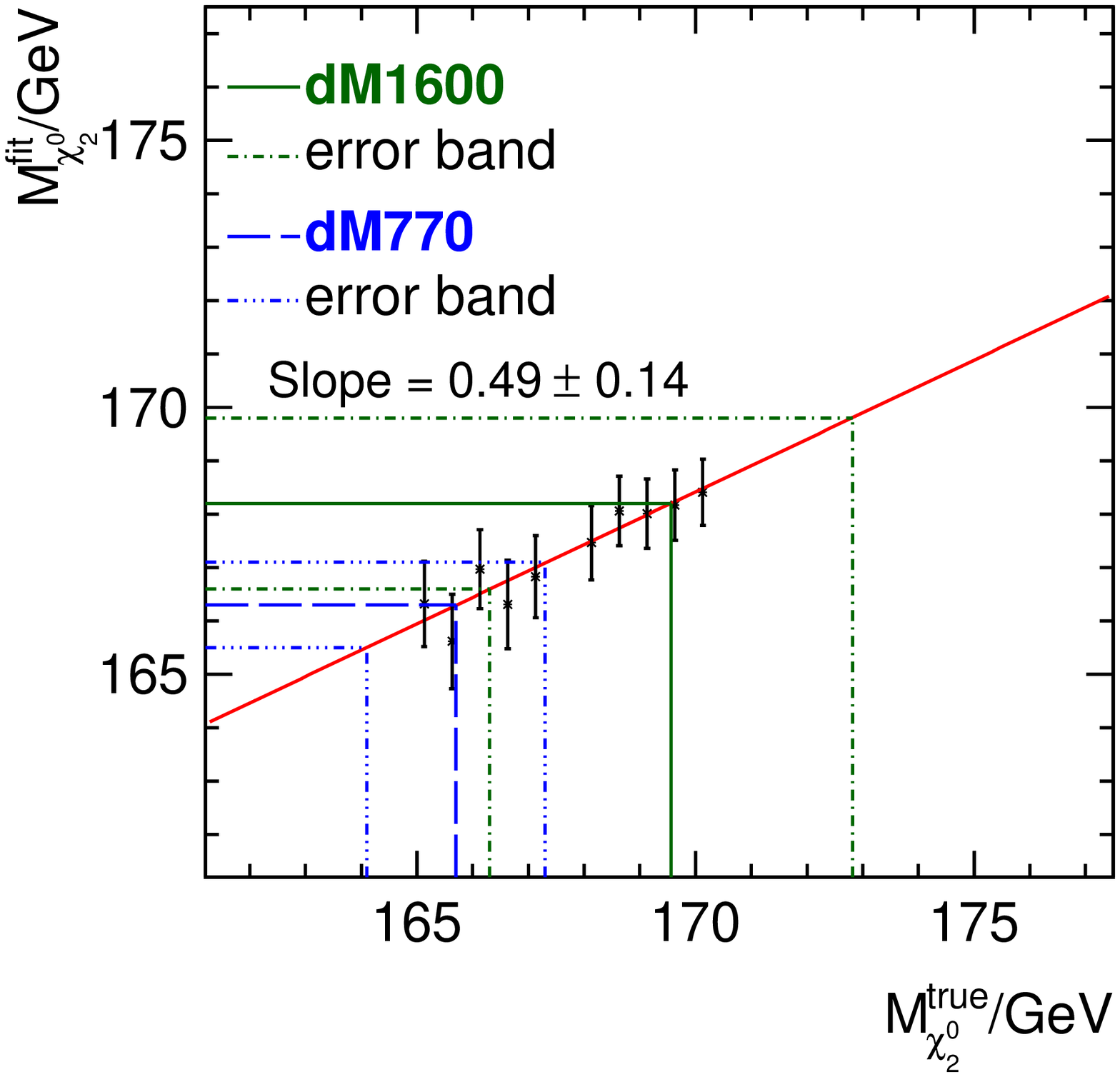}
\hspace{0.1cm}
\includegraphics[width=0.49\textwidth]{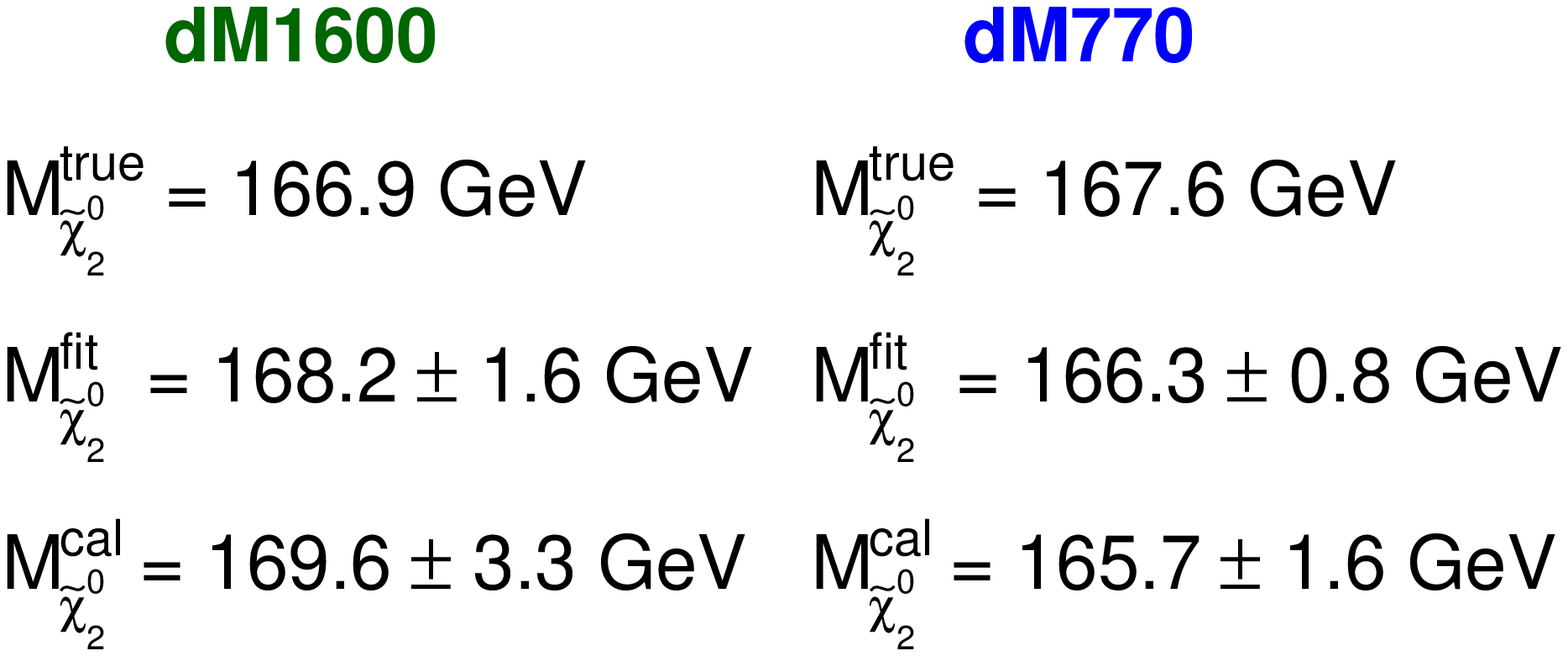}
%  \end{center}
  \caption{Calibration of the $\sqrt{s'}$ method for the neutralino mass determination. Since here
  the same decay mode is selected for both scenarios, the same calibration curve can be used.}
\label{fig:recoilMcalneu2}
\end{figure}
%%%%%%%%%%%%%%%%%%%%%%%%%%%%%%%%%%%%%%%%%%%%%%%%%%%%%%%%%%%%

Again the correlation between fitted mass and input mass is investigated, resulting in the 
calibration curve displayed in Figure~\ref{fig:recoilMcalneu2}. Since in both scenarios the 
same decay modes are selected, there is no reason to assume a different calibration behaviour 
in the other scenario. Therefore the same calibration curve is used. Since its slope is only 
about $0.5$, the final statistical uncertainty on the calibrated neutralino masses grows by a 
factor of $2$. After applying the calibration, the final results are
\begin{eqnarray}
{\mathrm{dM1600}} &:& \, M_{\wt{\chi}^0_2}^{\mathrm{cal}}\,=\,169.6\pm 3.3\,\GeV \quad (M_{\wt{\chi}^0_2}^{\mathrm{true}}\,=\,166.9\,\GeV)\,,\label{mass-neut-124}\\
{\mathrm{dM770}} &:& \, M_{\wt{\chi}^0_2}^{\mathrm{cal}}\,=\,165.7\pm 1.6\,\GeV \quad (M_{\wt{\chi}^0_2}^{\mathrm{true}}\,=\,167.6\,\GeV)\,. \label{mass-neut-127}
\end{eqnarray}

\subsubsection*{Measurement of the polarised neutralino cross sections}
The achievable precision on the polarised cross sections is estimated like in the chargino case, 
based on the event counts given in  Table~\ref{tab:neu2sel_cutflow}. Table~\ref{tab:polXsecNeu2} 
displays the corresponding efficiencies and purities as well as the resulting precision on the 
cross sections assuming an integrated luminosity of $500\,$fb$^{-1}$ for each polarisation 
configuration. Since the efficiencies in this case refer to the total number of produced neutralino 
events regardless of their decay channel, the branching ratios of the radiative decay  are displayed 
in addition for information, as well as the assumed production cross sections.
\begin{table}[ht!]
\renewcommand{\arraystretch}{1.25}
\centering
%\footnotesize
\begin{tabular}{l cc cc} \toprule
            & \multicolumn{2}{c}{$P(e^{+},e^{-})=(+30\%,-80\%)$} &  \multicolumn{2}{c}{$P(e^+,e^-)=(-30\%, +80\%)$} \\ \midrule
            & \quad dM1600 \quad \quad & dM770 & \quad dM1600 \quad \quad & dM770  \\ \midrule
 $\int {\mathcal{L}} dt$ & $500\,$fb$^{-1}$ & $500\,$fb$^{-1}$ &  $500\,$fb$^{-1}$ &  $500\,$fb$^{-1}$ \\ 
 $\sigma$ & $49.0\,$fb  & $48.4\,$fb   &  $38.9\,$fb   &  $38.4\,$fb   \\ 
 BR of selected mode     & $23.6\%$    & $74.0\%$     &  $23.6\%$     &  $74.0\%$ \\ 
\midrule
efficiency  $\epsilon$ 		& $5.8\%$ & $17.1\%$ &  $6.0\%$  &  $17.2\%$\\ 
purity $\pi$     		& $67.4\%$ & $85.8\%$ &  $62.3\%$  &  $82.5\%$\\ 
\midrule
 $\delta \sigma/\sigma$ & $3.2\%$& $1.7\%$  & $3.7\%$  & $1.9\%$   \\ \bottomrule
 \end{tabular} 
\caption{\label{tab:polXsecNeu2} Efficiency, purity, and relative statistical precision on 
the visible cross section for associated neutralino production. The cross section
values given here are higher than those displayed in Fig.~\ref{fig:xsec_roots} and~\ref{fig:xsec_pol} since
in the event generation step no generator-level cut on the energy of the ISR photon was applied, 
cf. Sec.~\ref{sec:evtgeneration}. }
\end{table}

For neutralino production, the total cross section is much less sensitive to beam polarisation than for chargino production, cf.\ Figure~\ref{fig:xsec_roots}. Therefore, the precision depends only marginally on the beam polarisation. Due to 
the different branching ratios for the radiative decay, the cross sections can be determined more
 precisely, to about $1.7\%$ in the dM770 scenario, while the measurements are about a factor two 
 less precise in the  dM1600 case. 

\section{Parameter determination \label{sec:paradet} }

In the final step of the analysis, we estimate the accuracy to which the underlying MSSM parameters 
can be determined. As discussed in Section~\ref{sec:pheno}, there are four real parameters that 
define the chargino and neutralino sector at tree level:
\begin{equation} \label{eq:parameters}
  M_1, \quad  M_2, \quad \mu,  \quad \tan\beta\,.
\end{equation}
We consider both $M_1 > 0$ and $M_1 < 0$, while the sign of $\mu$ cannot be resolved due to its negligible impact on observables.
They can be extracted using the measurements discussed in the previous section: the masses of the light 
chargino $\widetilde{\chi}_1^\pm$  and the neutralino $\widetilde{\chi}_2^0$, the mass difference 
$M_{\widetilde{\chi}_1^\pm}-M_{\widetilde{\chi}_1^0}$, the chargino pair production cross section and the 
$\widetilde{\chi}_1^0\widetilde{\chi}_2^0$ production cross section. The respective values and 
uncertainties are summarised in Tab.~\ref{tab:input}. Since we have not performed a detector 
simulation for $\sqrt{s} = 350$~GeV, the uncertainties in this case are obtained by rescaling 
the $\sqrt{s} = 500$~GeV result. The ratio of the cross sections is $\sim 30$ and we conservatively scale the errors by $\sqrt{30}$. One should note 
that at $\sqrt{s}\,=\,350$~GeV the SM background contributions are expected to be smaller, therefore the errors 
are likely to be lower than our current estimate.

\begin{table} \renewcommand{\arraystretch}{1.4}  \begin{center}
 \begin{tabular}{l r r}\toprule
   Observable &  {dM1600} &{dM770} \\ \midrule
   $M_{\widetilde{\chi}_1^\pm}$ [GeV] &  $166.2 \pm 2.0$  & $167.3\pm 1.5$\\
   $M_{\widetilde{\chi}_2^0}$ [GeV]&  $169.6 \pm 3.3$  & $165.7 \pm 1.6$\\
   $M_{\widetilde{\chi}_1^\pm} - M_{\widetilde{\chi}_1^0}$ [GeV] & $1.63\pm0.27$ & $0.81 \pm 0.04$ \\
   $\delta\sigma/\sigma(\widetilde{\chi}_1^+ \widetilde{\chi}_1^-\gamma)^{500}_{(-0.3,0.8)}$  & $16\% $ & $4.3\% $ \\
   $\delta\sigma/\sigma(\widetilde{\chi}_1^+ \widetilde{\chi}_1^-\gamma)^{500}_{(0.3,-0.8)}$  & $15\% $ & $2.6\% $ \\
   $\delta\sigma/\sigma(\widetilde{\chi}_1^+ \widetilde{\chi}_1^-\gamma)^{350}_{(-0.3,0.8)}$  & $ 33\%$ & $ 20\% $ \\
   $\delta\sigma/\sigma(\widetilde{\chi}_1^+ \widetilde{\chi}_1^-\gamma)^{350}_{(0.3,-0.8)}$  & $ 18\% $ & $ 9.0\% $ \\
   $\delta\sigma/\sigma(\widetilde{\chi}_1^0 \widetilde{\chi}_2^0\gamma)^{500}_{(-0.3,0.8)}$  & $4.0\%$ & $2.4\%$ \\
   $\delta\sigma/\sigma(\widetilde{\chi}_1^0 \widetilde{\chi}_2^0\gamma)^{500}_{(0.3,-0.8)}$  & $3.5\%$ & $ 2.3\% $\\
   $\delta\sigma/\sigma(\widetilde{\chi}_1^0 \widetilde{\chi}_2^0\gamma)^{350}_{(-0.3,0.8)}$  & $20\%$ & $10\%$ \\
   $\delta\sigma/\sigma(\widetilde{\chi}_1^0 \widetilde{\chi}_2^0\gamma)^{350}_{(0.3,-0.8)}$  & $18\% $ & $ 9.4\% $\\ \bottomrule
 \end{tabular}
\caption{Observables used in the fit: masses with uncertainties taken from Eqs.~\eqref{mass-char-124}, \eqref{mass-char-127}, \eqref{diff-124}, \eqref{diff-127},
\eqref{mass-neut-124}, \eqref{mass-neut-127}; relative  cross sections uncertainties from Tabs.~\ref{tab:polXsecCha1} and~\ref{tab:polXsecNeu2} including systematic uncertainty due to branching ratios as discussed in text. The subscript on $\delta\sigma/\sigma$ denotes the beam polarisation  
$P(e^+,e^-)$ and the superscript the centre-of-mass energy in GeV.\label{tab:input}}
 \end{center}
\end{table}

The cross section errors obtained in Sec.~\ref{sec:expprec} did not take into account the uncertainty 
on the branching ratios. The branching ratios  only depend on the available phase space for the decay 
of the virtual $W$ boson for charginos, given by the chargino--LSP mass difference.  With the 
previously obtained precision 
on the mass difference, the uncertainty on the BR of the selected decay modes
has to be taken into account. 
In case of the dM1600 scenario, due to the significant uncertainty on the chargino--LSP mass 
difference, it amounts to $15\%$ for the chargino cross section measurement. For the dM770 scenario 
it is $2\%$, comparable to the experimental uncertainty. These additional uncertainties are 
included in Table~\ref{tab:input}.

Using \texttt{Minuit}~\cite{James:1975dr} we minimise a $\chi^2$ function defined as
\begin{equation}
\chi^2 = \sum_i \left| \frac{{\mathcal{O}}_i - \bar{{\mathcal{O}}}_i }
                { \delta {\mathcal{O}}_i } \right|^2 ,
\end{equation}
where the sum runs over the input observables ${\mathcal{O}}_i$, with their corresponding experimental 
uncertainties
$\delta {\mathcal{O}}_i$. The theoretical values calculated using the
fitted MSSM parameters, Eq.~\eqref{eq:parameters}, are denoted by
$\bar{{\mathcal{O}}}_i$. Note that throughout the simulation the one-loop corrected masses, given by Eqs.~\eqref{124} and \eqref{127}, were used. The difference between the tree-level values and the ones obtained from \texttt{SOFTSUSY} is accordingly taken into account in the fit in order to ensure consistency. The variation of the observables, with respect to the central values, on the SUSY parameters, Eq.~\eqref{eq:parameters}, is in the following calculated at the tree level. Therefore, the results of the fit are to be understood as the tree-level fit. 

It turns out that using the experimental information one cannot constrain $\tan\beta$. This is not 
very surprising, as with the values of parameters used here, there is very little mixing between 
higgsinos and gauginos which could have provided some sensitivity to $\tan\beta$. As can be seen from Eq.~\eqref{eq:masses}, the chargino and neutralino masses depend only on $\sin2\beta$ and the sensitivity is rather weak, especially in the large $\tan\beta$ limit.   In order to have a converging 
fit we therefore fix $\tan\beta$ to values in the range $\tan\beta \in [1, 60]$ and fit the remaining 
three parameters of Eq.~\eqref{eq:parameters}. In principle, the cross sections depend also on the 
sneutrino and selectron masses but since the couplings to higgsinos are negligible, no dependence on these 
parameters enters our observables.

\begin{figure}[tp]
%  \begin{center}
\centering
\includegraphics[width=1.0\textwidth]{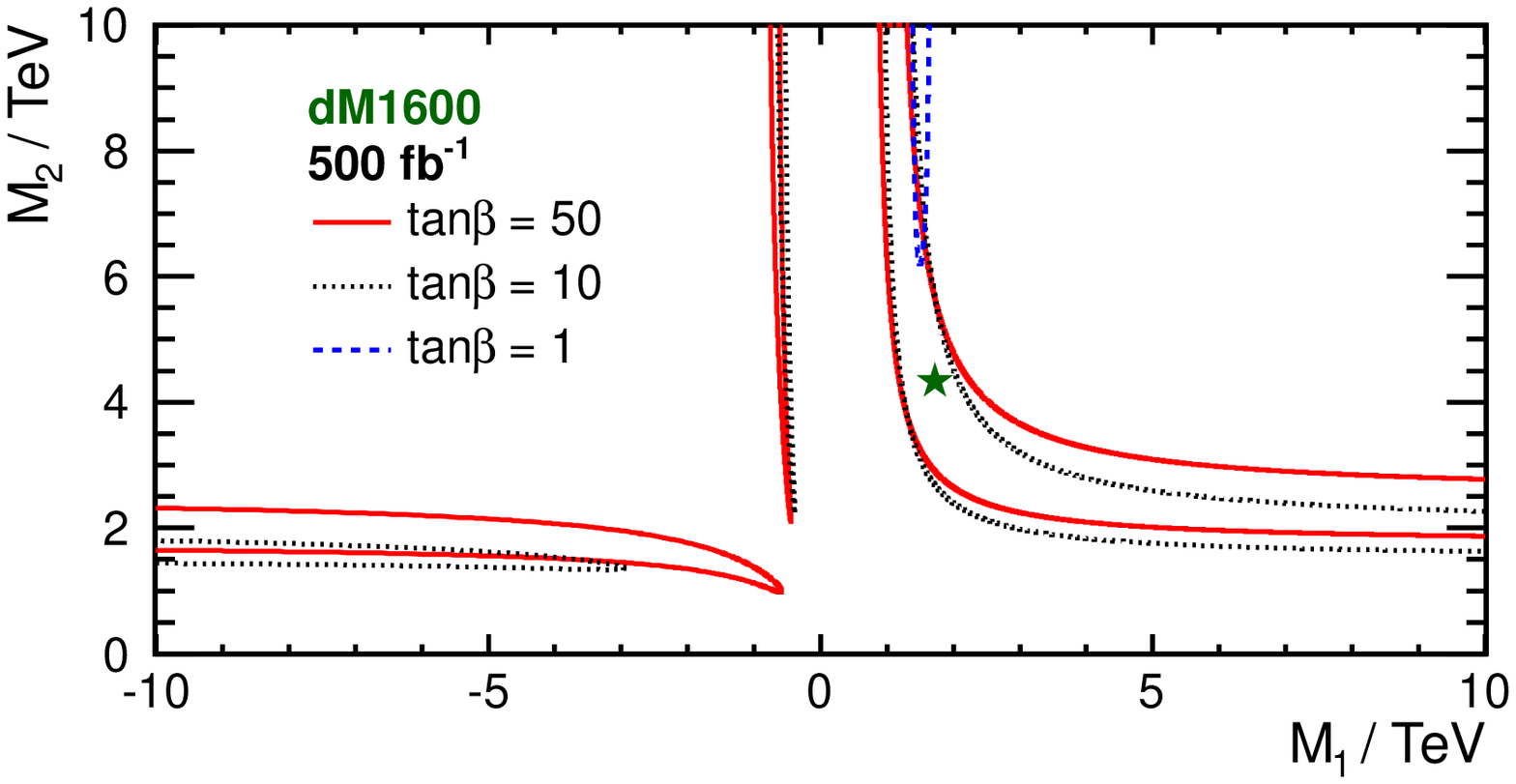}
\includegraphics[width=1.0\textwidth]{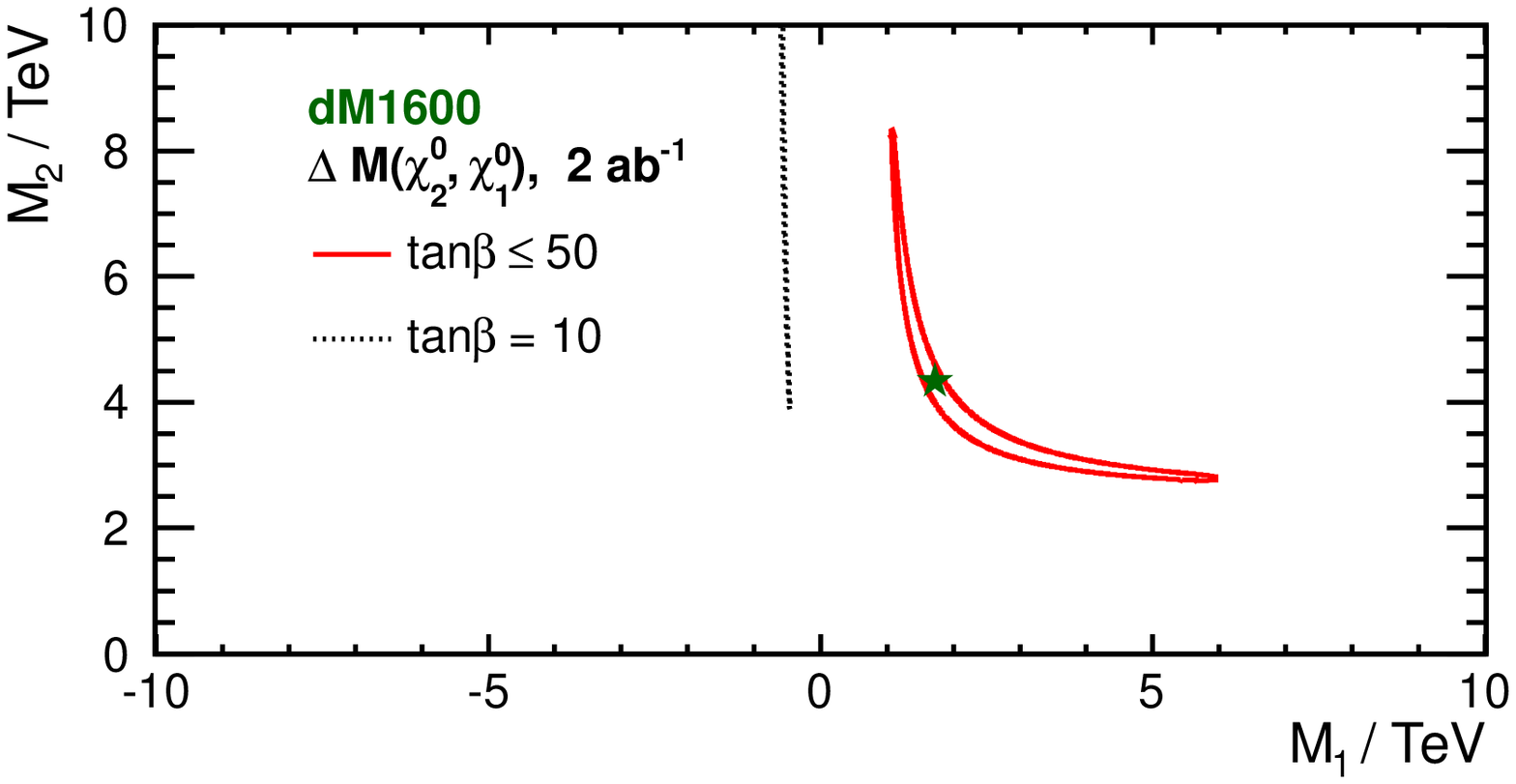}
%  \end{center}
 \caption{The 1-$\sigma$ contours for determination of $M_1$ and $M_2$ in scenario 
 dM1600. The input values of $M_1$ and $M_2$ are indicated by the star, and $(\tan\beta)_\text{true}\,=\,44$. \textbf{Top:} The fit using input values listed in Tab.~\ref{tab:input} for three different values of $\tan\beta$. 
% Black (dotted), red (solid), and blue (dashed) contours are for $\tan\beta = 50$, $10$ and $1$, respectively. 
\textbf{Bottom:} Projected fit results after a high-luminosity run, 
 $\int\,\mathcal{L}\,dt = 2~\mathrm{ab}^{-1}$, with experimental errors improved by factor 2 and with 
 the mass difference $M_{\widetilde{\chi}^0_2} - M_{\widetilde{\chi}^0_1}$ measurement included for a fixed $\tan\beta = 50$ and $\tan\beta = 10$. For $M_1>0$, other values of $\tan\beta$ do not further extend allowed regions and the respective contours are inside the one for $\tan\beta = 50$. For $M_1<0$, 
 solutions exist only for $8 \leq \tan\beta \leq 16$ near $M_1 \sim -0.5\ \mathrm{TeV}$.}
\label{fig:deltaMcal124}
\end{figure}

\begin{figure}[tp]
%  \begin{center}
\centering
\includegraphics[width=1.0\textwidth]{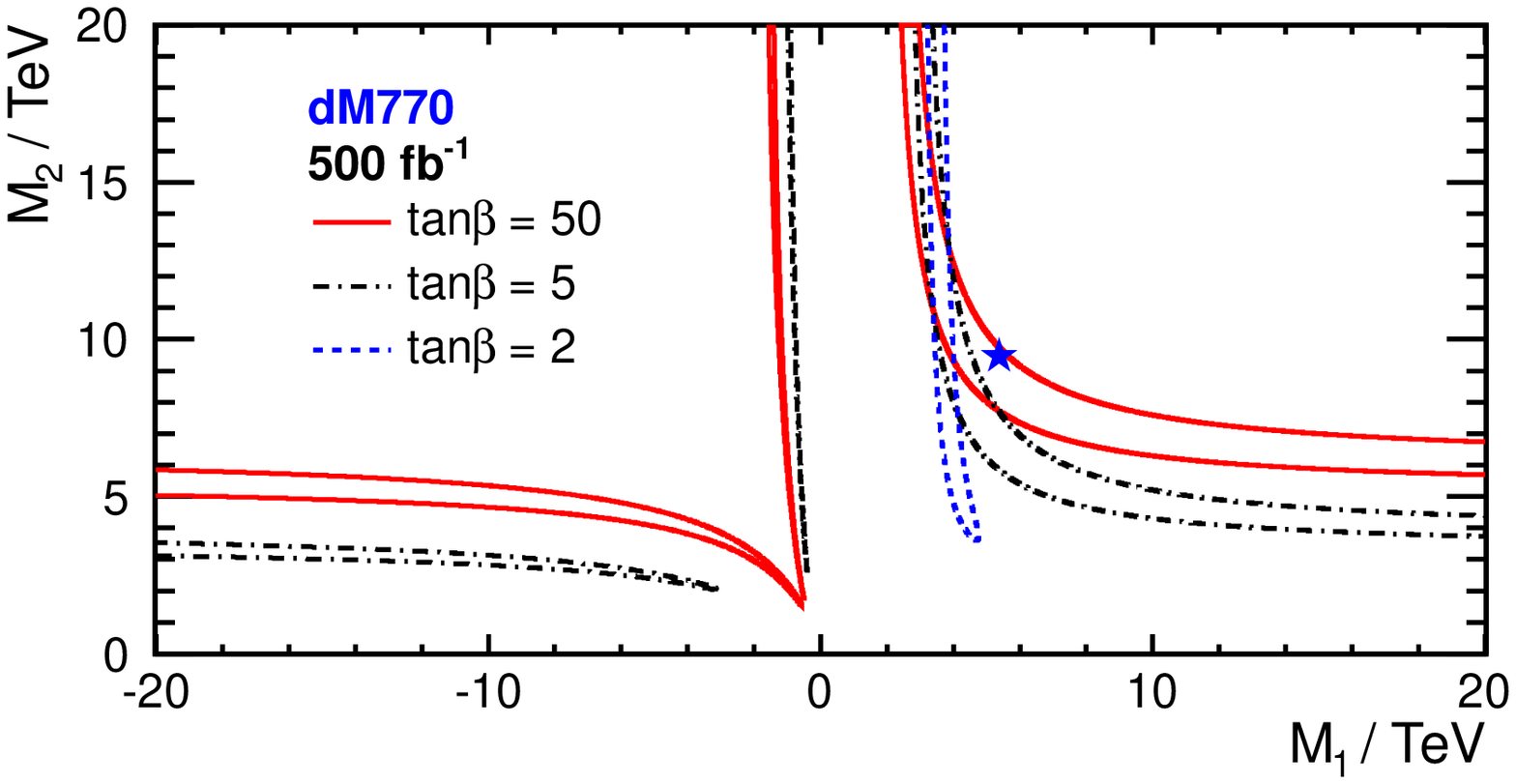}
\includegraphics[width=1.0\textwidth]{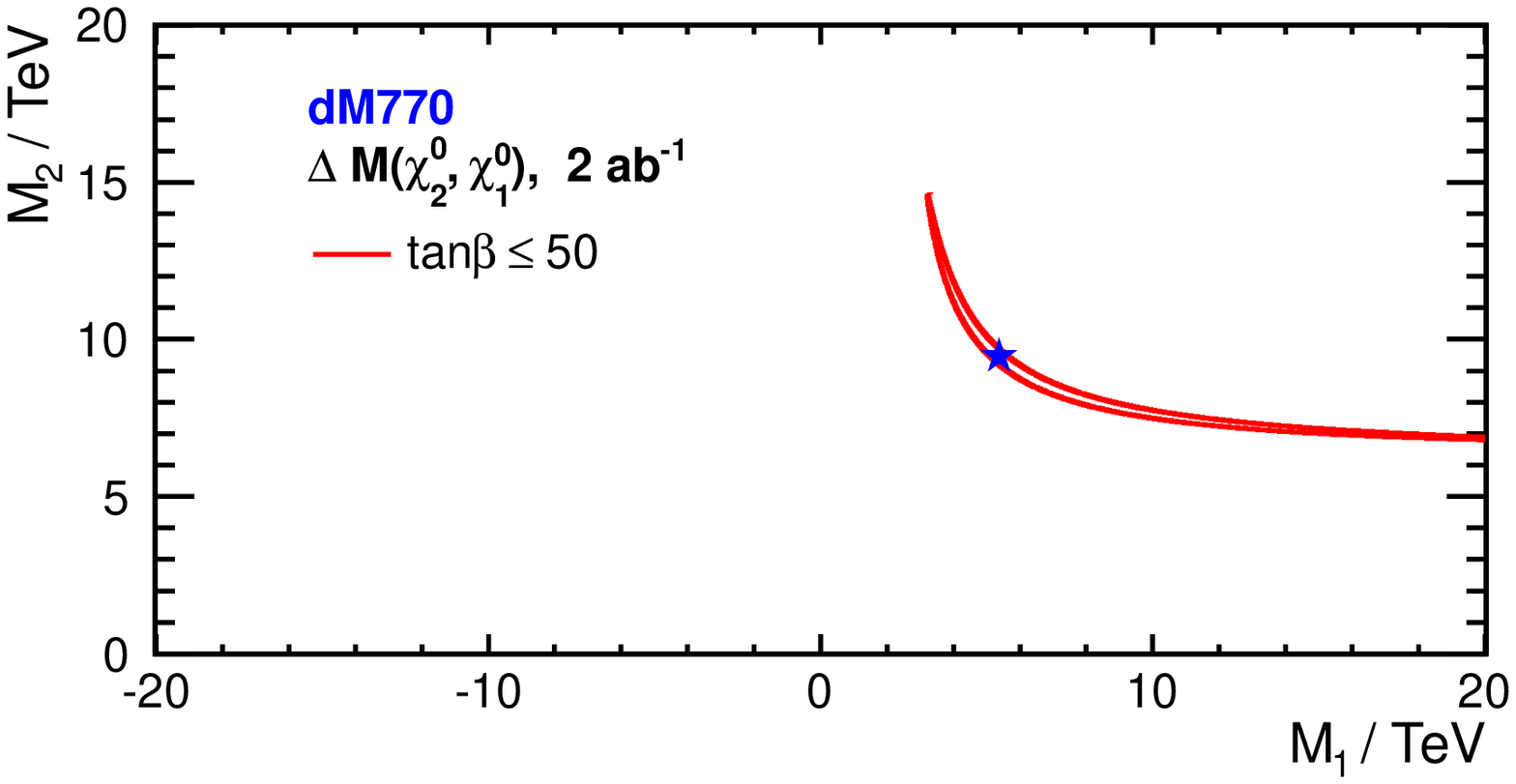}
%  \end{center}
\caption{The 1-$\sigma$ contours for determination of $M_1$ and $M_2$ in scenario
 dM770. The input values of $M_1$ and $M_2$ are indicated by the star, and  $(\tan\beta)_\text{true}\,=\,48$. \textbf{Top:} The fit using input values listed in Tab.~\ref{tab:input} for three different values of $\tan\beta$. 
%Black (dotted), red (solid) and blue (dashed) contours are for $\tan\beta = 50$, $5$ and $2$, respectively. 
\textbf{Bottom:} Projected fit results after a high-luminosity run, 
 $\int\,\mathcal{L}\,dt = 2~\mathrm{ab}^{-1}$, with experimental errors improved by factor 2 and with 
 the mass difference $M_{\widetilde{\chi}^0_2} - M_{\widetilde{\chi}^0_1}$ measurement included for 
 fixed $\tan\beta = 50$. Other values of $\tan\beta$ do not further extend allowed regions 
 and the respective contours are inside the shown one. }\label{fig:deltaMcal127}
\end{figure}

The results of the fit are shown in Tables~\ref{tab:fit124} and \ref{tab:fit127}.
The crucial observable for the determination of $M_1$ and $M_2$ is the mass difference between the LSP 
and the light chargino. On the other hand, the cross section measurements improve the determination of the 
$\mu$ parameter. In general, only lower limits can be obtained for $|M_1|$ and $M_2$, which are given in  Tables~\ref{tab:fit124} and \ref{tab:fit127} for positive and negative $M_1$. However, it turns 
out that $M_1$ and $M_2$ are strongly correlated and the allowed regions are only narrow strips in the
$M_1$--$M_2$ parameter space, see Figs.~\ref{fig:deltaMcal124} and \ref{fig:deltaMcal127}. The $\mu$ 
parameter can be determined with an accuracy $\sim 2.5$~GeV in dM770 scenario and $6.8$~GeV in 
dM1600 scenario.

\begin{table}[t]
\renewcommand{\arraystretch}{1.25}
% \begin{center}
\centering
 \begin{tabular}{cr >{\raggedleft}m{1.5cm} >{\raggedleft}m{2cm} >{\raggedleft}m{2cm} >{\raggedleft}m{2cm} R{2cm}  }\toprule
 & &  & \multicolumn{2}{r}{\phantom{}$\sqrt{s} = 500$ GeV\phantom{0\&}} & \multicolumn{2}{r}{$\sqrt{s} = 350\ \&\ 500$ GeV}  \tabularnewline \midrule
 & & input & lower & upper  & lower & upper   \tabularnewline \midrule
    \multirow{3}*{\begin{sideways} $500/\mathrm{fb}$ \end{sideways}}
 &  $|M_1|$ [TeV] & $1.7$ & $\sim 0.8$ $(0.4)$ & no & $\sim 0.8$ $(0.4)$ & no  \tabularnewline
 &  $M_2$ [TeV]& $4.36$ & $\sim 1.5$ $(1.0)$& no & $\sim 1.5$ $(1.0)$ & no  \tabularnewline
 &  $\mu$ [GeV]& $165.66$ & $165.2$ & $172.5$  & $165.4$ & $170.2$   \tabularnewline \midrule
   \multirow{3}*{\begin{sideways}$2/\mathrm{ab}$ \end{sideways}}
 &  $|M_1|$ [TeV]& $1.7$  & $\sim 1.0$ $(0.4)$& $\sim 6.0$ $(0.6)$& $\sim 1.0$ $(0.4)$ & $\sim 6.0$ $(0.6)$\tabularnewline 
 &  $M_2$ [TeV]& $4.36$ & $\sim 2.5$ $(3.5)$& $\sim 8.5$ (no)& $\sim 2.5$ $(3.5)$ & $\sim 8.5$ (no)   \tabularnewline
 &  $\mu$ [GeV]& $165.66$ & $166.2$ & $170.1$ & $166.4$ & $170.0$   \tabularnewline \bottomrule
 \end{tabular}
% \end{center}
 \caption{The 1-$\sigma$ allowed ranges for the parameter fit in scenario dM1600.  $\tan\beta$ is allowed to vary in the range $[1,60]$. Values in parentheses are for $M_1 < 0$.
 Determination of the parameters for integrated luminosities of $ \int {\mathcal{L}}\, dt = 500\ \mathrm{fb}^{-1}$ and $ 2\ \mathrm{ab}^{-1}$ per polarisation configuration is shown. The high-luminosity fit also includes the mass difference between neutralinos, $M_{\widetilde{\chi}^0_2} - M_{\widetilde{\chi}^0_1}$. For an exact shape of the allowed $M_1$--$M_2$ region, see Fig.~\ref{fig:deltaMcal124}.  The input values, see Eq.~\eqref{124}, are also shown.  \label{tab:fit124}}
\end{table}

\begin{table}[t]
% \begin{center}
\renewcommand{\arraystretch}{1.25}
\centering
 \begin{tabular}{cr >{\raggedleft}m{1.5cm} >{\raggedleft}m{2cm} >{\raggedleft}m{1.5cm} >{\raggedleft}m{2cm} R{1.5cm}  }\toprule
  & & & \multicolumn{2}{r}{\phantom{35}$\sqrt{s} = 500$ GeV\phantom{0\&}} & \multicolumn{2}{r}{$\sqrt{s} = 350\ \&\ 500$ GeV}  \tabularnewline\midrule
  &   & input    & lower       & upper     & lower        & upper  \tabularnewline \midrule
    \multirow{3}*{\begin{sideways} $500/\mathrm{fb}$ \end{sideways}}      
  & $|M_1|$ [TeV]& $5.3$  &$\sim 2$ $(0.3)$ & no & $\sim 2$ $(0.3)$ & no          \tabularnewline
  & $M_2$ [TeV]& $9.51$  &$\sim 3$ $(1.2)$ & no & $\sim 3$ $(1.2)$ & no          \tabularnewline
  & $\mu$ [GeV]& $167.22$ &$164.8$      & $167.8$   & $165.2$       & $167.7$     \tabularnewline \midrule 
  \multirow{3}*{\begin{sideways}$2/\mathrm{ab}$ \end{sideways}}
  & $|M_1|$ [TeV]& $5.3$  & $\sim 3$ & no & $\sim 3$ & no          \tabularnewline
  & $M_2$ [TeV]& $9.51$  & $\sim 7$ & $\sim 15 $ & $\sim 7$ & $\sim 15$          \tabularnewline
  & $\mu$ [GeV]& $167.22$ & $165.2$ & $167.4$ & $165.3$ & $167.4$    \tabularnewline \bottomrule
 \end{tabular}
% \end{center}
 \caption{The 1-$\sigma$ allowed ranges for the parameter fit in scenario dM770.  $\tan\beta$ is allowed to vary in the range $[1,60]$. Values in parentheses are for $M_1 < 0$ if solutions exist.
 Determination of the parameters for integrated luminosities of $ \int {\mathcal{L}}\, dt = 500\ \mathrm{fb}^{-1}$ and $  2\ \mathrm{ab}^{-1}$ per polarisation configuration is shown.  The high-luminosity fit also includes the mass difference between neutralinos, $M_{\widetilde{\chi}^0_2} - M_{\widetilde{\chi}^0_1}$. For an exact shape of the allowed $M_1$--$M_2$ region, see Fig.~\ref{fig:deltaMcal127}. The input values, see Eq.~\eqref{127}, are also shown. \label{tab:fit127}}
\end{table}

Inclusion of the cross section measurements at $\sqrt{s}=350$~GeV only affects the determination of 
the $\mu$ parameter. The low cross section results in a large error, larger by a factor of $5.5$ 
than for $\sqrt{s} = 500$~GeV. However, being close to the production threshold and on the 
quickly rising slope of the cross section, see Fig.~\ref{fig:xsec_roots}, helps to improve the
determination of the $\mu$ parameter in both scenarios. Taking the cross section at a higher centre-of-mass 
energy, e.g.\ $\sqrt{s}=370$~GeV, does not yield a significant improvement. The increased 
statistics does not compensate the advantage of being close to the production threshold. 
The masses and mass differences require larger statistics, and therefore, measurements at 
higher centre-of-mass energies (e.g.\ at $500\,$GeV) are indispensable. Finally, we note that the main driver 
for the rather high $\mu$ error is the variation of $\tan\beta$, which shifts the central values 
of the $\mu$ parameter by approximately $\sim 1$~GeV. This additional dependence will remain 
regardless of additional 
measurements and/or increased accuracy.

In order to better understand the interdependence of the $M_1$ and $M_2$ determination, in the 
top panels of Figs.~\ref{fig:deltaMcal124} and \ref{fig:deltaMcal127} we show the 
1-$\sigma$ contours in the $M_1$--$M_2$ plane for an integrated luminosity of $500\ \mathrm{fb}^{-1}$. Different contours correspond to 
different values of $\tan\beta$. For low values of either $M_1$ or $M_2$ the allowed region 
for the other parameter extends to arbitrary high values and cannot be reliably constrained. 
Significant variation with $\tan\beta$ is also visible, especially in case of scenario dM770. Both cases of the sign of the $M_1$ parameter are considered. In general, for $M_1 < 0$ lower absolute values of gaugino masses are consistent with data. In the most extreme cases, when $M_1 \gtrsim -800 \gev$, a direct production of $\wt \chi^0_3$ could be possible at the ILC with an increased centre-of-mass energy. We also note that for negative $M_1$ low values of $\tan\beta$ are experimentally excluded since they lead to an inverted mass hierarchy between the light chargino and the lightest neutralino, $m_{\wt \chi^\pm_1} < m_{\wt \chi^0_1}$.

Finally, we assess how much of an improvement can be expected after a high-luminosity run with a total of 
$ \int {\mathcal{L}} dt = 2\ \mathrm{ab}^{-1}$ integrated luminosity at $\sqrt{s} = 500$~GeV. We assume that with the increased luminosity the experimental errors would be reduced 
by a factor of 2. We furthermore include the measurement of the mass difference 
$M_{\widetilde{\chi}^0_2} - M_{\widetilde{\chi}^0_1}$, which was not studied here. For the $\mu$ parameter the increased luminosity narrows the allowed region by 2--3.5~GeV, as can be seen in the last rows of Tabs.~\ref{tab:fit124} and \ref{tab:fit127}. In case of the $M_1$ and $M_2$ parameters the results are shown in the bottom panels 
of Figs.~\ref{fig:deltaMcal124} and \ref{fig:deltaMcal127} for scenarios dM1600 and dM770, 
respectively. The neutralino mass difference, $M_{\widetilde{\chi}^0_2} - M_{\widetilde{\chi}^0_1}$, helps to break the 
correlation between $M_1$ and $M_2$ from the chargino--LSP mass difference, since it exhibits a different dependence on the fundamental parameters, in particular by introducing $\tan\beta$, see 
Eqs.~\eqref{eq:chadiff} and \eqref{eq:neudiff}. Hence, the inclusion of the neutralino mass difference helps to constrain the parameters in the low $\tan\beta$ regime, where the $\tan\beta$ dependence is numerically most significant. This effect would not be achieved by solely increasing the luminosity. The absolute mass measurements, which are obviously improved with higher statistics, have only a very limited impact on determination of the multi-TeV parameters, $M_1$ and $M_2$.  The neutralino mass difference, due to its different dependence on the fundamental parameters, provides important additional information to the fit. This underlines the importance of including this measurement in future analyses.

For the scenario dM1600, $M_2$ is now constrained 
in the $2.5$--$8.5$~TeV range, while $M_1 > 0$ in  the 1--6~TeV range. If $M_1 <0$ its value is confined in a small range, $-700\gev \lesssim M_1 \lesssim -400\gev$ for moderate values of $\tan\beta$, $8 \leq \tan\beta \leq 16$. Again, such a scenario could in principle be probed with an upgraded centre-of-mass energy. 
Also for dM770 one obtains a visible improvement, but still
with rather broad ranges of allowed values for $M_1$ and $M_2$; however, we only found valid solutions for $M_1 > 0$. In both cases, the contours for other 
values of $\tan\beta$ are mostly covered by the high $\tan\beta$ contour and are therefore not 
shown separately. This result is due to the weak dependence on $\tan\beta$ of the complete 
neutralino/chargino sector, cf.\ Section~\ref{sec:pheno}. Additional measurements, involving e.g.\ third-generation sfermions, would be needed to obtain meaningful constraints on $\tan\beta$. 

\section{Conclusions}
The current LHC results provide strong bounds on constrained supersymmetric models. In particular, the measured Higgs mass of
$m_h \approx 126$~GeV results in a tension with the naturalness argument. However,
in a specific class of scenarios, a rather light $\mu$-parameter in conjunction with a 
heavy coloured SUSY and gaugino spectrum can be well-motivated from a model-building perspective.
Such scenarios result in a strong mass degeneracy between the higgsino-like light
neutralinos and chargino, $\widetilde{\chi}^0_{1,2}$ and $\widetilde{\chi}^{\pm}_1$.
Due to the close mass degeneracies, three-body decays as well as
radiative decays are dominant leading to extremely challenging experimental
signatures.    

We therefore performed a simulation for the ILD detector in two different benchmark
scenarios at a centre-of-mass energy of $500\,$GeV, assuming an integrated luminosity of  $500\,$fb$^{-1}$ each for the two relevant polarisation configurations, which corresponds
to about $4$ years of ILC operation with design accelerator parameters.
In both scenarios, the $\mu$-parameter is around $160\,$GeV, 
but the SUSY gaugino parameters $M_1$ and $M_2$ are in the
 multi-TeV region. This leads to mass differences between the light chargino and 
 lightest neutralino of about $1.6\,$GeV and $770\,$MeV in the two benchmarks, respectively.
 These small mass differences lead to a very soft transverse momentum spectrum of the visible decay products, which ends at $5$ and $2\,$GeV for the two different scenarios. In this kinematic regime, 
 background processes from collisions involving the photon component of the electron and positron beams
 are about $5$ orders of magnitude larger than the signal. However, they can be heavily reduced by requiring the signal to be accompanied by
 a hard ISR photon. 
 
 In conjunction with the known beam energy, the $\widetilde{\chi}^{\pm}_1$ and $\widetilde{\chi}^0_{2}$ masses can be determined with a precision between $1$--$2\%$ from the 
 reduced centre-of-mass energy of the system recoiling against the photon. The mass difference with the LSP can be reconstructed by boosting the visible decay products into the rest frame of the produced gaugino pair. 
 This allows one to clearly resolve the lightest and next-to-lightest SUSY particles and to determine 
 the corresponding mass difference in the
two benchmarks with a relative accuracy of 17\% and 5\%, respectively. 
Finally, the polarised cross sections can be measured with an accuracy of a few percent.
The polarisation dependency, in particular the scaling factor between 
polarised and unpolarised cross sections, 
gives a clear indication of the higgsino nature.

In the last step of this analysis, we investigated whether the relevant SUSY parameters, namely 
the gaugino mass parameters  $M_1$ and $M_2$, the $\mu$-parameter and $\tan{\beta}$ can be
extracted from the experimental observables. We find that the
$\mu$-parameter can be determined to $\pm 4\%$. In case of $M_1$ and $M_2$, lower 
bounds in the multi-TeV range can be set, which depend slightly on the value
of $\tan{\beta}$, which cannot be fixed from these measurements alone. Additional solutions arise if one allows the bino mass parameter to be negative, $M_1 < 0$. Some of these, leading to a light $\wt \chi^0_3$, could be probed by direct production of the additional light neutralino.
If the measurement uncertainties could be reduced by a factor of 2, e.g.\ by
increasing the luminosity, or by improving the analysis, and if a measurement
of the mass difference between $\widetilde{\chi}^0_{2}$ and $\widetilde{\chi}^0_{1}$
could be performed with the same precision as for the chargino mass difference,
the constraints on $M_1$ and $M_2$ would be significantly more restrictive and become
less dependent on $\tan{\beta}$.

Our results demonstrate the strong physics potential of the ILC. High precision
is mandatory in order to resolve such challenging scenarios 
that could be easily missed at the LHC. The high experimental
precision of thr ILC of course has to be balanced by suitable higher
order predictions implemented in a Monte Carlo generator on a fully
differential level. First steps in this direction have already been
presented in~\cite{Kilian:2006cj,Robens:2006np}, but a full study
including decays would be needed and is in the line of future
work.

\section*{Acknowledgements}

We thank Frank G\"ade for providing the tracking efficiency histograms in presence of pair background. 
We thank the ILC Generator Group for providing the SM Whizard samples. We thank Aoife Bharucha for 
discussions on loop corrections. We thankfully acknowledge the support by the DFG through
 the SFB 676 ``Particles, Strings and the Early Universe''. This work has 
been partially supported by the MICINN, Spain, under contract FPA2010-17747; Consolider-Ingenio  
CPAN CSD2007-00042. We thank as well the Comunidad de Madrid through Proyecto HEPHACOS S2009/ESP-1473 
and the European Commission under contract PITN-GA-2009-237920.

%\section{Bibliography}

% ****************************************************************************
% BIBLIOGRAPHY AREA
% ****************************************************************************

\begin{footnotesize}
% IF YOU DO NOT USE BIBTEX, USE THE FOLLOWING SAMPLE SCHEME FOR THE REFERENCES
% ----------------------------------------------------------------------------

% ----------------------------------------------------------------------------

\end{footnotesize}

% ****************************************************************************
% END OF BIBLIOGRAPHY AREA
% ****************************************************************************

\end{document}